\documentclass[useAMS,usenatbib,a4]{mn2e}
\usepackage{amsmath,fleqn,graphicx,amssymb,epsfig}
\usepackage{multirow}
\renewcommand{\[}{\begin{equation}}
\renewcommand{\]}{\end{equation}}
\def\p{\partial}

\def\ex#1{\left\langle#1\right\rangle}
\newcommand{\ra}{\mathrm{a}}

\newcommand{\md}{\mathrm{d}}

\newcommand{\mi}{\mathrm{i}}

\usepackage[usenames]{color}
\usepackage[]{hyperref}

\definecolor{grey}{rgb}{0.75,0.75,0.75}
\definecolor{Orange}{rgb}{1.0,0.5,0.15}
\definecolor{brown}{rgb}{0.7,0.25,0.0}
\definecolor{pink}{rgb}{1.0,0.5,0.5}
\definecolor{darkerred}{rgb}{0.8,0,0}
\definecolor{darkerblue}{rgb}{0,0,0.8}
\definecolor{Blue}{rgb}{0,0.08,0.65}
\definecolor{Red}{rgb}{0.65,0.08,0.05}
\definecolor{Green}{rgb}{0.15,0.45,0.25}

\let\boldgrk=\gkvecten
\let\boldgrksc=\gkvecseven

\def\gkthing#1{{\mathchoice%
	{\hbox{{\boldgrk\char#1}}}
	{\hbox{{\boldgrk\char#1}}}
	{\hbox{{\boldgrksc\char#1}}}
	{\hbox{{\boldgrksc\char#1}}}}}

\def\vepsilon{\gkthing{15}}

\def\vtheta{\gkthing{18}}

\def\vxi{\gkthing{24}}

\def\rhop{\rho^{(p)}}
\def\Phip{\hat\Phi^{(p)}}\def\Phipp{\hat\Phi^{(p')}}

\def\Phips{\Phi^{(p)*}}
{\newif\ifnotend
\notendtrue
\def\veclist{ABCDEFGHIJKLMNOPQRSTUVWXYZabcdefghijklmnopqrstuvwxyz.}
\def\top#1#2.{#1}
\def\tail#1#2.{#2.}
\loop\expandafter\xdef\csname v\expandafter\top\veclist\endcsname%
{{\noexpand\bf\expandafter\top\veclist}}
\edef\veclist{\expandafter\tail\veclist}
\if\veclist.\notendfalse\fi\ifnotend\repeat}
\def\cE{{\cal E}}
\def\ra{{\rm a}}\def\rp{{\rm p}}\def\rt{{\rm t}}\def\d{{\rm d}}

\def\cF{{\cal F}}

\def\bolOm{\mbox{\boldmath$\Omega$}}
\def\vOmega{\bolOm}
\def\vDelta{\mbox{\boldmath$\Delta$}}

\def\e{\mathrm{e}}

\def\fracj#1#2{{\textstyle{#1\over#2}}}

\title[Revisiting  relaxation in globular clusters]
{Revisiting  relaxation in globular clusters}

\author[C. Hamilton, J-B. Fouvry, J. Binney and C. Pichon ]{
  Chris Hamilton$^{1,2}$\thanks{E-mail: ch783@cam.ac.uk},
Jean-Baptiste Fouvry$^3$\thanks{Hubble Fellow}, James Binney$^1$, Christophe Pichon$^{4,5,6}$\\
  $^1$Rudolf Peierls Centre for Theoretical Physics, 1 Keble Road,
  Oxford, OX1 3NP, UK\\
$^2$Department of Applied Mathematics and Theoretical Physics, University of
Cambridge, Wilberforce Road, Cambridge CB3 0WA, UK\\
$^3$Institute for Advanced Study, Einstein Drive, Princeton NJ 08540, USA\\
$^4$Institut d'Astrophysique de Paris and UPMC, CNRS (UMR 7095), 98 bis
Boulevard Arago, 75014, Paris, France\\
$^5$Korea Institute of Advanced studies (KIAS), 85 Hoegiro, Dongdaemun-gu,
Seoul, 02455, Republic of Korea\\
$^6$Institute for Astronomy, University of Edinburgh, Royal Observatory, Blackford Hill, Edinburgh, EH9 3HJ, UK
}

\begin{document}
\maketitle

\begin{abstract}
The classical theory of cluster relaxation is unsatisfactory because it
involves the Coulomb logarithm. The Balescu--Lenard (BL) equation provides a
rigorous alternative that has no ill-defined parameter. Moreover, the BL
equation, unlike classical theory, includes the cluster's self-gravity. A
heuristic argument is given that indicates that relaxation does not occur
predominantly through two-particle scattering and is enhanced by self-gravity. The BL
equation is adapted to a spherical system and used to estimate the flux
through the action space of isochrone clusters with different velocity
anisotropies. 
A range of fairly different secular behaviours is found depending on the fraction of radial orbits.
Classical theory is also used to compute the corresponding classical
fluxes. 
The BL and classical fluxes are very different because  (a)
 the classical theory materially under-estimates the impact of large-scale collectively amplified
fluctuations and (b)  only the
leading terms in an infinite sum for the BL flux are computed.
A complete theory of cluster relaxation likely requires that the sum in the
BL equation  be decomposed into a sum over a finite number of small wavenumbers
  complemented by   an integral over large wavenumbers
analogous to classical theory.
\end{abstract}


\section{Introduction}
\label{sec:intro}

As systems of almost coeval stars, globular clusters have been vital to the
development and testing of the theory of stellar structure and evolution. As
wonderfully clean dynamical systems that have enough stars to allow precise
statistical characterisation but few enough stars to evolve significantly by
stellar-dynamical relaxation within a Hubble time, their study has been
important for the development of stellar dynamics. Yet notwithstanding
intensive study for more than a century, aspects of these system remain
mysterious. In particular we don't understand how or where they formed.
Moreover, the efficiency with which gas must have been converted into stars
when they formed becomes more surprising as our knowledge of the general
star-formation process grows.

In this paper we argue that the dynamical evolution of globular clusters is
also less well understood than it should be with a critique of the standard
theory of two-body relaxation. Then we re-formulate, in the context of a
model globular cluster, the theory of the long-time (secular) evolution of a
stellar system using a sophisticated approach that has recently explained the
secular evolution of collisionless stellar discs. 

The conceptual foundation of the theory of secular evolution is that the system
evolves through a series of steady-state solutions of the collisionless
Boltzmann equation.  By Jeans' theorem, any such solution can be described by
a distribution function (DF) that depends on position and velocity
$(\vx,\vv)$ only through constants of stellar motion. Since we focus on
spherical systems, we know that the system's mean-field Hamiltonian
$H=\fracj12v^2+\Phi(\vx)$ admits three action integrals $J_i$, and we may
assume that at each time the DF takes the form $f(\vJ)$. On a timescale much
longer than the system's crossing time, small fluctuations of the potential
$\Phi$ around the slowly evolving mean spherical form for which the $J_i$ are the action
integrals cause the actions of stars to change, and thus the functional form
of $f$ to evolve.

The most obvious difference between the real potential $\Phi_N(\vx)$ and the
spherical approximation for which the $J_i$ are computed, is the local
singularities $\Phi_N(\vx)\sim -Gm/|\vx-\vx_i|$ near the location $\vx_i$ of
each star. The classical theory of relaxation~\cite[e.g.][chapter 7]{Binney2008} is
based on the idea that as a star moves through the cluster it is repeatedly
scattered by the singularities that lie along its path. Following
\cite{Chandra1943} one calculates as a
function of impact parameter $b$ and relative velocity the velocity change
$\delta\vv$ that an encounter causes, and then adds these changes by
integrating over impact parameters and relative velocities to obtain
diffusion coefficients for the Fokker-Planck equation that describes the
diffusion of stars through phase space~\citep{Chandra1949}:
\begin{align}
{\p f\over\p
t}=-{\p\over\p\vv}\cdot\biggl(\vDelta_1f-\fracj12{\p\over\p\vv}\cdot\Bigl[\vDelta_2f\Bigr]\biggr).
\end{align}
Here the first-order diffusion
coefficient $\vDelta_1$ is the average
$\ex{\delta\vv}$ of the $\delta\vv$, while the second-order diffusion
coefficient $\vDelta_2$ is obtained by averaging the $\delta\vv$ in quadrature. 

The calculation is made tractable by neglecting the acceleration caused by
the mean-field potential $\Phi$. That is, one assumes that in the absence of
a nearby singularity, the star's trajectory would be a straight line, and in
its presence it is a Keplerian hyperbola. This approximation is only valid when
the impact parameter $b$ is much less than the distance to the cluster's
centre.  The inconvenient truth is that the integral for $\ex{\delta\vv}$
over impact parameters diverges at large $b$, signalling that encounters with
distant stars contribute significantly to $\ex{\delta\vv}$.

Fortunately, the divergence is only logarithmic and the text-book work-around
for this conceptual difficulty is to cut the integral off at a distance
$b_{\rm max}$ comparable to the system's half-mass radius, and report $b_{\rm
max}$ through the value of the ``Coulomb logarithm'', which is the natural
logarithm of the ratio of $b_{\rm max}$ to the impact parameter that gives
rise to scattering by 90 degrees at the mean relative velocity. The venerable
age of this work-around should not blind us to its unsatisfactory nature: the
divergence of the integral for $\ex{\delta\vv}$ is no mathematical nicety but
reflects a real physical issue -- the significant contribution of
fluctuations in the potential generated by distant rather than nearby stars
\citep{Weinberg1993}.
Clearly these stars do not cause Rutherford-like scattering, and they do not
act in isolation. In so far as long-range encounters contribute significantly
to the secular evolution of the system, the classical
formulation of the problem is flawed.

The purpose of this paper is first to argue that the theory of the secular
evolution of globular clusters should be reformulated,  second
to explain what the reformulated theory looks like, and finally to present a
worked example of the reformulated theory in action. In Section
\ref{sec:Nfluct} we give an order-of-magnitude calculation that implies that
relaxation is driven not by individual encounters but by fluctuations in the
number of particles in different sub-regions of the system. This conclusion
suggests that the self-gravity of the system, which is completely neglected
in the traditional approach, can play a significant role. In Section
\ref{sec:BL} we give the governing equations and explain their physical
content.  Section~\ref{sec:sphere} deals with the application of these
equations to spherical systems.  Sections~\ref{sec:BL} and~\ref{sec:sphere}
are rather technical in nature and readers may like to skip straight from
Section~\ref{sec:Nfluct} to Section~\ref{sec:isochrone}, which examines the
diffusive fluxes in a family of clusters that have the spatial structure of
Henon's isochrone, but have differing degrees of velocity anisotropy. We show
that the flux computed from the BL equation is larger when self-gravity is
included than it is otherwise. We explore the extent to which the BL flux
grows as the velocity anisotropy increases towards the value at which the
cluster is prone to the radial-orbit instability.
In Section~\ref{sec:compare}, we compare the BL flux to the classical
theory. We discuss these results in Section~\ref{sec:discuss},
and conclude in Section~\ref{sec:conclude}.

\section{What drives secular evolution}
\label{sec:Nfluct}

A simple back-of-the-envelope calculation confirms the importance of
large-scale fluctuations that fall outside the scope of a Rutherford
scattering calculation.  We consider a system of mass $M$ with $N$ stars and
characteristic scale $R$, in which the characteristic internal speed is
$\sigma=\sqrt{GM/R}$. A subregion of size $r=xR$ contains mass $M_r\simeq
x^3M$ and $n\simeq x^3N$ stars, so on account of Poisson noise $M_r$
fluctuates by $\delta M_r=M_r/\surd n=x^{3/2}M/\surd N$ during times $\delta
t=r/\sigma$.  Consider a point that is distance $yR$ from our subregion. At
this point a single fluctuation in the subregion's gravitational attraction
will change the velocity of a test star by
\[\label{eq:dvPoisson}
\delta v={G\delta M_r\over (yR)^2}\delta t={GMx^{3/2}\over (yR)^2 \surd
N}{xR\over\sigma} ={\sigma x^{5/2}\over y^2\surd N}.
\]
 This formula states that for given $y$, large volumes $x\simeq 1$ perturb
$v$ very much more strongly than small volumes $x\ll1$. Against this trend we
must bear in mind that (a) $y\ge x$, (b) as $x$ decreases the number of
subregions perturbing increases as $x^{-3}$, and (c) the time within which
the contribution (\ref{eq:dvPoisson}) comes about decreases with $x$, so in a
given time each small subregion makes many more contributions to $v$ than
does a large subregion.

For the moment we assume that the contributions to $v$ from different subregions are
statistically independent, so it's appropriate to add the $\delta v$ in
quadrature. There are $\sim4\pi (y/x)^2$ subregions of scale $x$ that are
distance $yR$ from our point, and in a
global crossing time $t_{\rm cross}=R/\sigma$ each such subregion contributes
$x^{-1}$ times. So in a crossing time all these subregions
change $v^2$ by
\[
(\Delta v)^2=4\pi{y^2\over x^3}(\delta v)^2=4\pi{\sigma^2x^2\over y^2N}.
\]
 Now we have to sum over  $y=x,2x,3x,\dots,1$. We convert the sum to an
 integral using $\d y=x$ and have
\[
\sum{1\over y^2}\simeq{1\over x}\int_x^1{\d y\over y^2}
={1\over x}\left({1\over x}-1\right)\simeq{1\over x^2}.
\]
 Hence in a crossing time the subregions of scale $x$ change $v^2$ by 
\[\label{eq:dvsq}
(\Delta v)^2\simeq4\pi\sigma^2/N.
\]
 Remarkably, this is independent of $x$, so by this reckoning regions of each
scale $xR$ contribute equally to changing $v^2$. 

At this point it is
important to consider correlations between the fluctuations in subregions. On
first examining a cluster, the fluctuations between the masses in different
subregions characterised by a given value of $x$ will be negligible on all
but the largest scales: the basis of Poisson statistics is the selection of
items from an essentially infinite pool of objects and this model will begin
to fail only when the number of items selected, $x^3N$, becomes comparable to
$N$. When $x\sim1$ and the assumption of statistical independence fails, it
does so because fluctuations in one half of the cluster are {\it
anti}-correlated with fluctuations in the other half. Consequently, as
$x\to1$ we should be adding  values of $\delta\vv$ {\it linearly} rather
than quadratically because the downward fluctuation in one hemisphere pushes
a test star in the same direction as the upward correlation in the other
hemisphere. Hence, by neglecting correlations between subregions we are
merely under-estimating the importance of large scales relative to small scales.

We must also consider temporal fluctuations: since an over-density in one
subregion must run out through its neighbours, the overdensity in adjacent
cells of a given scale are temporally correlated. Our calculation allows for
this correlation by taking the persistence time of an overdensity to be
$\delta t=xR/\sigma$, but to incorporate this insight into our final value
for $(\Delta v)^2$ we must add the contributions of all relevant scales $x$:
the contribution from the smallest values of $x$ are based on a picture in
which after a time $\delta t=xR/\sigma$ mass is reassigned to cells from some
fictional outside reservoir rather than by passing mass between adjacent
cells. Only by including independently the contributions to $(\Delta v)^2$
from larger values of $x$ do we arrive at a value that reflects temporal
correlations between over-densities. Since the number of scales between the
smallest one, $x_{\rm min}$ and the whole cluster is $-\ln x_{\rm min}$, we
multiply equation (\ref{eq:dvsq}) by $-\ln x_{\rm min}$.
The smallest subregion it's sensible to consider cannot
be smaller than a decent multiple of the inter-particle distance $\sim
R/N^{1/3}$. When it is equal to this distance, we have
\[\label{eq:DVsq}
(\Delta v)^2_{\rm t_{\rm cross}}\simeq{4\pi\sigma^2\ln N\over 3N}.
\]
 The {relaxation time} is the time required for fluctuations to change
any velocity by order of itself, thus for $(\Delta v)^2$ to accumulate to
$\sigma^2$. From equation (\ref{eq:DVsq}) it follows that
\[\label{eq:tRelax}
t_{\rm relax}\simeq {N\over4\ln N}t_{\rm cross}.
\]
This result is essentially identical with equation (7.108) for the half-mass
relaxation time in \cite{Binney2008}.

In an ideal gas the number of molecules in a given volume is given by Poisson
statistics as was assumed above, and the time evolution of fluctuations can
be computed by considering them to arise from thermally excited sound waves.
The self-gravity of a stellar system makes the system more compressible on
large scales than on small scales, where self-gravity is unimportant and an
ideal gas provides a valid model. Gravity reduces the energy density
associated with a sound wave of given amplitude -- as the wave's wavelength
rises towards the Jeans length, the energy density falls to zero. Hence, when
thermally excited, waves with longer wavelengths will have larger amplitudes
than Poisson statistics would imply.  It follows that in a real system
fluctuations near the size of the system (the Jeans length) dominate,
contrary to our finding above of equal contributions from all scales
\citep{Weinberg1993}. 

In summary, we can recover the traditional formula for the ``two-body''
relaxation time by computing the effects of Poisson fluctuations in the mass
density within the system. Given that self-gravity makes a gas more
compressible than a classical ideal gas, on the largest spatial scales, we
expect the actual density fluctuations in a cluster to have larger amplitudes
than simple shot noise predicts. It follows that the classical relaxation
time is too long by a factor that depends on the amount by which self-gravity
enhances fluctuations.

\cite{Weinberg1993} showed that popular models of clusters have a weakly
damped mode in which the core oscillates along some line in antiphase to
the envelope that surrounds it. The physics of this dipole mode, in which the
core exchanges momentum with the halo, is heavily dependent on the fact that the
cluster generates its own gravitational field rather than being confined with
an external field from which it could draw momentum at will. In
Section~\ref{sec:isochrone} we will show that in the self-gravitating case
a cluster's relaxation is significantly accelerated as a result of this mode
being stimulated by Poisson noise.
 
Clusters with strongly radially biased velocity dispersions are subject to
the radial-orbit instability
\citep[][\S5.5.2]{Fridman1984,PalmerPap1987,Binney2008}. In such a cluster a
quadrupolar distortion of the cluster's initial spherical shape grows
exponentially. In essence, when pressure in the tangential directions falls
below a certain threshold, the cluster becomes Jeans unstable in these
directions. A cluster with a less radially biased velocity distribution may
not be Jeans unstable but the frequencies of quadrupolar modes will be low
because the system is highly compressible in the tangential directions
\citep{MayB1986}. In such a case we have to expect enhanced stellar diffusion
in angular momentum and a shorter relaxation time than~\eqref{eq:tRelax}. In
Section~\ref{sec:isochrone} we quantify this prediction.

\section{The Balescu--Lenard equation}
\label{sec:BL}

Over the last 20 years an approach to the secular evolution of stellar systems that
differs radically from that of Spitzer--Chandrasekhar has emerged.
In an important series of papers Weinberg demonstrated the significance of
self-gravity in the responses of stellar systems to perturbation.
\cite{Weinberg1989} showed that the time required for dynamical friction to
drag a satellite to the centre of its host galaxy is 2--3 times longer when
self-gravity is included in the calculation than when it is not. This conclusion was obtained by extending to spherical systems
the approach to stellar dynamics based on angle-action coordinates and
orthonormal potential-density pairs that~\cite{Kalnajs1976} introduced in the
context of discs.  This apparatus enabled for the first time a satisfactory
treatment of the large-scale distortions of a stellar system.
\cite{Weinberg1991} went on to determine the frequencies of the fundamental
modes of anisotropic stellar systems and thus to determine the level of
radial anisotropy at which the radial-orbit instability sets in. Using the
same apparatus~\cite{Weinberg1994} showed that popular models of globular
clusters have very weakly damped modes, a result that will prove crucial
below. 

Whereas in his previous papers the focus had been on the impact of an
externally applied perturbation,
\cite{Weinberg1993} for the first time
discussed the impact of Poisson noise.  He did so in the context of the
periodic cube of stars, and for this idealised system he was able to derive a
well defined collision integral for the Boltzmann equation. By varying the
ratio $r$ of the Jeans length to the size of the cube, he could show that
self-gravity, by amplifying the excitation of large-scale fluctuations by Poisson
noise, dramatically accelerates the system's relaxation when $r$ is close to
unity.
\cite{Weinberg1998} extended this analysis from the periodic cube to
real systems. Using angle-action coordinates and orthonormal
potential-density pairs he computed the dressed response of the system to a
single particle, and the mean energy invested in thermally excited low-order
modes.

\cite{Heyvaerts2010}, in the spirit of Weinberg's analysis of the periodic
cube, approached the problem of secular evolution by returning
to the BBGKY hierarchy of equations for the $n$-particle distribution
functions $f^{(n)}$ that one obtains by integrating the $6N$ dimensional
Liouville equation for the dynamics of an N-particle system over the
coordinates of all but the first $n$ particles. Working in angle-action
coordinates to order $1/N$, and taking full account of the system's
self-gravity, he obtained an expression for the 2-particle distribution
function $f^{(2)}$. Inserting  this into the BBGKY
equation for the 1-particle DF, he obtained a Fokker-Planck equation together
with expressions for its diffusion coefficients. This equation resembles an
equation derived by~\cite{Balescu1960} and~\cite{Lenard1960} for the secular
evolution of an electrostatic plasma, and in their honour we shall refer to it
as the `BL equation'.  In Heyvaerts' derivation of the BL equation there is no
suggestion of particles scattering one another.  Instead, particles 1 and 2
interact when their frequency vectors $\vOmega_i$ satisfy a resonance condition
$\vn_1\cdot\vOmega_1+\vn_2\cdot\vOmega_2=0$, where the $\vn_i$ are vectors
with integer components. 

\cite{Rauch1996} introduced the concept of `resonant relaxation' in the
context of star clusters that are dominated by a central massive black hole.
In these systems the mean potential is nearly Keplerian, so each star can be
replaced by the elliptical, Gaussian wire one obtains by averaging its motion
around a Kepler orbit. The wires apply torques to each other, with the
consequence that they precess. Unless the precession frequencies are closely
matched, differential precession causes the sign of each angular momentum
exchange to change quite rapidly, and hence average to zero.
Consequently, the important interactions are those between stars that have
commensurable precession frequencies -- hence the term \textit{resonant} relaxation.
Heyvaerts' work implies that similar physics applies to any stellar system.

\cite{Chavanis2012} derived the same equation as Heyvaerts by a different,
perhaps more physically intuitive, route that does not involve the BBGKY
hierarchy. He expressed the 1-particle DF as a sum of a part $f_0$ that
evolves only on a secular timescale, and a fluctuating part $f_1$ that
averages to zero on an orbital timescale. The potential was similarly
decomposed into mean and fluctuating parts, $\Phi_0$ and $\Phi_1$. Then he
could easily show that $f_0$ satisfies
\[\label{eq:beforeBL}
{\p f_0\over\p t}=-\ex{[f_1,\Phi_1]},
\]
 where $[.,.]$ is a Poisson bracket and $\ex{..}$ is an ensemble average.
Equations for $f_1$ and the potential $\Phi_1$ to which it gives rise are
readily found, and Chavanis solved them for initial conditions characteristic
of Poisson statistics.  When the solutions are inserted into equation~\eqref{eq:beforeBL},
the BL equation is recovered. 

The BL equation can be written
\[
{\p f_0(\vJ,t)\over\p t}=-{\p\over\p\vJ}\cdot\vF,
\]
 where $\vF(\vJ)$ is the flux of stars through action space. This flux falls naturally
into two parts:
\[\label{eq:FfromD}
\vF(\vJ)=-\vD_1(\vJ)f_0(\vJ)-\vD_2(\vJ)\cdot{\p f_0\over\p\vJ},
\]
where for brevity we have suppressed the slow time dependence of all
quantities. The term proportional to the first-order diffusion coefficient $\vD_1$
describes dynamical friction: the constant drift back to lower actions and
lower energy that in thermal equilibrium perfectly balances the stochastic
drive away from the origin of action space. The second-order diffusion
coefficient $\vD_2$ is analogous to an anisotropic thermal conductivity,
which permits heat to diffuse from hotter (large $f_0$) to colder regions at
a particular speed in each direction. If $\vD_1$ vanished, the only
steady-state solution would be $f=0$, which would be established after the
term proportional to $\vD_2$ had driven stars out into the infinite
phase-space volume that exists at large $|\vJ|$.

Heyvaerts' derivation of the BL equation explicitly focuses on two-particle
effects in as much as he starts by solving for the function that describes
two-particle correlations. Chavanis' derivation arrives at a picture in which
secular evolution occurs through interactions between pairs of resonantly
coupled particles as a consequence of the mathematical accident that the
collision term on the r.h.s. of equation~\eqref{eq:beforeBL} is non-vanishing
only because the fluctuations in the DF and in the potential are correlated.
In Chavanis' derivation the conceptual focus is on the impact of fluctuations
in density and potential, and such fluctuations are by no means confined to
pairwise interactions.
The system's self-gravity plays a role because the fluctuating part of the
DF, $f_1$ evolves in a potential to which it contributes through Poisson's
equation.

In detail, with $f_0$ normalised such that
\[
(2\pi)^3\int\d^3\vJ\,f_0=N\mu,
\]
 with $\mu$ the mass of a single particle and $N$ the number of particles, we have the
following expressions for the diffusion coefficients
\begin{align}\label{eq:DiffCoeffs}
\vD_1(\vJ)&=-\fracj12(2\pi)^4\mu\sum_{\vn\vn'}
\int\d^3\vJ'
\bigl|E_{\vn\vn'}(\vJ,\vJ',\vn\cdot\vOmega)\bigr|^2\cr
&\times 
\vn'\cdot{\p
f_0\over\p\vJ'}\delta(\vn'\cdot\vOmega'-\vn\cdot\vOmega)\,\vn,\cr
\vD_2(\vJ)&=\fracj12(2\pi)^4\mu\sum_{\vn\vn'}
\int\d^3\vJ'
\bigl|E_{\vn\vn'}(\vJ,\vJ',\vn\cdot\vOmega)\bigr|^2\cr 
&\times f_0(\vJ')\delta(\vn'\cdot\vOmega'-\vn\cdot\vOmega)\,\vn\otimes\vn.
\end{align}
 In these formulae $\vn,\vn'$ are 3-vectors with
integer components, and $\vOmega\equiv\vOmega(\vJ)$ is the vector formed by
the frequencies of the star with actions $\vJ$. Analogously
$\vOmega'\equiv\vOmega(\vJ')$. The occurrence of Dirac delta functions in
equations~\eqref{eq:DiffCoeffs} tells us that diffusion is possible only to
the extent that stars have resonating frequencies: the rate at which stars
with actions $\vJ$ diffuse is proportional to a sum over all resonating
stars, regardless of their location within the system. The impact of self
gravity is encoded in the complex \textit{susceptibility coefficients}
$E_{\vn\vn'}(\vJ,\vJ',\omega)$. 
To explain the content of the susceptibility coefficients
we have to introduce additional apparatus in the next section.

On account of the extensive similarities between the expressions for $\vD_1$
and $\vD_2$, the action-space flux of stars~\eqref{eq:FfromD} can be written
in a relatively compact form
\begin{align}\label{eq:Ffromff}
\vF&(\vJ)=\fracj12(2\pi)^4\mu\sum_{\vn\vn'}\vn
\int\d^3\vJ'
\bigl|E_{\vn\vn'}(\vJ,\vJ',\vn\cdot\vOmega)\bigr|^2\cr
&\!\times\! 
\biggl(\vn'\!\cdot{\p\over\p\vJ'}-\vn\cdot{\p\over\p\vJ}\biggr)f_0(\vJ)f_0(\vJ')
\delta(\vn'\!\cdot\vOmega'-\vn\cdot\vOmega).
\end{align}
 The factor $\mu$ before the summation sign above signals that the flux is
 proportional to $1/N$.

\subsection{Potential-density pairs}

To solve Poisson's equation for the fluctuating potential $\Phi_1$ given the
fluctuating DF $f_1$ one makes use of orthonormal potential-density pairs
\citep{Kalnajs1976}.
Let $(\Phi^{(p)},\rho^{(p)})$ be such a pair. Then
\[
\nabla^2\Phi^{(p)}(\vx)=4\pi G\rho^{(p)}(\vx)\,,
\]
and
\[\label{eq:PDnorm}
\int\d^3\vx\,(\Phi^{(q)})^*\rho^{(p)}=-\cE\delta_{pq},
\]
 where $(\Phi^{(q)})^*$ is the complex conjugate of $\Phi^{(q)}$ and $\cE$ is a convenient constant with the dimensions of energy. 
These pairs make it straightforward to compute the potential generated by
any given density distribution:
\[
\rho_1(\vx,t)=\sum_p A_p(t)\rhop(\vx)
\]
implies that
\[
\Phi_1(\vx,t)=\sum_p A_p(t)\Phi^{(p)}(\vx),
\]
with
\[
A_p(t)=-{1\over\cE}\int\d^3\vx\,\Phi(\vx)^*\rho_1(\vx,t).
\]

When
expressed in action-angle variables, the basis functions depend on all
six coordinates. We work with their Fourier transforms with respect to the
angle variables, which are defined by
\begin{align}
f_1(\vtheta,\vJ,t)&=\sum_\vn\hat
f_1(\vn,\vJ,t)\e^{\mi\vn\cdot\vtheta},\cr 
\Phi_1(\vtheta,\vJ,t)&=\sum_\vn\hat
\Phi_1(\vn,\vJ,t)\e^{\mi\vn\cdot\vtheta}.
\end{align}
A straightforward calculation starting from the linearised collisionless
Boltzmann equation shows  that the Laplace transform 
\[
\widetilde f_1(\vn,\vJ,\omega)=\int_0^\infty\d t\,\hat f_1(\vn,\vJ,t)\e^{\mi\omega t}
\]
of $\hat f_1(\vn,\vJ,t)$ is given by
\[\label{eq:firstftoPhi}
\widetilde f_1(\vn,\vJ,\omega)=  {\displaystyle \vn\cdot{\p
f_0\over\p\vJ}\widetilde\Phi_1(\vn,\vJ,\omega)-\mi\hat f_1(\vn,\vJ,0)
\over \vn\cdot\vOmega-\omega}.
\]
This equation gives the dynamical relationship between an initial condition
$\hat f_1(\vn,\vJ,0)$ and both the disturbances to the DF and the potential that it
provokes.  

The disturbed DF and potential are also linked by Poisson's
equation $\nabla^2\Phi_1=4\pi G\int\d^3\vv\,f_1$.  We multiply equation~\eqref{eq:firstftoPhi}
by $\Phips\sum_\vn\e^{\mi\vn\cdot\vtheta}$ and integrate over phase space.
Then $\widetilde A_p(\omega)$, the Laplace  transform that describes $\Phi_1$,
appears on both sides. Gathering on the left the terms proportional to 
$\widetilde A_p(\omega)$, we obtain the matrix equation
\begin{align}\label{eq:firstA}
\sum_{p'}\epsilon_{pp'}(\omega)\widetilde A_{p'}(\omega)
=\mi{(2\pi)^3\over\cE}\!\int\!\d^3\vJ\sum_\vn
{\hat f_1(\vn,\vJ,0)[\Phip(\vn,\vJ)]^*
\over \vn\cdot\vOmega-\omega},
\end{align}
 where
\[\label{eq:defseps}
\epsilon_{pp'}(\omega)\equiv\delta_{pp'}-M_{pp'}
\]
with
\[\label{epsilon}
M_{pp'}\equiv{(2\pi)^3\over\cE}\!\int\!\d^3\vJ\sum_\vn
{\vn\cdot{\p f_0\over\p\vJ}\over \vn\cdot\vOmega-\omega}
[\Phip(\vn,\vJ)]^*\Phipp(\vn,\vJ).
\]
The matrix of susceptibility coefficients $E_{\vn\vn'}(\vJ,\vJ',\omega)$
that appears in equations~\eqref{eq:DiffCoeffs} for the diffusion coefficients is the inverse of the
matrix $\vepsilon(\omega)$ when the latter is written in the $(\vn,\vJ)$
coordinate system:
\[\label{eq:defE}
E_{\vn\vn'}(\vJ,\vJ',\omega)
\equiv{1\over\cE}\sum_{pp'}\Phip(\vn,\vJ)\vepsilon^{-1}_{pp'}(\omega)
[\Phipp(\vn',\vJ')]^*,
\]
It allows us to obtain from equation~\eqref{eq:firstA} the key relation
\begin{align}\label{eq:runE}
\widetilde\Phi_1(\vn,\vJ,\omega)&=
(2\pi)^3
\mi\int\!\!\d^3\vJ' \sum_{\vn'} E_{\vn\vn'}(\vJ,\vJ',\omega)
{\hat f_1(\vn',\vJ',0)\over \vn'\cdot\vOmega'-\omega}. 
\end{align}

\subsection{Role of self-gravity}

There are two terms in the numerator of equation~\eqref{eq:firstftoPhi} for
the disturbance to the DF. The first describes the impact of
self-gravity, and disappears in the limit $G\to0$, while the second describes the 
evolution of the initial disturbance in the mean-field potential. The former term
gives rise to the `response matrix' integral $M_{pp'}$ in the definition~\eqref{eq:defseps} of
$\vepsilon$. In the
limit $G\to0$, the integral's prefactor $\cE^{-1}\sim G^{-1}$ diverges, but
each of the potential basis functions in the integrand vanishes like $G$, so
the self-gravitating response vanishes with $G$ as we expect.  We can switch off self-gravity in the BL equation by simply putting $M_{pp'}=0$, thereby recovering the inhomogeneous Landau equation \citep{Chavanis2013}.

According to equation~\eqref{eq:runE}, $E_{\vn\vn'}(\vJ,\vJ',\omega)$ gives
the change in the potential at frequency $\omega$ that is provoked by a change in
the initial conditions.   Schematically we have an equation for the potential response of the form $\mathcal{L} \widetilde\Phi_1 = \mathcal{S}$, with $\mathcal{L}(\omega)$ a linear operator and $\mathcal{S}$ a source term.  If the system has a normal mode at frequency $\omega$ (that is, a solution to $\mathcal{L}(\omega) \widetilde\Phi_1 = 0$) there isn't a unique relationship between the response of the system and the driving perturbation $\mathcal{S}$, because to any
solution of the governing equation we can add a multiple of the normal mode.  It follows that when $\omega$ is an eigenfrequency of the system, $\vepsilon$ is a singular matrix.  Its inverse $\vE$ therefore diverges as $\omega$ tends to the frequency of the normal mode.

The sums over $\vn$ and $\vn'$ in equation~\eqref{eq:Ffromff} suggest
that diffusion is simply driven by pairs of resonating particles. This is a
serious over-simplification, however, because the contribution of each pair
is proportional to $|E_{\vn\vn'}|^2$.  This factor is large if the frequency at which
the pair is communicating lies near an eigenfrequency of the whole system.
One star perturbs a star at some distance from it by exciting oscillations in
the entire cluster. If the cluster is responsive at the frequency in
question, the two resonating stars can communicate effectively and rather
quickly exchange significant energy and angular momentum, so we have rapid
diffusion through phase space.  In a tepid stellar disc the impact of the
$|E_{\vn\vn'}|^2$ factor can be enormous, because the first star may launch a leading
spiral wave, which is swing amplified as it is reflected off the forbidden
zone around corotation before it is absorbed by the second star near the
wave's inner Lindblad resonance.
\cite{Fouvry2015b} showed that on account of
swing amplification the relaxation time in a realistically cool disc can be
$\sim1000$ times shorter than one predicts when swing amplification is
neglected. Below we show that self-gravity can significantly
shorten the relaxation time of a non-rotating spherical system.

\section{Application to spherical systems}
\label{sec:sphere}

In the case of a spherical system, convenient action-space coordinates are
the radial action $J_r$, the total angular momentum $L$ and the component
$L_z$ of angular momentum parallel to the $z$ axis.  We confine ourselves to
systems with no special axis; that is, we consider systems for which the DF depends
only on $J_r$ and $L$, so $f_0$ is independent of $L_z$, which at given $L$
merely encodes the inclination of the angular momentum vector with respect to
the $z$ axis.

Before the foregoing apparatus can be applied to any spherical system, a
conceptual difficulty has to be finessed.  The problem is that in a spherical
system stars have only two independent frequencies. That is, there is a
one-parameter family of vectors $\vn$ such that $\vn\cdot\vOmega=0$ $\forall\
\vJ$.\footnote{In the Kepler potential of a massive black hole there is only
\textit{one} independent frequency and this problem is still more acute
\citep{Fouvry2017}.} This is a significant issue for the derivation of the BL
equation since the derivations of~\cite{Heyvaerts2010} and
\cite{Chavanis2012} assume that in each patch of action space the frequencies
$\Omega_i$ can be used as coordinates instead of the actions $J_i$, and when
the frequencies are systematically  degenerate this is not the case. It is
also a problem as regards equations~\eqref{eq:DiffCoeffs} for the diffusion
coefficients. Indeed, in the $(J_r,L,L_z)$ coordinate system, any vector
$\vn=(0,0,n)$ yields $\vn\cdot\vOmega=0$ because $\Omega_3=\p H/\p L_z=0\
\forall\ \vJ$.  Physically, the vanishing of $\Omega_3$ expresses the fact
that in a spherical potential orbital planes do not precess. Mathematically
vectors of the form $\vn=(0,0,n)$ are problematic because they make the
arguments of the Dirac delta functions vanish throughout action space, so the
corresponding integral over actions becomes undefined.  

To resolve the above problem, we define the 2-vector
\[
\widetilde\vJ\equiv(J_r,L).
\]
We similarly define integer 2-vectors $\widetilde\vn=(n_1,n_2)$ and
frequency 2-vectors $\widetilde\vOmega=\p H/\p\widetilde\vJ$ and observe that
\[\label{eq:tildeOmega}
\vn\cdot\vOmega=\widetilde\vn\cdot\widetilde\vOmega
\]
 because $\Omega_3=0$.  It will be convenient to quantify $J_3=L_z$ through the
inclination $\beta$ defined by
\begin{equation}
\cos \beta \equiv \frac{J_3}{J_2}\quad(0\leq \beta \leq \pi) .
\end{equation}
In terms of $\beta$ the action-space volume element is 
\begin{align}
\label{volelem}
\d^3\vJ=\d J_r\, L\d L\,\d(\cos\beta)=\d^2\widetilde\vJ\,J_2\,\d(\cos\beta).
\end{align}

In the spherical case, it is natural to take the potential-density pairs to
be proportional to spherical harmonics $Y_\ell^m$. Thus we write\footnote{The indices on the spherical harmonics run over $\ell = 0, 1, 2,...$ and $m=-\ell, -\ell+1, ..., \ell-1, \ell$.} 
\begin{align}
\label{mercury}
\Phi^{(p)}(\vx) \equiv \Phi_{\ell m n}(r,\vartheta,\phi) 
&= Y_\ell^m(\vartheta, \phi)\,U_n^\ell(r),
\nonumber\\
\rho^{(p)}(\vx) \equiv \rho_{\ell m n}(r,\vartheta,\phi) 
&= Y_\ell^m(\vartheta, \phi)\,D_n^\ell(r),
\end{align}
where the ${(U_n^\ell, D_n^\ell)}$ are real radial functions.  Several
authors~\citep[e.g.][]{Clutton-Brock1973,Weinberg1989,Hernquist1992,Rahmati2009}
have proposed choices for $U_n^\ell(r)$. In Section~\ref{sec:isochrone}, we give an
explicit example of such basis functions.  Then one can show
\citep[see][]{Tremaine1984} that
\begin{equation}
\widehat{\Phi}^{(p)}(\vn,\vJ) \!=\!
\delta_{m^p}^{n_3}\,\mi^{m^p-n_2}Y_{\ell^p}^{n_2}(\pi/2,0)\,
R_{n_2m^p}^{\ell^p}(\beta)\,W_{\ell^pn^p}^{\widetilde\vn}(\widetilde\vJ),
\nonumber
\end{equation}
where\footnote{In equation~\eqref{eq:defsR} the sum over $t$ is restricted
such that all arguments of the factorial operators are $\ge0$.}
\begin{align}\label{eq:defsR}
R_{n_2m}^{\ell}(\beta) =&\sum_t  (-1)^t \frac{\sqrt{(\ell+n_2)!(\ell-n_2)!
(\ell+m)!(\ell-m)!}}{(\ell-m-t)!(\ell+n_2-t)!t!(t+m-n_2)!}
\nonumber\\
& \times [\cos(\beta/2)]^{2\ell+n_2-m-2t}[\sin(\beta/2)]^{2t+m-n_2}
\end{align}
is the spin-$\ell$ Wigner rotation matrix, and
\begin{equation}
\label{wmat}
W_{\ell n}^{\widetilde\vn}(\widetilde\vJ) \!=\! \frac{1}{\pi} \!
\int_0^\pi \!\!\d\theta_1 U_{n}^{\ell}(r(\theta_1)) \cos \left[ n_1
\theta_1 \!+\!n_2(\theta_2\!-\!\psi)\right].
\end{equation}
Here $\psi$ is the angular coordinate within the orbital plane. It depends
on both $\theta_1$ and $\theta_2$, but the difference $\theta_2-\psi$ is a
function of $\theta_1$ alone, so the integral is well defined.

When $f_0$ is independent of $L_z$, the only $\beta$
dependence in the response matrix from equation~\eqref{epsilon} comes from
the rotation matrices $R$.  Given the orthogonality of these matrices when
integrated over ${ \cos \beta }$~\citep[e.g.][]{Edmonds1996}
\begin{equation}\label{eq:Rorthog}
\int_{-1}^1 \!\!\!\! \d(\cos \beta') \, R^{\ell^p}_{n_2n_3}(\beta') \,
R^{\ell^q *}_{n_2n_3}(\beta') = \frac{2 \delta_{\ell^p}^{\ell^q}}{2 \ell^p + 1},
\end{equation}
we can write the response matrix as
\[
\label{M} M_{pq}(\omega) =
\delta_{\ell^p}^{\ell^q}\delta_{m^p}^{m^q} 
\xi_{\ell^p n^pn^q}(\omega),
\]
where
\begin{align}\label{eq:gives_xi}
\xi_{\ell^p n^pn^q}&(\omega)\equiv {(2\pi)^3\over\cE} \!\!\!\!\!\! \sum_{\substack{n_1 \\
|n_2|\leq \ell^p \\ (\ell^p - n_2) \,\mathrm{even}}} \!\!\!\!\!\!\!
 \frac{2}{2\ell^p + 1} \left| Y_{\ell^p}^{n_2}(\pi/2,0)\right|^2
\nonumber
\\
& \times \!\int\! \d^2 \widetilde{\vJ} \, L \,
\frac{\widetilde{\vn}\cdot\partial f_0/\partial
\widetilde{\vJ}}{\omega-\widetilde{\vn}\cdot \widetilde  {\vOmega}}\,
W^{\widetilde{\vn}}_{\ell^pn^p}(\widetilde{\vJ})
\,W^{\widetilde{\vn}}_{\ell^pn^q}(\widetilde{\vJ}).
\end{align}
It follows that the matrix $\vepsilon=\vI-\vM$ defined by equation~\eqref{eq:defseps} is
\[\label{eq:gives_eps}
\epsilon_{pq} = \delta_p^q - \delta_{\ell^p}^{\ell^q}
\delta_{m^p}^{m^q} \,  \xi_{\ell^p n^p n^q}(\omega).
\]
Then putting ${ \delta_p^q = \delta_{\ell^p}^{\ell^q}  \delta_{m^p}^{m^q}\delta_{n^p}^{n^q} }$ we arrive at
\begin{align}
\label{clearthen}
\epsilon^{-1}_{pq} = \delta_{\ell^p}^{\ell^q}\delta_{m^p}^{m^q} N_{\ell^p n^p n^q}(\omega),
\end{align}
where ${ N_{\ell^p n^p n^q}(\omega) }$ is the inverse of ${ [\delta_{n^p}^{n^q}- \xi_{\ell^p n^p n^q}(\omega)] }$.

We insert this expression for $\vepsilon^{-1}$ into equation~\eqref{eq:defE}
for the susceptibility coefficients and use the reality of  $R$, $W$ and ${
Y_\ell^m(\pi/2,0) }$, to obtain
\begin{align}
E_{\vn\vn'}(\vJ&,\vJ',\omega) 
= {1\over\cE}\sum_{ \substack{\ell^p n^p\\ \ell^q n^q}}\sum_{\substack{|m^p|\leq\ell^p\\  
  |m^q|\leq \ell^q}} \delta_{m^p}^{n_3}\delta_{m^q}^{n'_3}
\delta_{\ell^p}^{\ell^q} \delta_{m^p}^{m^q}
\nonumber
\\
&\times \left[ \mi^{(n_2-n'_2)} \, \mi^{(m^p-m^q)}\right] 
 \left[Y_{\ell^p}^{n_2}(\pi/2,0) \, Y_{\ell^q}^{n'_2} (\pi/2,0)  \right]
\nonumber
\\
&\times \left[R^{\ell^p}_{n_2m^p}(\beta) \, R^{\ell^q}_{n'_2m^q}(\beta')\right]
\nonumber
\\
&\times \left[W^{\widetilde{\vn}}_{\ell^pn^p}(\widetilde{\vJ}) \, 
W^{\widetilde{\vn}'}_{\ell^q n^q}(\widetilde{\vJ}')\right]  N_{\ell^p n^p n^q}(\omega).
\label{1/Dlong}
\end{align}
The sums on $\ell^q$, $m^p$ and $m^q$ are now trivially executed. 
The non-trivial part of $E_{\vn\vn'}$ that is independent of $\beta$ is
\begin{align}
\label{ifwedefine}
\Lambda_{\widetilde{\vn}\widetilde{\vn}'}^{\ell}  (\widetilde{\vJ}, & \,
\widetilde{\vJ}' ,  \omega)
 \equiv Y_{\ell}^{n_2}(\pi/2,0) \, Y_{\ell}^{n'_2}(\pi/2,0)
\nonumber
\\
& \, \times \sum_{n^pn^q} W^{\widetilde{\vn}}_{\ell n^p}(\widetilde{\vJ})
\,  W^{\widetilde{\vn}'}_{\ell n^q}(\widetilde{\vJ}') \, N_{\ell n^p
n^q}(\omega),
\end{align}
and with this definition $|E_{\vn\vn'}|^2$ can be written
\begin{align}
&\left| E_{\vn\vn'}(\vJ,\vJ',\omega)\right|^2  ={1\over\cE^2} \delta_{n_3}^{n'_3} 
\nonumber
\\
&\hskip.75cm\times\sum_{ \substack{\ell^p}} \sum_{\substack{\ell^q}}  
\Lambda_{\widetilde{\vn}\widetilde{\vn}'}^{\ell^p}(\widetilde{\vJ},\widetilde{\vJ}',\omega)
 \,
 \Lambda_{\widetilde{\vn}\widetilde{\vn}'}^{\ell^q *} (\widetilde{\vJ},\widetilde{\vJ}',\omega)
\nonumber
\\&\hskip.75cm \times 
R^{\ell^p}_{n_2 n_3} \!(\beta) \, 
R^{\ell^q}_{n_2n_3} \!(\beta) \, 
R^{\ell^p}_{n'_2n'_3} \!(\beta') \, 
R^{\ell^q}_{n'_2n'_3} \!(\beta') .
\label{1/D_full}
\end{align} 

The diffusion coefficients~\eqref{eq:DiffCoeffs}, for which we require
$|E_{\vn\vn'}|^2$, involve integrals over the
actions, whose volume element is given by~\eqref{volelem}.  Given that the DF $f_0$ does not depend on $J_3$, we can execute
the integral over $\cos \beta$ up front.
In view of the orthogonality relation~\eqref{eq:Rorthog}
we have
\begin{align}
\int \d J_3'\,&\big|E_{\vn\vn'}(\vJ,\vJ',\vn\!\cdot\!\vOmega)
\big|^2  ={1\over\cE^2}\delta_{n_3}^{n'_3}   \,J_2'
\nonumber
\\
& \, \times \!\! \sum_{ \substack{\ell}} \frac{2}{2 \ell \!+\! 1}  
\left| \Lambda_{\widetilde{\vn}\widetilde{\vn}'}^{\ell} \!
(\widetilde{\vJ},\widetilde{\vJ}',\widetilde{\vn}\!\cdot\!\widetilde\vOmega)\right|^2 
\left| R^{\ell}_{n_2 n_3} \!(\beta) \right|^2 \!.
\label{integralz}
\end{align}

We now turn to evaluating the flux $\vF$ from equation~\eqref{eq:Ffromff}.
The assumed form $f_0(\widetilde\vJ)$ of the DF together with equation~\eqref{eq:tildeOmega}
enables us to simplify equation~\eqref{eq:Ffromff} to
\begin{align}\label{eq:Ffromff2}
\vF&(\vJ)=\fracj12(2\pi)^4\mu\sum_{\vn\vn'}\vn
\int\!\d^2\widetilde\vJ'\!\int\!\d J_3'
\bigl|E_{\vn\vn'}(\vJ,\vJ',\vn\cdot\vOmega)\bigr|^2\nonumber \\
&\hskip -0.5cm \times\! 
\biggl(\widetilde\vn'\cdot{\p\over\p\widetilde\vJ'}-\widetilde\vn\cdot{\p\over\p\widetilde\vJ}\biggr)f_0(\widetilde\vJ)f_0(\widetilde\vJ')
\delta(\widetilde\vn'\cdot\widetilde\vOmega'-\widetilde\vn\cdot\widetilde\vOmega).
\end{align}
Consequently  the only dependence of $\vF$ on the
third components of $\vn$ and $\vn'$ is given by equation~\eqref{integralz},
which states that
non-zero contributions to $\vF$ arise only when $n_3=n'_3$. For given
values of $\widetilde\vn$ and $\widetilde\vn'$ we can sum over all values of
$n_3=n_3'$ and take advantage of the 
identities
\begin{align}
\sum_{n_3} \left|R^{\ell}_{n_2n_3}(\beta)\right|^2 &=
1,\nonumber\\
\sum_{n_3} n_3 \left|R^{\ell}_{n_2n_3}(\beta)\right|^2&=
n_2\cos\beta=n_2L_z/L,
\end{align}
to find
\[\label{eq:Fform}
\vF=\begin{pmatrix}F_1\\F_2\\F_3\end{pmatrix}=\sum_{\widetilde\vn}
\begin{pmatrix}n_1\\n_2\\n_2J_3/J_2\end{pmatrix}\cF_{\widetilde\vn}(\widetilde\vJ),
\]
where
\begin{align}
\cF_{\widetilde\vn}&(\widetilde\vJ)=(2\pi)^4{\mu\over\cE^2}\sum_{\widetilde\vn'}
\int\!\d^2\widetilde\vJ'\,J_2'
\sum_{\ell}\frac{\bigl|\Lambda^{\ell}_{\widetilde\vn\widetilde\vn'}
(\widetilde\vJ,\widetilde\vJ',\widetilde\vn\cdot\widetilde\vOmega)\bigr|^2}
{2\ell+1}\nonumber \\
&\hskip -0.5cm\times\! 
\biggl(\widetilde\vn'\cdot{\p\over\p\widetilde\vJ'}-\widetilde\vn\cdot{\p\over\p\widetilde\vJ}\biggr)f_0(\widetilde\vJ)f_0(\widetilde\vJ')
\delta(\widetilde\vn'\cdot\widetilde\vOmega'-\widetilde\vn\cdot\widetilde\vOmega).
\end{align}

Computing the divergence of $\vF$ from equation~\eqref{eq:Fform} we find
\[
\hbox{div}\,\vF={\p F_1\over\p J_1}+{\p F_2\over\p J_2}+{F_2\over J_2}.
\]
Since $F_1$ and $F_2$ only depend on $\widetilde\vJ$, it follows  that
div$\,\vF$ does not depend on $J_3$. Hence, \textit{a DF that initially is independent
of $L_z=J_3$ will remain so as diffusion proceeds.} This result ensures that
an initially stable spherical cluster remains spherical. 

We will work with the integral with respect to $J_3$ of the BL equation. We
define
\[\label{eq:def_fbar}
\overline{f}(\widetilde\vJ)\equiv\int_{-J_2}^{J_2}\d
J_3\,f_0(\widetilde\vJ)=2J_2f_0(\widetilde\vJ),
\]
and note that since $\hbox{div}\,\vF$ is independent of $J_3$,
\begin{align}
\int_{-J_2}^{J_2}\d J_3\,\hbox{div}\vF&=2J_2\hbox{div}\vF\cr
&=2\biggl({\p (J_2F_1)\over\p J_1}+{\p (J_2F_2)\over\p J_2}\biggr).
\end{align}
Hence the integral of the BL equation with respect to $J_3$ is
\[\label{eq:Fbar1}
{\p\overline{f}\over\p
t}=-{\p\over\p\widetilde\vJ}\cdot\overline{\vF}(\widetilde\vJ),
\]
where
\[\label{eq:Fbardef}
\overline{\vF}\equiv\sum_{\widetilde\vn}\widetilde\vn\overline{\cF}_{\widetilde\vn}(\widetilde\vJ)
\]
with
\begin{align}\label{eq:Fbar2}
\overline{\cF}_{\widetilde\vn}&(\widetilde\vJ)\equiv\fracj12(2\pi)^4
{\mu\over\cE^2}\sum_{\widetilde\vn'}
\int\!\d^2\widetilde\vJ'\,J_2J_2'
\sum_{\ell}\frac{\bigl|\Lambda^{\ell}_{\widetilde\vn\widetilde\vn'}
(\widetilde\vJ,\widetilde\vJ',\widetilde\vn\cdot\widetilde\vOmega)\bigr|^2}
{2\ell+1}\nonumber \\
&\hskip -0.5cm \times\! 
\biggl(\widetilde\vn'\cdot{\p\over\p\widetilde\vJ'}-\widetilde\vn\cdot{\p\over\p\widetilde\vJ}\biggr)
{\overline{f}(\widetilde\vJ)\over J_2}{\overline{f}(\widetilde\vJ')\over
J'_2}\,
\delta(\widetilde\vn'\cdot\widetilde\vOmega'-\widetilde\vn\cdot\widetilde\vOmega).
\end{align}
Equations~\eqref{eq:Fbar1} to~\eqref{eq:Fbar2} describe diffusion of stars
in the $J_rL$ plane; no references to the $L_z$ coordinate remain. The
equations are formally similar to those that govern the relaxation of a
two-dimensional stellar disc~\citep{Fouvry2015b}. {In equation~\eqref{eq:Fbar2}
the dressed friction was already computed by~\cite{Weinberg1989} in his equation~(53). Self-gravity can be switched off by setting $M_{pq}=0$ when computing the coefficients $\Lambda^{\ell}_{\widetilde\vn\widetilde\vn'}$.  Fluxes computed with self-gravity switched off are referred to as `bare', whereas fluxes which include self-gravity are `dressed'.

Since $L$ and $J_r$ are inherently positive, it is a logical necessity that
the component of $\overline{\vF}$ perpendicular to the $L$ and $J_r$ axes
should vanish at those axes, so stars cannot diffuse to negative values of $L$ or $J_r$. The
factor $J_2=L$ after the infinitesimal in the definition~\eqref{eq:Fbar2} of
$\overline{\cF}_{\widetilde\vn}(\widetilde\vJ)$ guarantees that
$\overline{\vF}$ vanishes as $L\to0$. For the case $J_r\to0$, we can change integration variable in the coefficients
$W^{\widetilde\vn}_{\ell n}(\widetilde\vJ)$ defined by equation~\eqref{wmat}, so that they are integrals over $r$ from $r_\mathrm{p}$ to $r_\mathrm{a}$.  They will vanish as $J_r \to 0$ because then $r_\mathrm{p} \to r_\mathrm{a}$ so the limits of integration become identical. By equation~\eqref{ifwedefine}
$\Lambda^\ell_{\widetilde\vn\widetilde\vn'}$ vanishes with
$W^{\widetilde\vn}_{\ell n}$ causing $\overline{\cF}_{\widetilde\vn}$ to
vanish also. Hence there can be no flux across the $L$ axis either.

\section{Application to the isochrone}
\label{sec:isochrone}

The natural testbed for the formalism presented above is the isochrone model
since for it alone we have analytic expressions
$\vtheta(\vx,\vv)$ and  $\vJ(\vx,\vv)$ for the angle-action variables.

\cite{Henon1959} derived the isochrone potential
\begin{equation}
\Phi_0(r) \equiv \Phi_\mathrm{iso}(r) = - \frac{GM}{b+\sqrt{b^2+r^2}},
\end{equation}
where $b$ sets the model's linear scale and $M$ is the model's mass, as
the only potential in which $\Omega_r$ is independent of $L$ at given energy $E$.
The Hamiltonian for motion in this potential
is
\[\label{hiss}
H_0(\widetilde\vJ) = - \frac{(GM)^2}{2\left[ J_r \!+\! \frac{1}{2}\left( L\! +\! \sqrt{L^2\!+\!4GMb}\right)\right]^2}, 
\]

\cite{Henon1960} gave the ergodic DF $f(E)$ that self-consistently generates
the isochrone potential. Since we wish to probe the impact of velocity
anisotropy on the rate of relaxation, we take the DF $f_0$ to be a member of
the  Osipkov-Merritt family of
DFs~\citep{MayB1986,Binney2008}. Specifically,
\begin{align}
f_0 & \, (Q) = \frac{(1-Q)^{-4}M}{128\sqrt{2}\pi^3(GMb)^{3/2}} \Big(\! \sqrt{Q}\big[ 27 \!+\! 77\gamma \!-\! (66 \!+\! 286 \gamma) Q +
\nonumber
\\
& \, (320 +136 \gamma )Q^2- (240+32\gamma)Q^3 + 64Q^4\big]  +
\nonumber
\\
& \hskip -0.25cm \, \frac{3 \arcsin \sqrt{Q}}{\sqrt{1\!-\!Q}} \big[ (17\gamma\!-\!9) \!+\! (28\!-\!44\gamma)Q  
 \!+\!(16\!-\!8\gamma)Q^2 \big] \Big),
\label{barrydf}
\end{align}
where
\begin{equation}
Q \equiv -\frac{b}{GM}\left( E+ \frac{L^2}{2R_{\ra}^2}\right),\quad
\mathrm{and} \quad  \gamma \equiv (b/R_\ra)^2 .
\end{equation}
With 
\begin{equation}
\beta = 1- \frac{\langle v_{\rt}^2\rangle}{2\langle v_{r}^2 \rangle},
\end{equation}
where $v_\rt=\sqrt{v_\vartheta^2+v_\phi^2}$ is the tangential speed, 
for any Osipkov-Merritt model we  have~\citep{Binney2008}
\begin{equation}
\beta(r) = \frac{1}{1+R_\ra^2/r^2}.
\end{equation}
Hence  $R_{\ra}$ is called the `anisotropy
radius' and is the radius near which the velocity distribution transitions
from isotropic at $r\ll R_\ra$ to radially biased at $r\gg R_\ra$. 
Fig.~\ref{fplotfQ} shows the function $f_0(Q)$ for several values of $R_\ra$,
while Fig.~\ref{betaplot} shows the corresponding runs of $\beta(r)$.
\begin{figure}
\centerline{\includegraphics[width=.9\hsize]{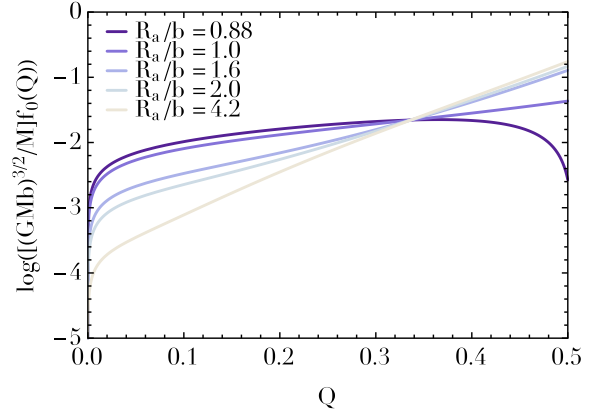}}
 \caption{\small{The Osipkov-Merritt DF $f_0(Q)$ (equation~\eqref{barrydf})
as a function of $Q$.  Note that for $f_0$ to be well defined we require ${
R_{\ra} > 0.874b }$.  }} \label{fplotfQ}
\end{figure}
%
\begin{figure}
\centerline{\includegraphics[width=.9\hsize]{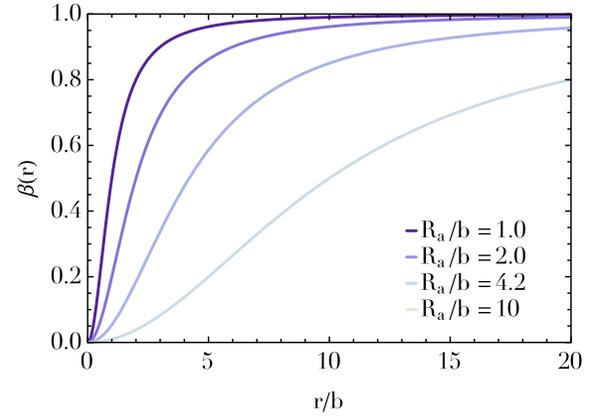}}
 \caption{\small{Anisotropy ${ \beta \equiv 1- \langle v_{\rt}^2 \rangle/
\langle v_{r}^2 \rangle }$ as a function of radius $r$ for various anisotropy
radii $R_{\ra}$.  }} \label{betaplot}
\end{figure}

The model's stars are confined to the part of  the $(E,L)$ plane that is
shaded in Figs.~\ref{fplotfQEL} and~\ref{fig:fbar}.
\begin{figure}
\centerline{
\includegraphics[width=.9\hsize]{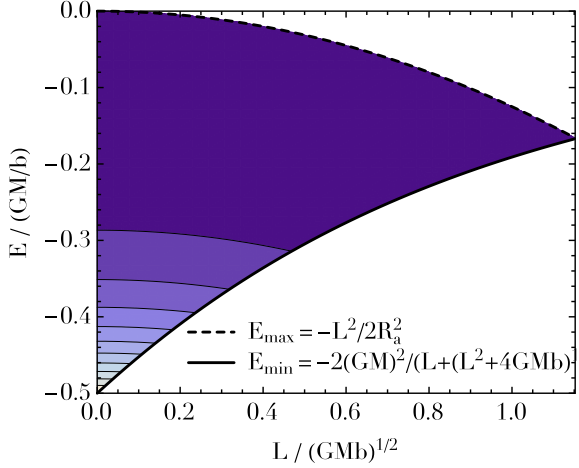}}
 \caption{\small{Contours of ${ f_0(Q(E,L)) }$ in the allowed part of ${
(E,L) }$ space, for ${ R_\ra = 2b }$.  Contours are spaced linearly from the
minimum (dark) to the maximum (light) values taken by ${ f_0(Q(E,L)) }$.  }}
\label{fplotfQEL}
\end{figure}
%
\begin{figure}
\centerline{
\includegraphics[width=.8\hsize]{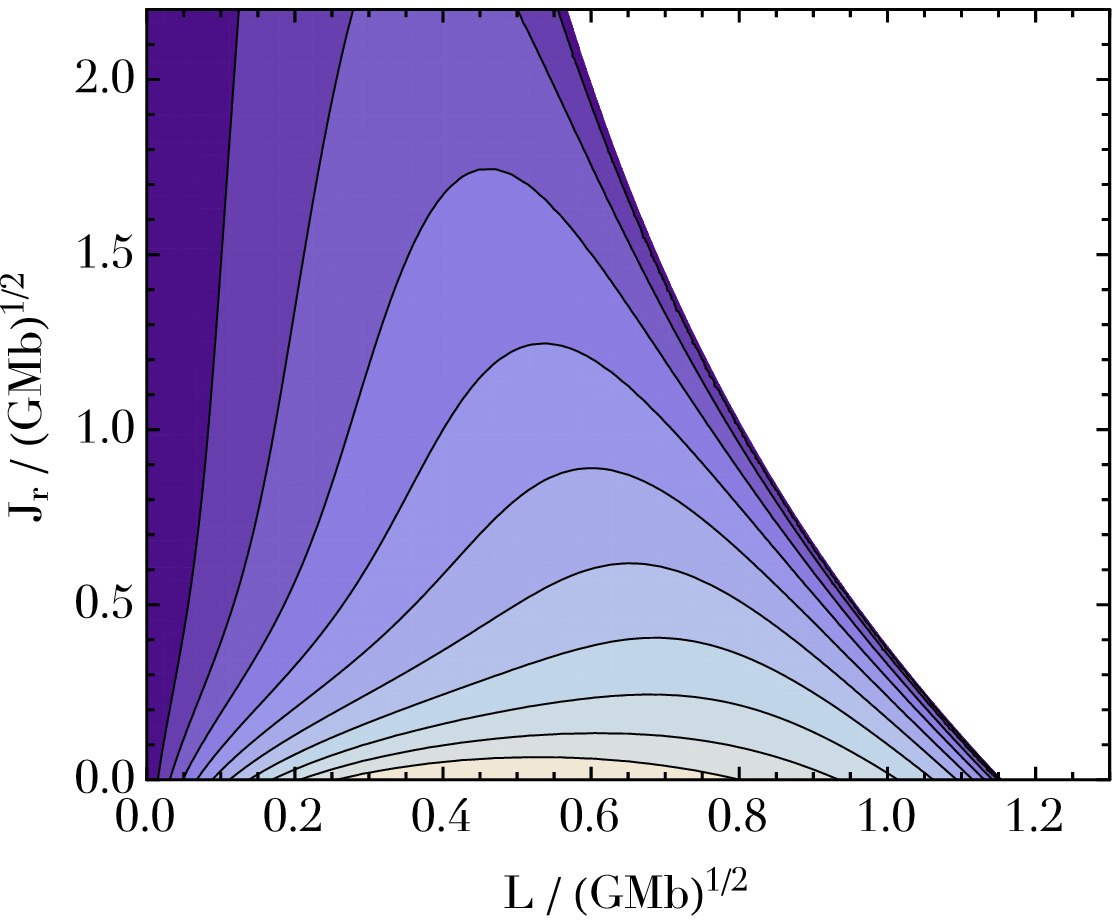}
}
\centerline{
\includegraphics[width=.8\hsize]{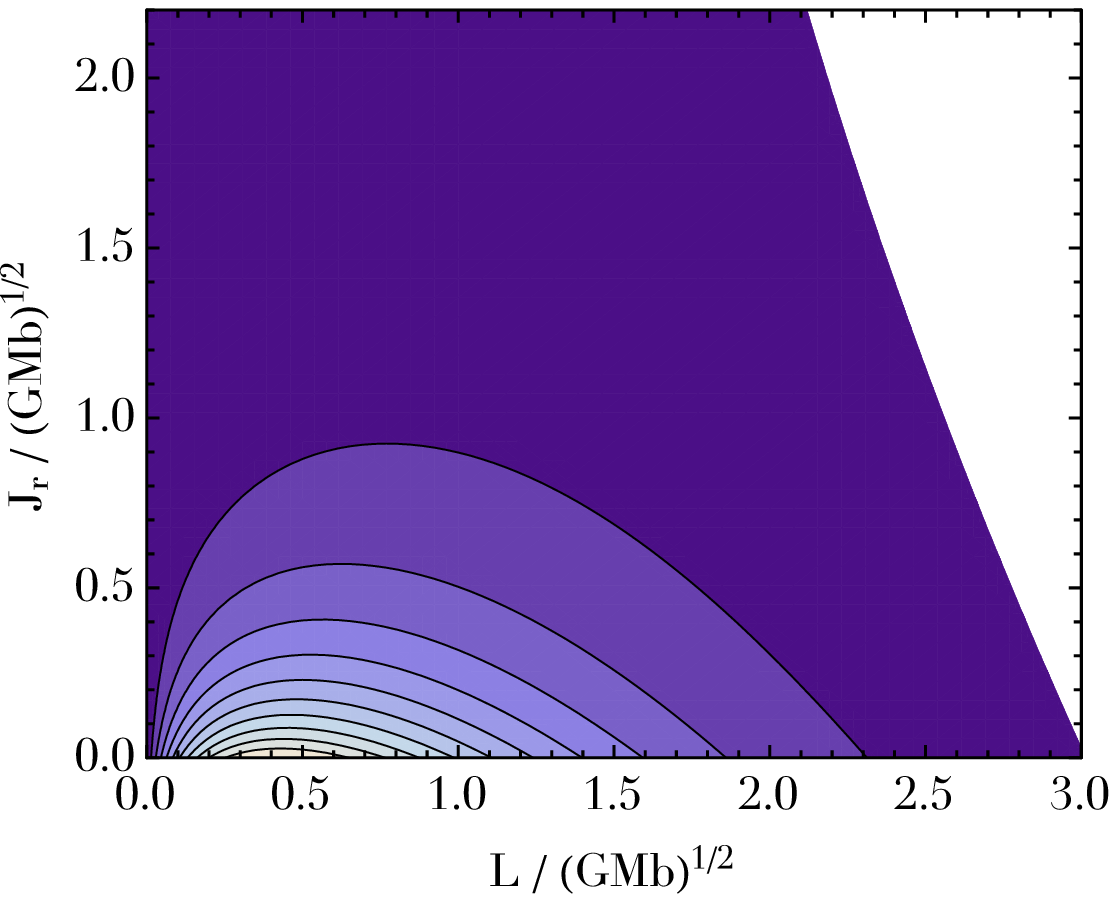}
}
\caption{Contour plots of the two-dimensional DF $\overline{f}=2Lf_0(Q)$ when $R_\ra=2b$
(above) and $10b$ (below).}\label{fig:fbar}
\end{figure}
The lower boundary of this region is set by
the requirement that $E$ be not smaller than the energy of a circular orbit
at the given value of $L$, which is
\[
E_\mathrm{min} = -\frac{2(GM)^2}{\left(L+\sqrt{L^2+4GMb} \right)^2}.
\]
The upper boundary is simply the curve $Q=0$.  In addition, to have ${ f_0 \geq 0
}$ everywhere, we require ${ R_{\ra} > 0.874b }$.  By the Doremus-Feix-Baumann theorem
\citep[][\S5.5]{Binney2008} the radial modes of a spherical model are all
stable if $\p f_0/\p E<0$ everywhere, and models with $\p f_0/\p L>0$ tend to
be unstable. We consider only models with ${ R_{\ra} > 0.97b }$ to  ensure that
$\partial f_0/\partial E<0$ and $\partial f_0 /\partial L <0$ everywhere. 

In significantly anisotropic models the boundary at $Q=0$ generates large
gradients in $f$ that are problematic numerically. Consequently, for models
with $R_\ra<10b$ we have smoothed $f$ by multiplying it by
$\e^{-0.01/Q}$.

We use the radial basis functions of~\cite{Weinberg1989} rescaled by a factor
$M/\surd R$, so they become
\begin{align}
\label{licence}
U_n^{\ell} (r) &=  -{GM\over R}
{4\pi\surd2\over\alpha_{\ell n}|j_\ell(\alpha_{\ell n})|}j_\ell(\alpha_{\ell n}r/R)
\\
D_n^{\ell} (r) &= {M\over R^3}
{\alpha_{\ell n}\surd2\over|j_\ell(\alpha_{\ell n})|}j_\ell(\alpha_{\ell n}r/R),
\end{align}
where the density is assumed to vanish beyond the truncation radius $R$,
$j_\ell$ is a spherical Bessel function of the first kind and
$\alpha_{\ell n}$ is the $n$th zero of  $j_\ell$. For these basis functions
the normalising constant in equation~\eqref{eq:PDnorm} is
\[
\cE=4\pi{GM^2\over R}.
\]

\subsection{Calculation of the response matrix}
\label{sec:response}

To ensure that our code is applicable to any spherical model and not just
isochrone models, we use as coordinates for orbit space the radii $r_\rp$
and $r_\ra$ of peri-
and apo-centre rather than $J_r$ and $L$.
These radii are the roots of the equation
\begin{equation}
E - \Phi(r) - \frac{L^2}{2r^2} = 0.
\end{equation}
Hence, given $r_\rp$ and $r_\ra$ it follows easily that
\begin{equation}
\label{rpraEL}
E = \frac{r_{\ra}^2 \Phi(r_{\ra}) - r_{\rp}^2\Phi(r_{\rp})}{r_{\ra}^2 - r_{\rp}^2}; \,\,\,\,\,\,\,\,\,\, L = \sqrt{\frac{2[\Phi(r_{\ra})-\Phi(r_{\rp})]}{r_{\rp}^{-2}-r_{\ra}^{-2}}}.
\end{equation}
Putting $H=E$ in the Hamiltonian~\eqref{hiss} we recover $J_r(r_\rp,r_\ra)$. Integrals
over $\d^2\widetilde\vJ$ can be expressed as integrals over $(r_\rp,r_\ra)$
using the Jacobian
\[
\frac{\p(J_r,L)}{\p(r_\rp,r_\ra)}=\frac{\p(J_r,L)}{\p(E,L)}\frac{\p(E,L)}{\p(r_\rp,r_\ra)}
={1\over\Omega_r}\frac{\p(E,L)}{\p(r_\rp,r_\ra)}.
\]
Since ${ f_0 = f_0 \left(Q(E,L) \right) }$, 
we have
\begin{equation}
\widetilde\vn\cdot{\p f_0\over\p\widetilde\vJ} 
= - \frac{b}{GM}\frac{\md f_0}{\md Q} \left[n_1\Omega_r + n_2\left( \Omega_\vartheta+\frac{L}{R_{\ra}^2}\right)\right].
\nonumber
\end{equation} 

To perform the integration numerically over the $(r_\rp,r_\ra)$ plane implied
by equation~\eqref{eq:gives_xi}, we divide the plane into regions labelled by
$i$. The $i$th region is centred at ${ (r_{\rp}^i, r_{\ra}^i) }$ and covers
the square ${ r_{\rp} \in [r_{\rp}^i - \Delta r/2, r_{\rp}^i + \Delta r/2],
\,\,\, r_{\ra} \in [r_{\ra}^i - \Delta r/2, r_{\ra}^i + \Delta r/2] }$.  Then
in each region we separately Taylor expand the frequency
$\widetilde\vn\cdot\widetilde\vOmega-\omega$ that appears as the denominator
in equation~\eqref{eq:gives_xi} and its numerator. The ratio of Taylor expansions is then
analytically integrated through the region. The  numerical steps
involved are exactly those that have been detailed extensively
in~\cite{Fouvry2015b} so we will not repeat the recipe
here.

The most computationally expensive part of this procedure is the
calculation from equation~\eqref{wmat} of the matrix elements $W$ at each point on the
${(r_{\rp},r_{\ra})}$ grid.  It is important that we use a large, dense grid,
as the biggest contributions to $\widetilde\vn\cdot\p f_0/\p\widetilde\vJ$
come from the edge of the model, where $|\md f_0/\md Q|$ is largest.
However, for small values of $R_\ra$ the requirement $Q>0$ usefully restricts
the portion of the $(r_\rp,r_\ra)$ grid within which the matrix elements of
$W$ need to be computed (Fig.~\ref{fplotfQEL}). For this reason fluxes are more easily computed for
models with small values of $R_\ra$ than large ones.

\subsection{Recovering unstable modes}
\label{sec:modes}

To demonstrate that our evaluation of the response matrix
$\mathbf{M}(\omega)$ is correct, we recover the known unstable modes of our
system.  For a spherical system the perturbations can be conveniently decomposed into spherical harmonics $Y_\ell^m$, and we can consider each harmonic $\ell$ separately. Modes correspond to frequencies ${ \omega = \omega_0 + \mi \eta }$
with $\eta>0$
for which the matrix $\vepsilon= \vI-\vM(\omega)$ has a zero eigenvalue; in
other words, we seek $\omega$ such that
\begin{equation}
\det\vepsilon = 0.
\end{equation}
The mode has frequency $\omega_0$ and growth
rate $\eta$. Since from equation~\eqref{eq:gives_eps}
\[
\epsilon_{pq} =\delta_{\ell}^{\ell^q}
\delta_{m^p}^{m^q}[\delta_{n^p}^{n^q}-   \xi_{\ell n^p n^q}(\omega)].
\]
for each fixed~$\ell$ we need to compute the determinant of the matrix
${\delta_{n^p}^{n^q} - (\vxi_{\ell})_{n^pn^q} }$.

\cite{Saha1991} used the Osipkov-Merritt DF~\eqref{barrydf} to study the
radial-orbit instability in anisotropic isochrone models. He
found that a mode with ${ \ell = 2}$ becomes unstable when
$R_\ra$ falls below $\sim4b$.  In terms of the natural unit of time $T_\mathrm{I} = \sqrt{b^3/GM}$, growth rates $\eta$ are very small (${ \eta
\ll 0.001/T_{\rm I} }$) when the mode first becomes unstable but grow towards
${ \eta \approx 0.025/T_{\rm I} }$ as $R_{\ra} \to b$.\footnote{Saha's Figure~2(b)
shows the growth rates $\eta$ for various anisotropies $\beta$.  To convert between
Saha's $\beta$ and $R_\ra$, use his Figure~1.}

The calculations in~\cite{Saha1991} used $10$ radial basis functions $U_n^{\ell}$
from a different family from that  used here, and focused on $\ell=2$.  Since in a
spherical system the growth rate must be independent of the azimuthal part of
the spherical harmonics, Saha set $m = 0$.  The unstable mode has vanishing
pattern speed, so in searching for it Saha set $\omega_0 = 0$.  We follow all
of these conventions. In our standard computation the truncation radius of the
basis functions is  $R
= 20b$, and
the grid in ${
\{(r^i_{\rp},r^i_{\ra})\} }$ covers the range ${ r^i \in [0.15b, 19.85b] }$
with grid spacing ${ \Delta r = 0.1b }$.  Finally, the maximum index $n_1$
in the sum of
equation~\eqref{eq:gives_xi} is $n_1= n^\mathrm{max} = 4$.  The justification for
this choice is that the isochrone's frequencies satisfy
\[\label{eq:iso_freqs}
{\Omega_2\over\Omega_1}=\fracj12\left(1+{J_2\over\sqrt{J_2^2+4GMb}}\right)\,,
\]
 so orbits with $J_2\ll2\sqrt{GMb}$ satisfy $2 \Omega_2 - \Omega_1
\approx 0$. Consequently, we expect equation~\eqref{eq:gives_xi} to be
dominated by the term with $(n_1,n_2) = \pm(1, -2)$.

Since $\eta$ is very small and all other contributions to the response matrix
are real, the imaginary part of ${ \det[\mathbf{I}-\vxi_{\ell=2}(\mi \eta)]
}$ is negligible.  Hence to locate the unstable modes, we need only plot ${
\mathrm{Re} \left(\det[\mathbf{I}-\vxi_{\ell=2}(\mi \eta)] \right) }$ for
various values of the growth rate $\eta$, and pick out the $\eta$ for which
${ \mathrm{Re} \left(\det[\mathbf{I}-\vxi_{\ell=2}(\mi \eta)] \right) = 0 }$.
Fig.~\ref{fig:draftdet} shows $\mathrm{Re}
\left(\det[\mathbf{I}-\vxi_{\ell=2}(\mi \eta)] \right)$ as a function of
$\eta$ for three values of $R_\ra$.
\begin{figure}
\centerline{\includegraphics[width=.9\hsize]{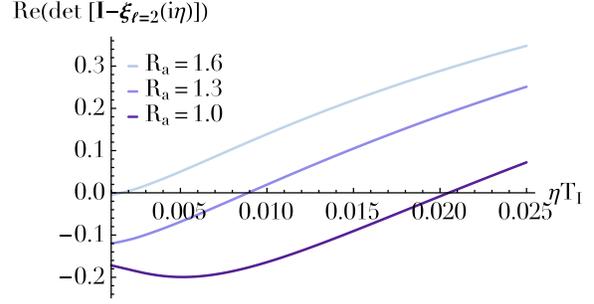}}
 \caption{\small{Plot of ${ \mathrm{Re}\left(\det[\mathbf{I}-\vxi_{\ell=2}(0
+ \mi \eta)]\right) }$ in units of $T_{\rm I}$ as a function of the growth
rate $\eta$ for
${ R_\ra/b = 1.0,1.3,1.6 }$ (from bottom to top). Crossing the $\eta$ axis indicates the
existence of the radial-orbit instability.  }} \label{fig:draftdet}
\end{figure}
Each curve crosses the $x$ axis, so all
these models are subject to the radial-orbit instability. However, the model
with $R_\ra=1.6b$ only just manages to cross the $x$ axis, suggesting that the
limiting value for instability, $R_{\rm a,max}$ lies close to $1.6b$. This
conclusion is consistent with the conclusion of~\cite{MayB1986} that
$R_{\rm a,max}\simeq1.67b$. The top row of
Table~\ref{tab:saha} gives our growth rates for these four models plus
indications that models with $R_\ra/b\ge2$ are stable.
\begin{table*}
\centering
\caption{Growth rates for the radial-orbit instability in the
isochrone model.  We compare values from our standard computation with values
from~\protect\cite{Saha1991}.
The notation
$\times_y$ means that we could not recover an unstable mode, and that the
value of ${ \mathrm{Re}\left(\det[\mathbf{I}-\vxi_{\ell=2}(\mi
\eta)]\right) }$ at $\eta=0$ was $y$.}\label{tab:saha}
	\begin{tabular}{lcccccc}
$R_\ra/b$&1.0&1.3&1.6&2.0&3.0&4.2\\
		\hline
Our $\eta T_{\rm I}$&0.021&0.009&0.002&$\times_{0.20}$&$\times_{0.43}$&$\times_{0.53}$\\
Saha linear $\eta T_{\rm I}$&0.024&0.009&0.004&$<0.001$&$\ll0.001$&$\lll0.001$\\
Saha N-body $\eta T_{\rm I}$&0.029&0.015&0.010&--&--&--\\
		\hline
	\end{tabular}
\end{table*}
The second and third
rows of Table~\ref{tab:saha} give corresponding results from
\cite{Saha1991}. Our growth rates tend to be slightly smaller than the rates
Saha obtained from linear theory, and significantly smaller than the rates
Saha inferred from N-body simulations, especially for the marginally unstable
model.

Linear theory enables us to predict the shape (but not the amplitude) of the
perturbation to  the potential associated with an unstable mode. Indeed, with $\eta$ the
growth rate of the unstable mode, $\vxi_{\ell = 2}(\mi\eta)$ is an Hermitian
matrix (to see this, take the Hermitian conjugate of equation \eqref{eq:gives_xi} and use $W^{\widetilde{\mathbf{n}}}_{\ell n}=W^{-\widetilde{\mathbf{n}}}_{\ell n}$).  It has an eigenvector $\vX$ with eigenvalue unity, and the
perturbation to the potential is
\begin{align}\label{eq:mode}
\Phi_1(\vx) &\propto  \mathrm{Re} \left[ \sum_{n=1}^{n^{\rm max}} X^n\Phi^{(n,2,0)}(\vx)\right]\\
&=\left[\sum_{n=1}^{n^{\rm max}} X^n  {j_2(\alpha_{2n}r/R)\over\alpha_{2 n}\left| j_2(\alpha_{2n}) \right|}
\right] P_2(\cos \theta).
\end{align}
The purple line in Fig.~\ref{radialphi} shows $\Phi_1$ when $R_\ra=b$, while
the black curve reproduces the corresponding plot from~\cite{Saha1991}.
\begin{figure}
\centerline{\includegraphics[width=0.9\hsize]{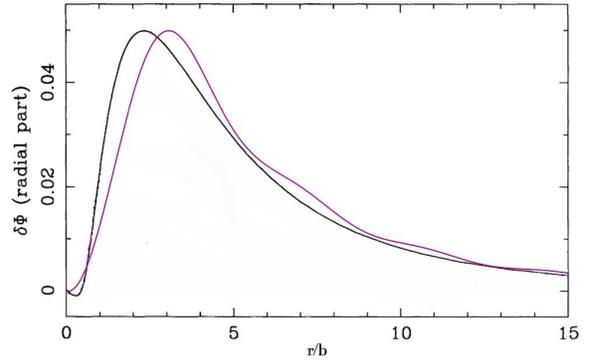}}
\caption{The radial part of our $\Phi_1$ (what Saha calls $\delta \Phi$) in purple, superimposed on Saha's
version of the same mode (black), for the unstable ${ \ell = 2 }$ mode
in the model with ${ R_\mathrm{a} = b }$. The normalisation of the vertical
axis is arbitrary. (Graphics edited from~\protect\cite{Saha1991}).}
\label{radialphi}
\end{figure}
Overall the agreement is good.

\subsubsection{Convergence study}

Table~\ref{tab:conv} shows how the recovered growth rates of the radial-orbit
instability ($\ell=2$) are affected by changing the four key parameters of
the computation.
\begin{table*}
\centering
\caption{Growth rates for the radial-orbit instability in the isochrone model
computed for various values of the parameters that control the precision of
the calculations. The tabulated values are of $\eta T_{\rm I}$ and the values
in bold are from our standard computation. The notation $\times_y$ means that
we could not recover an unstable mode, and that the value of ${
\mathrm{Re}\left(\det[\mathbf{I}-\vxi_{\ell=2}(\mi \eta)]\right) }$ at
$\eta=0$ was $y$.}\label{tab:conv}
\begin{tabular}{c c c c c c c} 
 \hline
 Anisotropy radius $R_\mathrm{a}/b$ & $1.0$ & $1.3$ & $1.6$ & $2.0$ & $3.0$ & $4.2$ \\ 
  \hline
 $\Delta r/b = 0.05$ & $0.0210$ & $0.0111$ & $\times_{0.01}$ & $\times_{0.19}$ & $\times_{0.43}$ & $\times_{0.53}$\\ 
 $\Delta r/b = 0.07$ & $0.0220$ & $0.0100$ & $\times_{0.02}$ & $\times_{0.19}$ & $\times_{0.43}$ & $\times_{0.53}$\\ 
$ \Delta r/b = {\bf0.1}$ & ${\bf0.0205}$ & ${\bf0.0089}$ & ${\bf0.0016}$ &  $\times_{0.20}$ &
$\times_{0.43}$ & $\times_{0.53}$\\  
$ \Delta r/b = 0.15$ & $0.0200$ & $0.0081$ & $0.0068$ & $\times_{0.17}$ & $\times_{0.44}$ & $\times_{0.54}$\\ 
$\Delta r/b = 0.2$ & $0.0245$ & $0.0097$ & $0.0057$ & $\times_{0.22}$ & $\times_{0.44}$ & $\times_{0.54}$\\ 
 $\Delta r/b = 0.3$ & $0.0222$ & $0.0085$ & $\times_{0.03}$ & $\times_{0.21}$ & $\times_{0.45}$ & $\times_{0.55}$\\ \hline
 $R/b=10$ & $0.0203$ & $\times_{0.01}$ & $\times_{0.15}$ & $\times_{0.27}$ & $\times_{0.39}$ & $\times_{0.45}$\\ 
 $R/b=16$ & $0.0216$ & $0.0084$ & $0.0016$ & $\times_{0.22}$ & $\times_{0.41}$ & $\times_{0.50}$\\ 
$ R/b={\bf20}$ & ${\bf0.0205}$ & ${\bf0.0089}$ & ${\bf0.0016}$ &
$\times_{0.20}$ & $\times_{0.43}$ & $\times_{0.53}$\\ 
  $R/b=25$ & $0.0180$ & $0.0081$ & $0.0026$ & $\times_{0.17}$ & $\times_{0.43}$ & $\times_{0.56}$\\   
 \hline
   $n_1^\mathrm{max} = 1$ & $0.0208$ & $0.0089$ & $0.0014$ & $\times_{0.19}$ & $\times_{0.43}$ & $\times_{0.54}$\\ 
 $n_1^\mathrm{max} = 2$  & $0.0204$ & $0.0088$ & $0.0015$ & $\times_{0.19}$ & $\times_{0.43}$ & $\times_{0.54}$\\ 
 $n_1^\mathrm{max} = {\bf4}$ & ${\bf0.0205}$ & ${\bf0.0089}$ & ${\bf0.0016}$ & $\times_{0.20}$
& $\times_{0.43}$ & $\times_{0.53}$\\  
  $n_1^\mathrm{max} = 6$ & $0.0205$ & $0.0089$ & $0.0017$ & $\times_{0.19}$ & $\times_{0.42}$ & $\times_{0.53}$\\ 
 $n_1^\mathrm{max} = 8$ & $0.0206$ & $0.0090$ & $0.0017$ & $\times_{0.19}$ & $\times_{0.42}$ & $\times_{0.52}$\\ 
 \hline
 $n^\mathrm{max} = 8$ & $0.0182$ & $0.0079$ & $0.0001$ & $\times_{0.22}$ & $\times_{0.47}$ & $\times_{0.58}$\\ 
$n^\mathrm{max} = {\bf10}$ & ${\bf0.0205}$ & ${\bf0.0089}$ & ${\bf0.0016}$ &  $\times_{0.20}$ & $\times_{0.43}$ & $\times_{0.53}$\\ 
  $n^\mathrm{max} = 13$ & $0.0220$ & $0.0096$ & $0.0020$ & $\times_{0.17}$ & $\times_{0.38}$ & $\times_{0.48}$\\
 $n^\mathrm{max} = 16$ & $0.0224$ & $0.0098$ & $0.0023$ & $\times_{0.15}$ & $\times_{0.34}$ & $\times_{0.43}$ \\[1ex] 
 \hline
\end{tabular}
\end{table*}
These are (i) the spacing $\Delta r$ of the grid in $r_\rp$
and $r_\ra$; (ii) the truncation radius $R$ of the basis functions (which
also sets the upper limit on values of $r_\rp$ and $r_\ra$); (iii) the
maximum included value of the quantum number $n_1$ associated with $J_r$;
(iv) the number $n^{\rm max}$ of basis functions summed over
(equation~\eqref{eq:mode}). The results for the significantly unstable models,
$R_\ra=1$ and 1.3, show gratifyingly little sensitivity to these parameters.
The most significant fact is that dropping $R$ to $10b$ killed the
instability in the model with $R_\ra=1.3b$. The results for the marginally
unstable model $R_\ra=1.6b$ do change significantly with the values of
parameters, but clear trends are not evident. It is notable that
experimenting with different parameter values does not change the conclusion
that the model with $R_\ra=2b$ is stable.  No modification produced a curve
of $\mathrm{Re}\left(\det[\mathbf{I}-\vxi_{\ell=2}(\mi \eta)]\right)$ that
differed significantly from those plotted in Fig.~\ref{fig:draftdet}.
 
Tables~\ref{tab:saha} and~\ref{tab:conv} in conjunction with Fig.~\ref{radialphi} give us
confidence that the response matrix is being calculated correctly.

\subsection{Calculation of the Balescu--Lenard flux} \label{sec:BLflux}

The magnitude of the diffusive flux $\overline\vF$ is a direct measure of a system's
relaxation rate: as the relaxation timescale tends to infinity, the
diffusive flux tends to zero. Hence we now compute from equations~\eqref{eq:Fbardef} and~\eqref{eq:Fbar2}
the diffusive flux
in the $LJ_r$ plane. In Section~\ref{sec:compare} we will compare this flux with
that predicted by classical theory.

The sum over $\ell$ in equation~\eqref{eq:Fbar2} can be pulled out front, so we obtain the flux
\[
\overline{\vF}=\sum_\ell \overline{\vF}_\ell,
\] 
as a sum of contributions from each multipole.
For given $\widetilde\vn,\widetilde\vn'$, the integrand on the right of
equation~\eqref{eq:Fbar2} can be straightforwardly computed from
equations~\eqref{ifwedefine} and~\eqref{clearthen}, so the only challenge is
the evaluation of the integral over $\widetilde\vJ'$, which has the general form
\[
\int \d^2\widetilde\vJ' \, g(\widetilde\vJ') \, \delta \big( h ( \widetilde\vJ') \big) .
\]
Following~\cite{Fouvry2015b}, this may be written
\begin{equation}
\label{resdom}
\int_{C} \md \sigma(\widetilde\vJ') \, \frac{g(\widetilde\vJ')}{\left| \nabla
h(\widetilde\vJ')\right|},
\end{equation}
where $C$ is the curve along which the resonant condition
$0=h(\vJ')=\widetilde\vn'\cdot\widetilde\vOmega'-\widetilde\vn\cdot\widetilde\vOmega$
is satisfied, and $\d\sigma(\widetilde\vJ')$ is the line element on $C$.
Details of the evaluation of the resulting line integral can be found in
\cite{Fouvry2015b}.

\subsubsection{Wavevectors to consider}

To evaluate the BL flux we have to perform a sum over all possible `pairs' $(\widetilde\vn,\widetilde\vn')$ of two-dimensional vectors with integer components. By a slight abuse of language, we shall refer to objects like
$\widetilde\vn$ as `wavevectors'. In principle there are infinitely many of these wavevector pairs, but some of them do not contribute to the flux. To eliminate pairs which do not contribute we first note that for each value of $\ell$ we can restrict ourselves to pairs for which the quantity
$\Lambda_{{\widetilde\vn\widetilde\vn'}}^\ell$ defined by equation~\eqref{ifwedefine} is
non-zero.  From the appearance in this equation of
$\widetilde n_2$ and $\widetilde n_2'$ as superscripts of spherical harmonics
it is clear that the sum can be restricted to $|\widetilde n_2|\le\ell$ and
$|\widetilde n_2'|\le\ell$. Moreover, the spherical harmonics vanish unless $\ell-\widetilde n_2$
and $\ell-\widetilde n_2'$ are even. Hence the values to be included are 
\[
\begin{matrix}
\ell=0&:&\quad\widetilde n_2&=&0\\
\ell=2&:&\quad\widetilde n_2&=&\pm2,0
\end{matrix}
\qquad
\begin{matrix}
\ell=1&:&\qquad\widetilde n_2&=&\pm1\\
\ell=3&:&\qquad\widetilde n_2&=&\pm3,\pm1\\
\ell=4&:&\qquad\widetilde n_2&=&0,\pm2,\pm4,
\end{matrix}
\]
and similarly for $\widetilde n_2'$.  Unfortunately, $\widetilde n_1$ and $\widetilde n_1'$ are unrestricted. However one can easily show from the definitions in Section \ref{sec:sphere} that the pair $(\widetilde\vn,\widetilde\vn')$
makes exactly the same contribution to the flux as the pair $(-\widetilde\vn,-\widetilde\vn')$, halving the computation time. In addition one can show that for $\ell \leq 2$ (and respecting the rules above) the quantity $\widetilde\vn\cdot\widetilde\vOmega$ has the same sign throughout action space (equation~\eqref{eq:iso_freqs}).  Since the resonant condition requires
$\widetilde\vn\cdot\widetilde\vOmega=\widetilde\vn'\cdot\widetilde\vOmega'$,
once $\widetilde\vn$ has been chosen, only one sign of $\widetilde n_1'$ can be of
interest, restricting the pair count further.  A corresponding statement does not hold for $\ell \geq 3$.

If for the sake of definiteness we consider only terms with
$|\widetilde n_1|,|\widetilde n_1'|\le2$, then when $\ell=0$ and thus $\widetilde n_2=\widetilde
n_2'=0$, we require $|\widetilde n_1|,|\widetilde n_1'|>0$ and have to include the four relevant pairings
from $(\widetilde n_1,\widetilde n_1')=(1,\pm1),(1,\pm2),(2,\pm1),(2,\pm2)$. However, the
corresponding pair count rises to 25 for $\ell=1$, to 49 for $\ell=2$, to well over 100
for $\ell=3$, etc. Clearly, increasing the upper bound
on values of $|\widetilde n_1|$ to be considered leads to still more intimidating pair
counts and we must seek to identify the wavevector pairs that make the
largest contributions to $\overline{\vF}$.  Since computing time scales as
$(\Delta r)^{-4}$, we do this by performing a preliminary computation on a
coarse grid $\Delta r\sim0.25$. 

The $\ell=1$ fluxes for models with $10^5$ stars and $R_\ra=2b$ (left),
$R_\ra=4.2b$ (centre) and for the isotropic model (right) are shown in the upper row of
Fig.~\ref{fig:ChrisFlux}.  
\begin{figure*}
\centerline{
\includegraphics[height=.3\hsize]{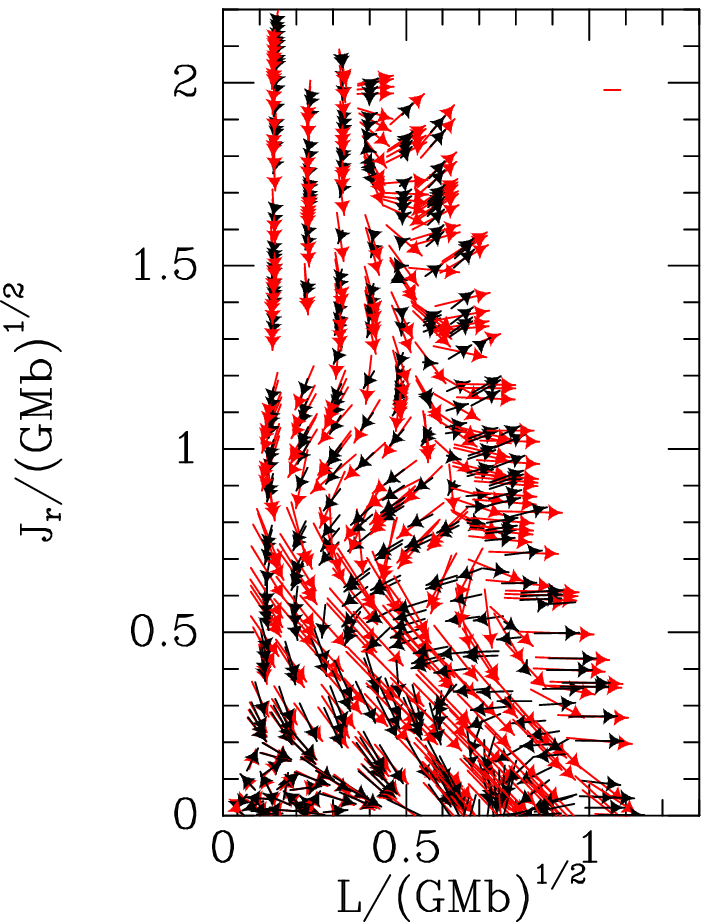}
\includegraphics[height=.3\hsize]{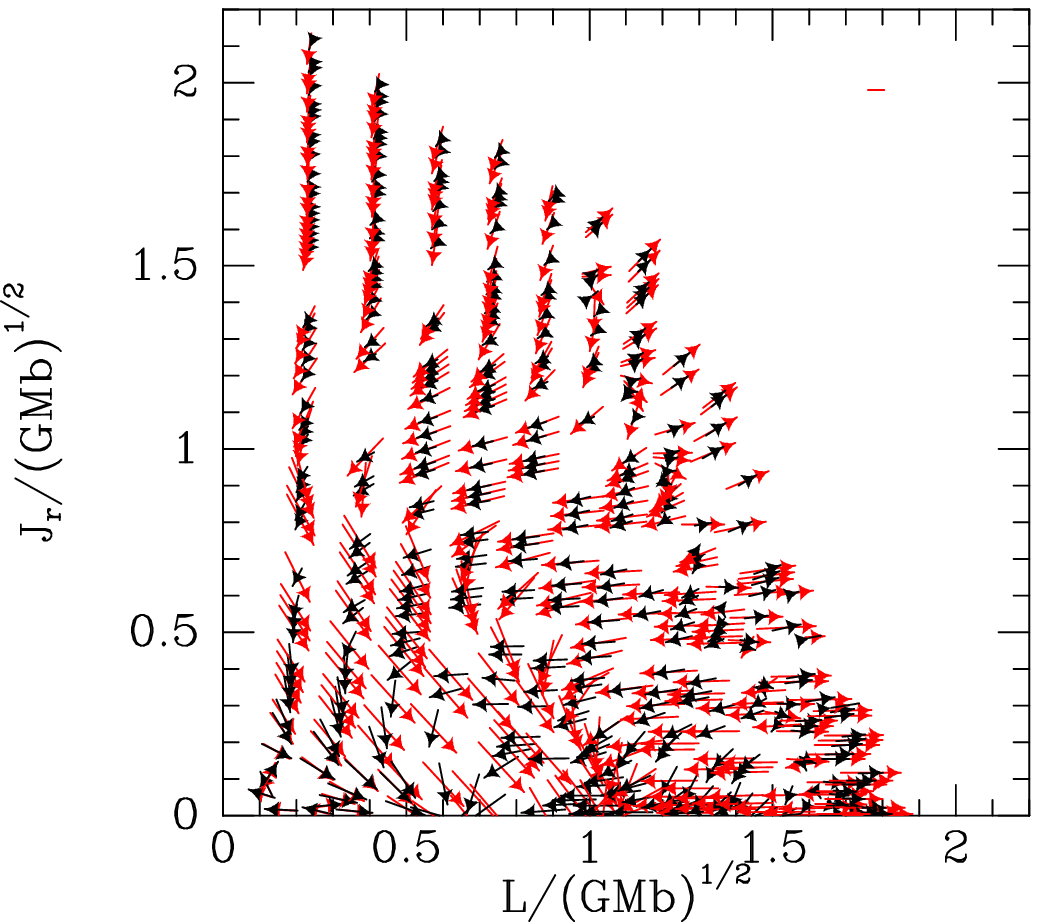}
\includegraphics[height=.3\hsize]{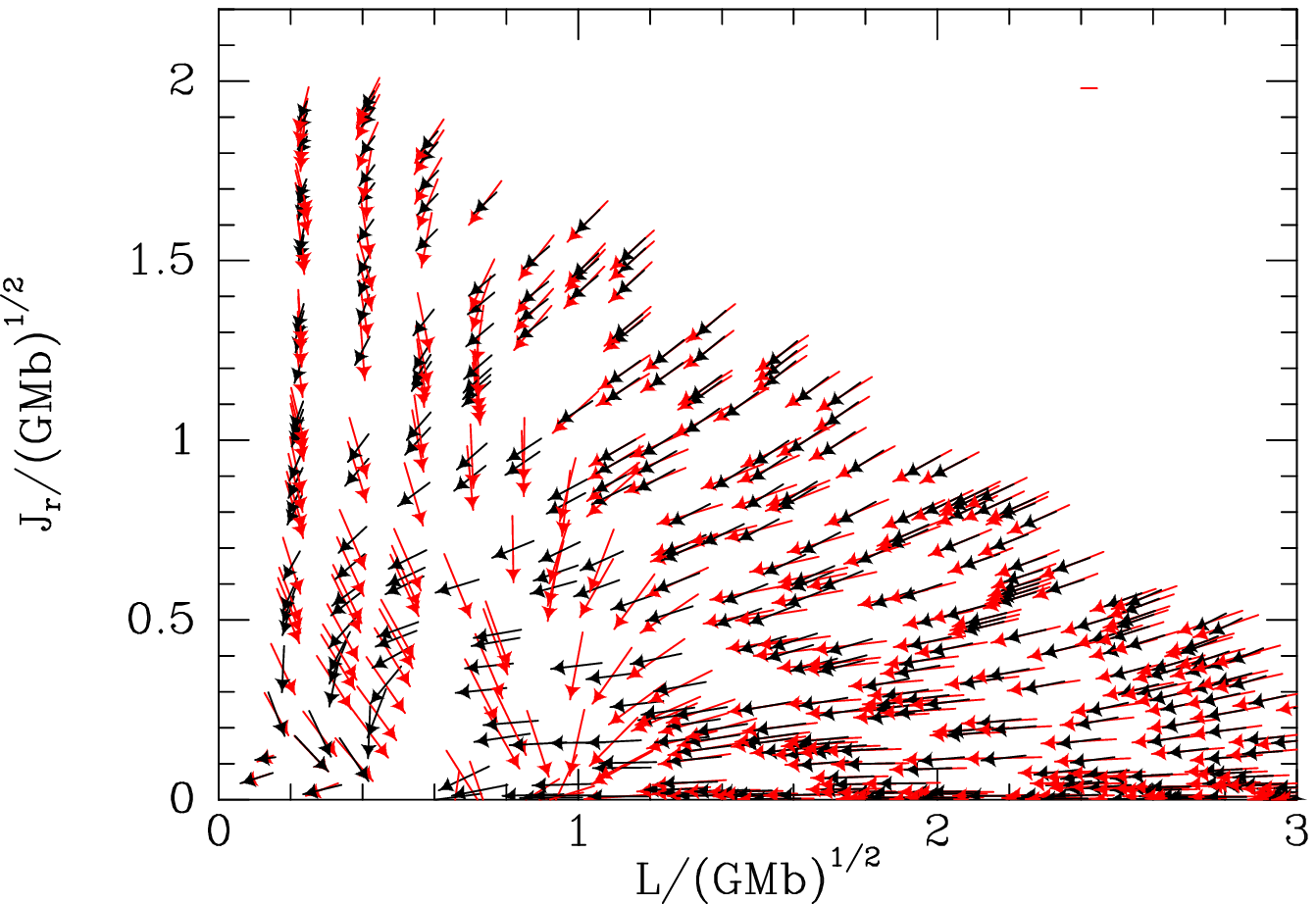}
}
\centerline{
\includegraphics[height=.3\hsize]{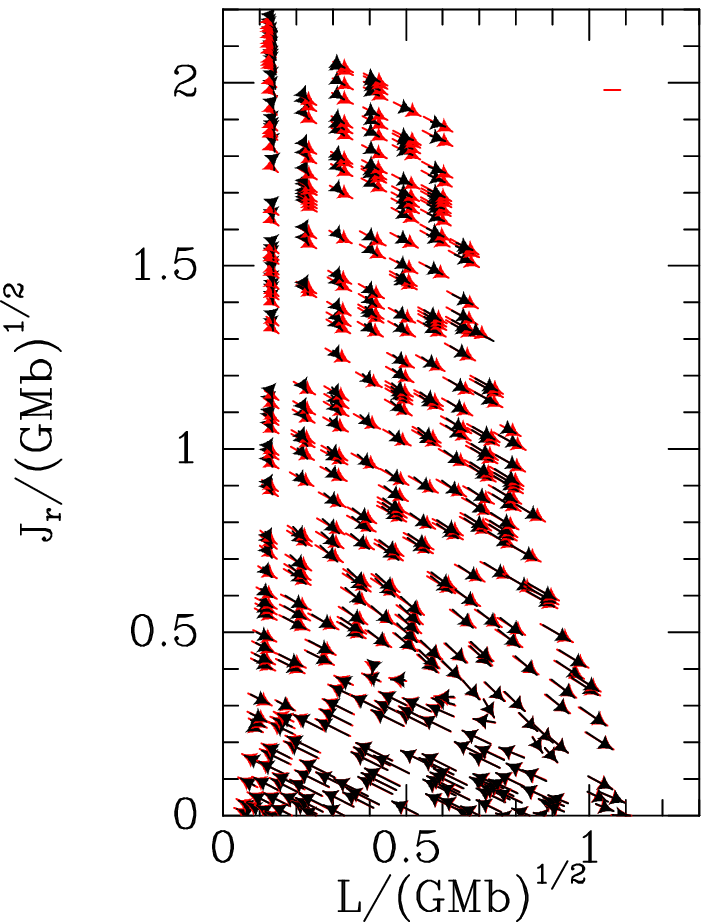}
\includegraphics[height=.3\hsize]{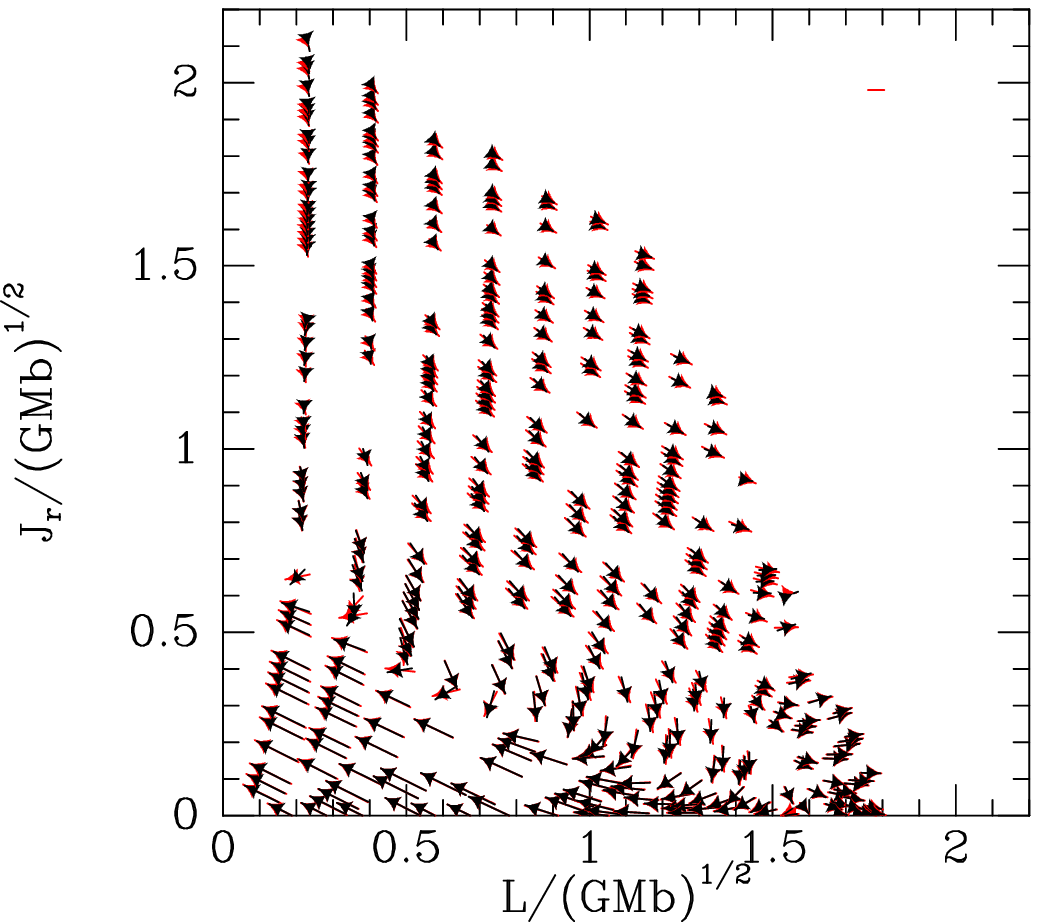}
\includegraphics[height=.3\hsize]{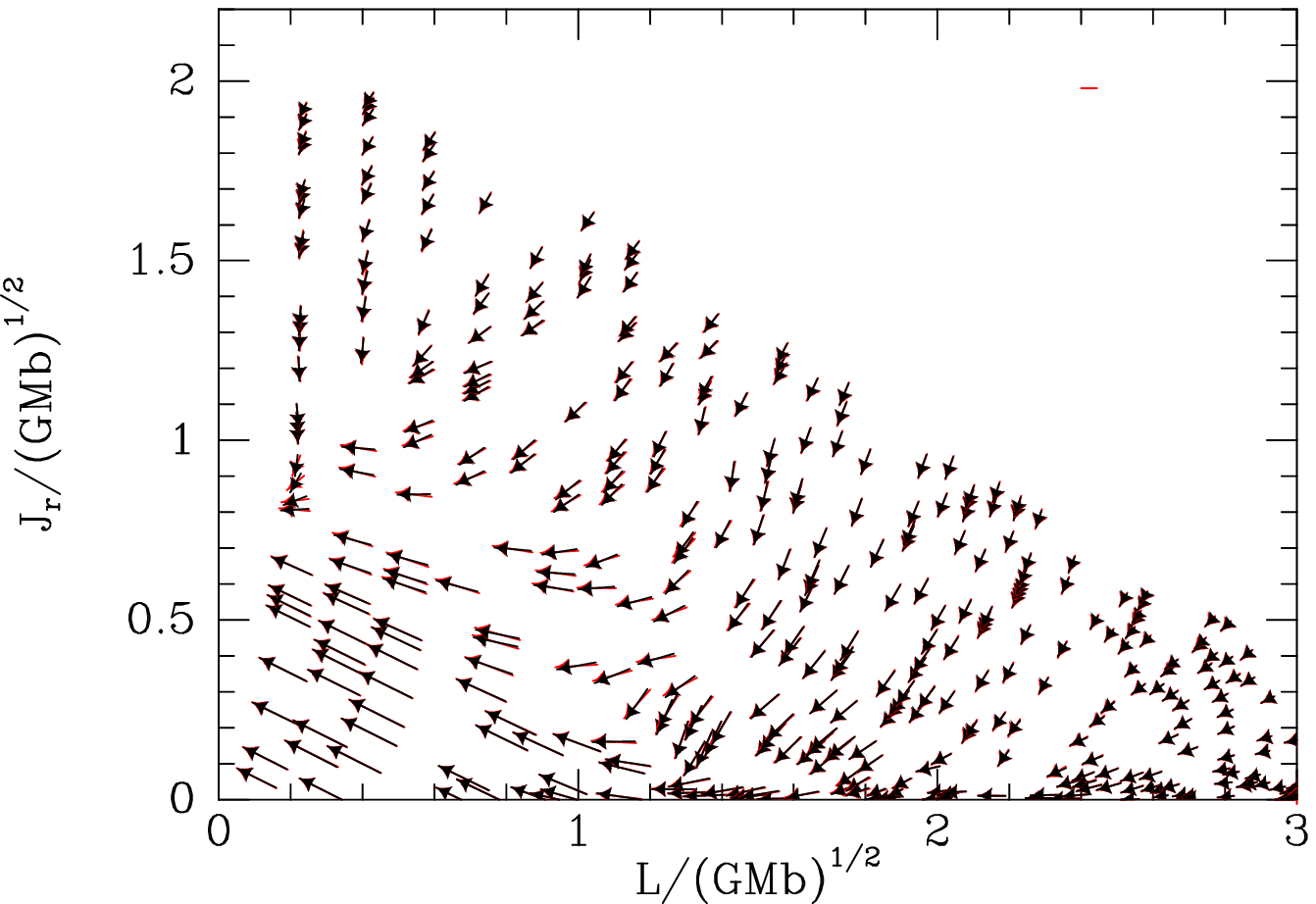}
}
 \caption{Flux vectors $\overline{\vF}_\ell$ in clusters with $N=10^5$
equal-mass particles with $R_\ra/b=2$ (left) and $4.2$ (centre), and in the
isotropic model 
(right).  The fluxes $\overline{\vF}_1$ are shown in the upper row, with the
fluxes $\overline{\vF}_2$ shown below. The dressed flow is shown by red
arrows on top of which are printed black arrows describing the bare flow. The length of an arrow is proportional to
$\log_{10}[|\overline{\vF}|/(10^{-12}Mb^{-2})]$, so that an arrow of zero length corresponds to $|\overline{\vF}_\ell|=10^{-12}Mb^{-2}$. The bar at top right corresponds to a hundred-fold increase in
$ |\overline{\vF}_\ell |$.
}\label{fig:ChrisFlux}
\end{figure*}
Red arrows show the dressed (self-gravitating) flux, and black arrows show the corresponding bare flux computed by setting the response matrix $M_{pq}=0$ (i.e. with self-gravity switched off). These fluxes are obtained by summing over all pairs
$(\widetilde\vn,\widetilde\vn')$ with $\widetilde n_1,\widetilde n_1'\in[-2,2]$. In all three
models, $\overline{\vF}_1$ generally shifts stars to lower $J_r$ but with a
characteristic swirling action that causes $L$ to sometimes decreases and
sometimes increase. In anisotropic models this swirling pattern is broken at the right-hand edge of
the populated part of the $LJ_r$ plane, where the restriction $Q>0$ gives
rise to a cliff-edge in $\overline{f}(\widetilde\vJ)$: stars naturally diffuse over this
edge notwithstanding the fact that a small distance to the left of this edge
stars are diffusing in almost the opposite direction. Since $\widetilde\vn$
sets the direction of its contribution to $\overline{\vF}$ (equation~\eqref{eq:Fform}),
the swirling pattern in the upper panels of Fig.~\ref{fig:ChrisFlux}
indicates that the dominant value of $\widetilde \vn$ changes as one moves through
action space.

In Fig.~\ref{fig:ChrisFlux} bare and dressed fluxes are plotted in black and
red arrows, respectively, with black over-plotting red where they coincide.
The length of an arrow is proportional to
$\log_{10}[|\overline{\vF}|/(10^{-12}Mb^{-2})]$. If
$|\overline{\vF}|<10^{-12}Mb^{-2}$ no arrow is plotted, hence the blank
regions in the panels. A red line at the upper right of each panel indicates
how the length of an arrow increases when $|\overline{\vF}|$ is multiplied by
100. The shortness of this line makes clear the large dynamical range covered
by $\overline\vF$, largely as a consequence of the large dynamic range in each model's
phase-space density $f_0$.  In much of the upper three panels of
Fig.~\ref{fig:ChrisFlux} the red and black arrows are aligned but the red
arrow is significantly longer.  In the lower left corner arrows of each
colour point in significantly different directions, indicating that the
contributions from some vectors $\widetilde\vn$ are amplified by self-gravity much more
than others.

The strong enhancement of $\overline{\vF}_1$ by self-gravity is a natural
consequence of the finding of~\cite{Weinberg1991} and~\cite{Weinberg1994} that the
$\ell=1$ modes of clusters can be very weakly damped. These modes involve the
cluster's core and halo moving in antiphase along a line, so the linear
momenta of the moving parts cancel. 
The impact of these modes on
$\overline{\vF}_1$ is much enhanced by including self-gravity.  Displacing the
core with respect to the halo obviously becomes harder as the extent of
radial anisotropy increases, and careful examination of the arrows in the
upper row of Fig.~\ref{fig:ChrisFlux} confirms that including self-gravity
enhances $\overline{\vF}_1$ less when $R_\ra=2b$ than in the isotropic model.
We note that for the secular evolution of discs the tapering of the inner and outer regions 
that was implemented in~\cite{Fouvry2015b} prevented any contributions from these $\ell =1$ modes.

The lower three panels of Fig.~\ref{fig:ChrisFlux} show the fluxes $\overline{\vF}_2$,
obtained by adding the contributions of the 49 pairs
$(\widetilde\vn,\widetilde\vn')$ with $\widetilde n_1\in[-2,2]$ and
$\widetilde n_2\in[-2,2]$.  In
these panels red is much less in evidence than in the upper panels,
indicating that couplings associated with $\ell=2$ are less strongly affected
by self-gravity than those associated with $\ell=1$. Red is most evident in
the panel for $R_\ra=2b$, as is to be expected given the vulnerability of
this model to quadrupole distortions~\citep{MayB1986,Saha1991}. In all three lower
panels by far the largest $\ell=2$ fluxes occur in the bottom left corner, so
physically in the cluster's core. There the flux shifts stars to larger $J_r$
and smaller $L$. In the nearly unstable model $R_\ra=2b$, the flux outside
this core region moves stars to smaller $J_r$ and larger $L$, and thus tends to
reduce the anisotropy that inclines the model to the radial-orbit
instability.

Fig.~\ref{fig:totFlux} shows sums of the $\ell=1$ and $\ell=2$ fluxes.
\begin{figure*}
\centerline{
\includegraphics[height=.3\hsize]{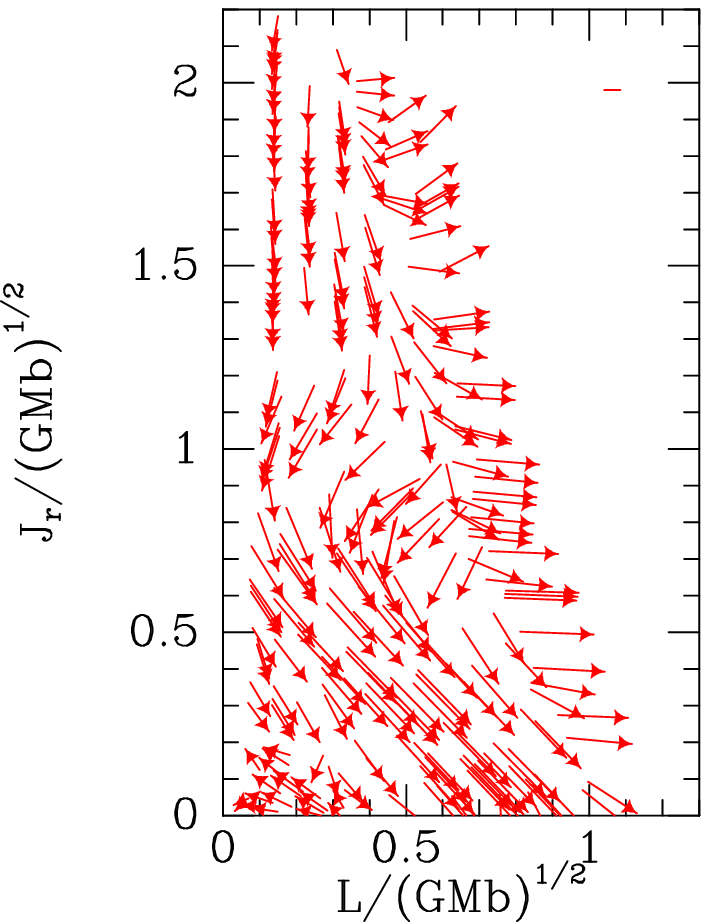}
\includegraphics[height=.3\hsize]{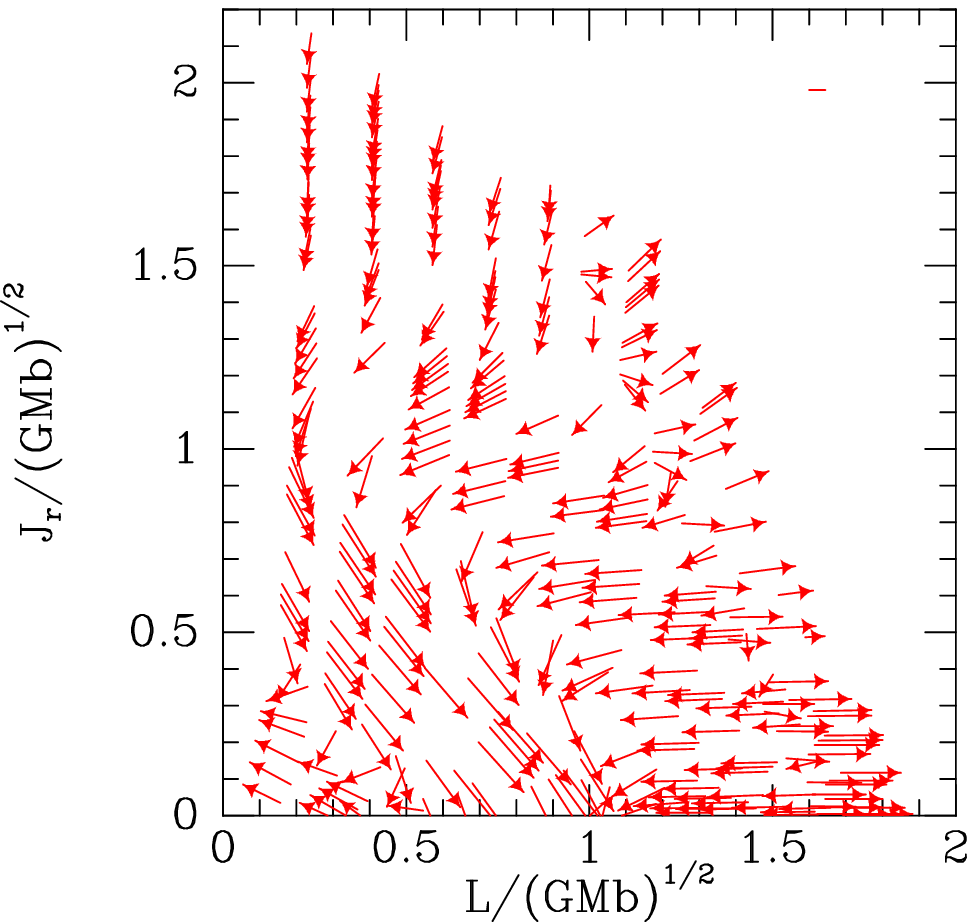}
\includegraphics[height=.3\hsize]{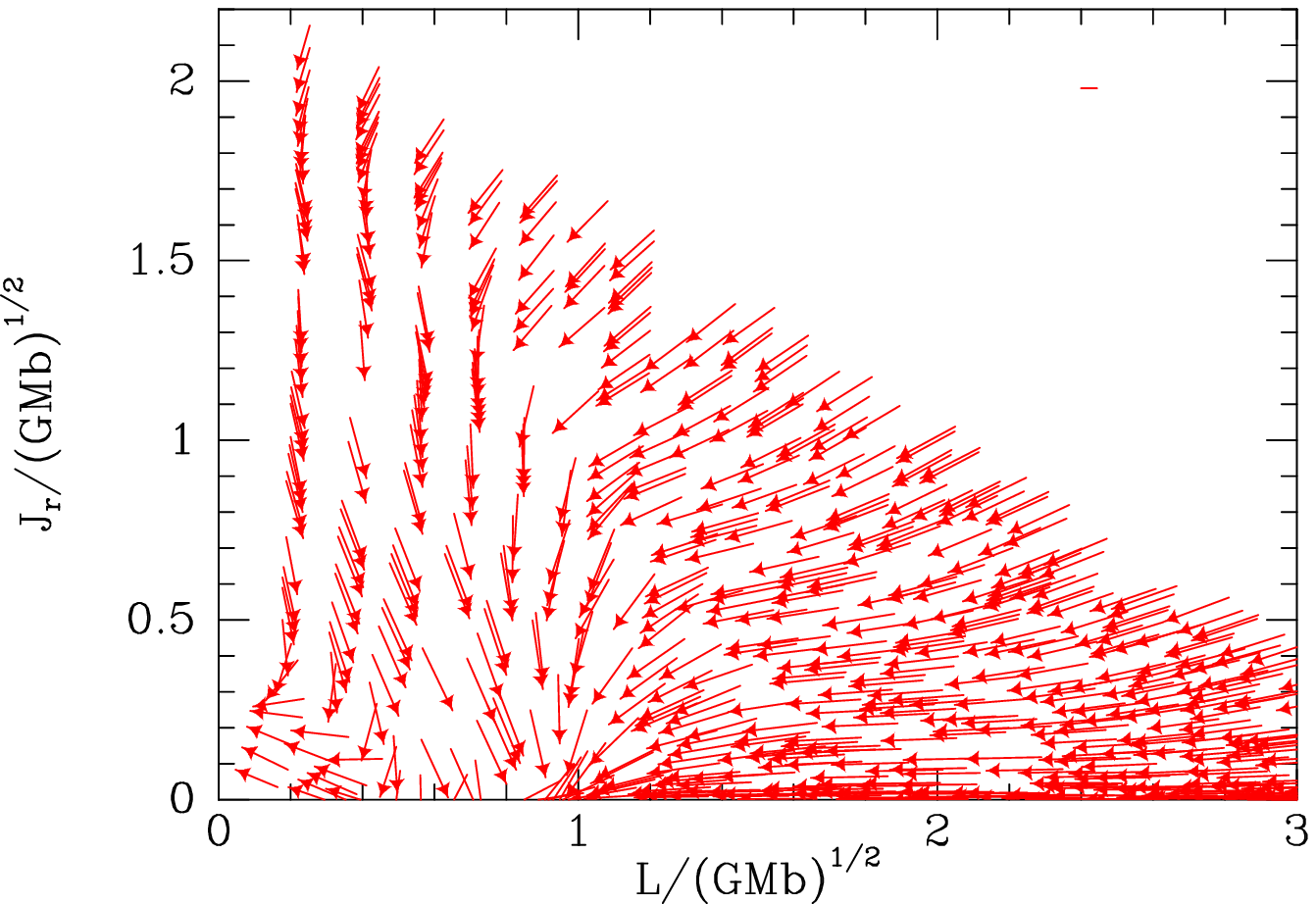}
}
\caption{The result of adding the $\ell=1$ and 2 contributions to
$\overline{\vF}$. As in Fig.~\ref{fig:ChrisFlux} panels are for models with
$R_\ra=2b$ (left), $R_\ra=4.2b$ (centre) and the isotropic model (right) and the arrows are scaled in the same
way.}\label{fig:totFlux}
\end{figure*}
In the region of action space that is populated in all three models, the
structure of the flow does not differ greatly between models.  As $R_\ra$
increases, the populated part of phase space extends into a region in which the
flow is towards smaller $L$.

Away from the cluster cores, the total fluxes are very similar to the
$\ell=1$ fluxes because the latter are significantly larger than the $\ell=2$
fluxes. In the core $\overline{\vF}_2$ dominates and pushes stars to smaller
$L$ and slightly larger $J_r$.  Around the core $\overline{\vF}_1$ and
$\overline{\vF}_2$ are in opposite directions but the net flow reduces $J_r$.
In the model with $R_\ra=2b$ the flow generally increases $L$, thus
diminishing the model's radial anisotropy.

We find the largest of $\overline{\vF}_0$, $\overline{\vF}_3$ and
$\overline{\vF}_4$ is always smaller than $\overline{\vF}_2$, so we do not
present our values for them.

Fig.~\ref{fig:divFBL} plots $\p \overline{f} /\p t$ for the models
with $R_\ra/b=2$ and $4.2$ (left and centre) and the isotropic model (right)
computed from the divergence of the dressed BL fluxes plotted in
Fig.~\ref{fig:totFlux}.
\begin{figure*}
\centerline{
\includegraphics[width=.3\hsize]{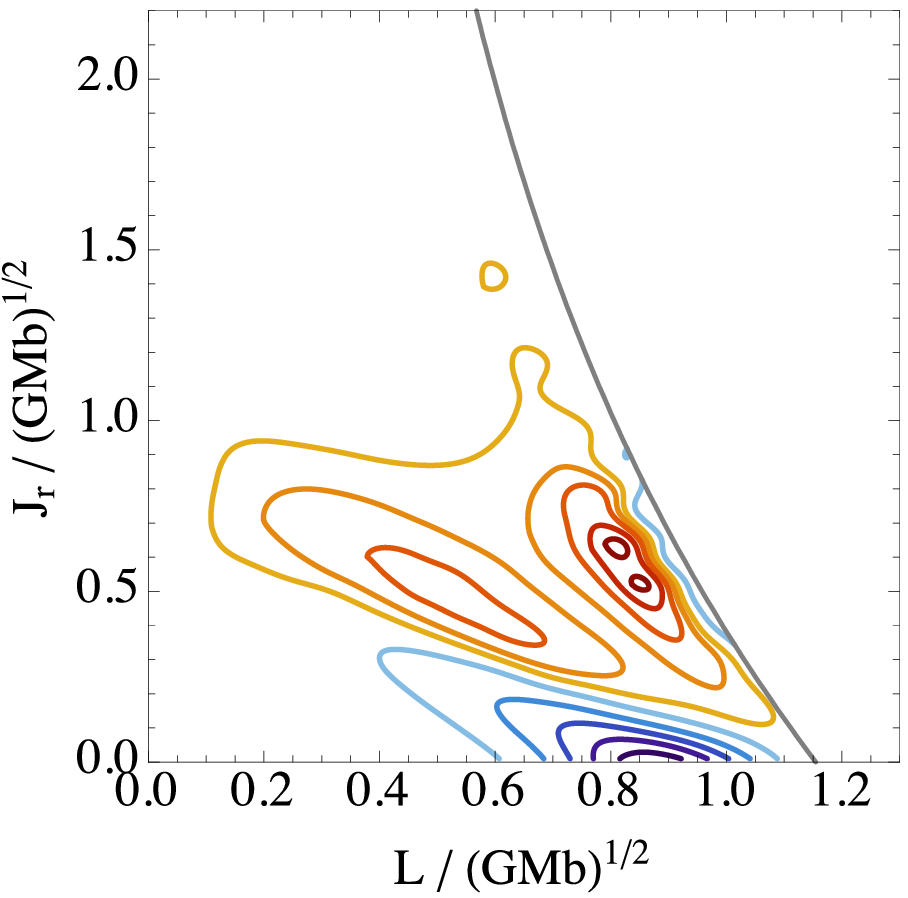}
\includegraphics[width=.3\hsize]{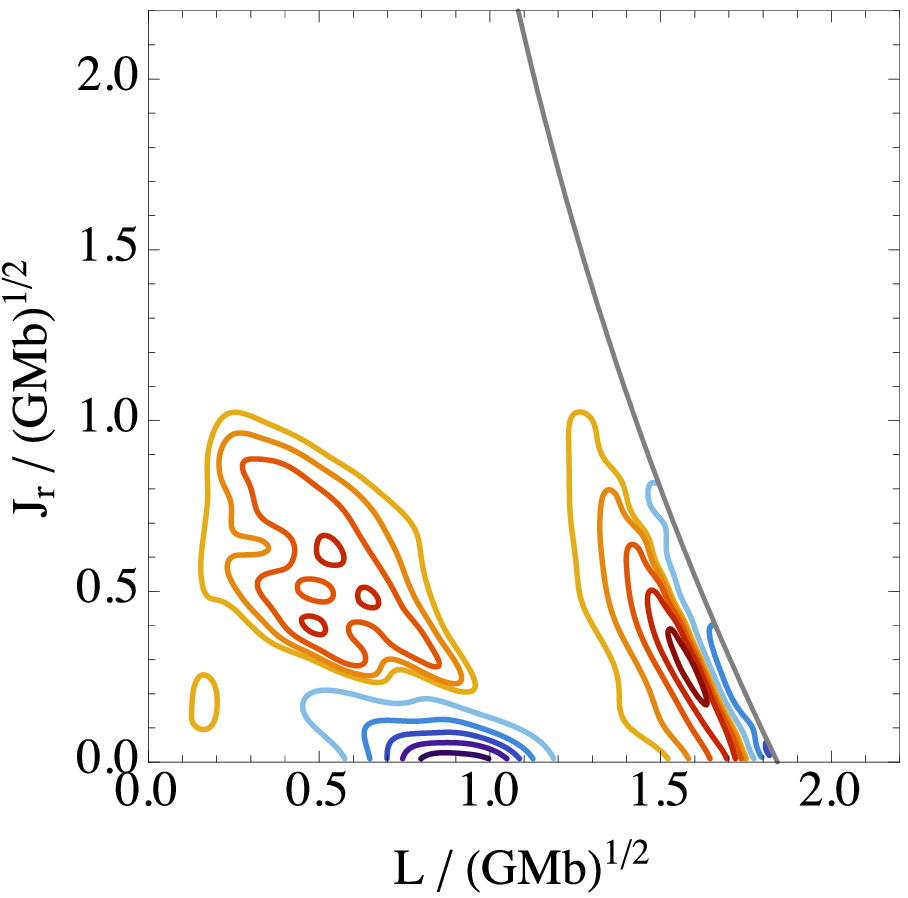}
\includegraphics[width=.3\hsize]{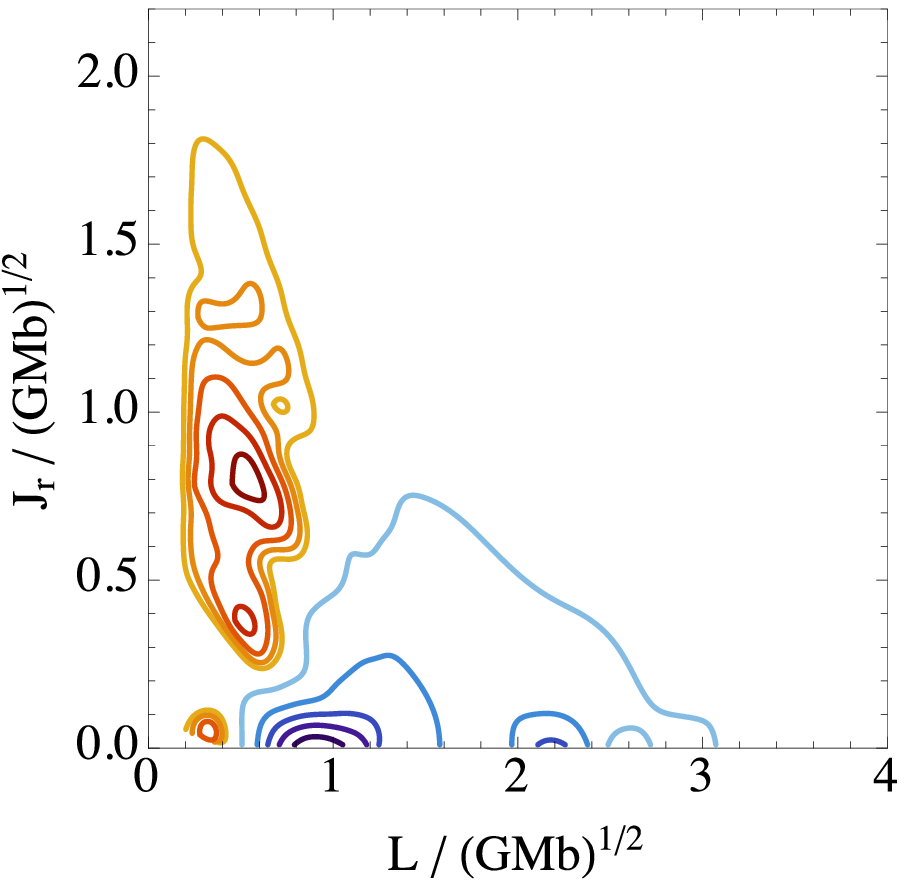}
}
\caption{The rate of change of the DF, $\p\overline{f}/\p t = -\p/\p\widetilde\vJ \cdot \overline\vF $
for the models with $R_\ra=2b$ (left), $R_\ra=4.2b$ (centre) and the isotropic model (right).
These rates are minus the divergence of the dressed BL fluxes.
Red contours indicate that $\overline{f}$ is decreasing, blue that it is increasing. The
boundary of the allowed action space domain is represented by the heavy black line.
}\label{fig:divFBL}
\end{figure*}
The red contours indicate regions in which
$\overline{f}$ is decreasing, while blue contours mark regions in which it is
increasing. The data are quite noisy because they
involve differentiation of the numerically computed fluxes. However, in all
three models $\overline{f}$ increases at $J_r\simeq0$ and $L\simeq\sqrt{GMb}$
and decreases at high $J_r$ and small $L$.
In order to illustrate the variety of diffusion features that can generically arise from the BL
equation, in Appendix~\ref{App:Tangential} we briefly consider the case of 
a tangentially anisotropic DF and illustrate again how self-gravity may impact
the properties of the orbital diffusion.

\section{Comparison with classical theory}
\label{sec:compare}

Let us now 
compare the results of the BL formalism to those of the classical (Spitzer--Chandrasekhar) theory of local
scattering.
In Appendix~\ref{App:class} we show that classical theory predicts
\[
{\p\overline{f}\over\p t}=-\sum_{i=1}^2{\p\over\p J_i}(\overline F_1^i + \overline F_2^i),
\]
 where the fluxes $\overline\vF_i$ are  given by
\begin{align}\label{eq:tildeF}
\overline F_1^i&=(2\pi)^{-2}\overline{f}
\int\d\theta_r\,\d\psi\,\Bigl[D^i-\fracj12\Gamma^i_{jn}D^{jn}\Bigr]\cr
\overline F_2^i&=-\fracj12(2\pi)^{-2}{\p\over\p J_j}\biggl(\overline{f}\int\d\theta_r\,\d\psi\,D^{ij}\biggr),
\end{align}
(here we are using the summation convention over repeated indices). Here $D^i$ and $D^{jn}$ are rates of change of expectation values that can
be obtained from the Rosenbluth potentials~\citep{RosenbluthMJ1957},
$\Gamma^i_{jn}$ is a Christoffel symbol that emerges from the functional
dependence of the actions on velocity, and $\theta_r$ and $\psi$ are angles
appearing in this dependence. 

Evaluation of the Rosenbluth potentials for an anisotropic velocity
distribution is costly, so, as is standard practice in studies of
globular-cluster evolution~\citep[e.g.][]{Cohn1979,Drukier1999}, we evaluate
the Rosenbluth potentials using an equivalent ergodic DF
\[
f_{\rm iso}(E,r)=\int_0^1\d c\,f\left(E,{r^2v^2_{\rm
E}\over2R_\ra^2}(1-c^2)\right),
\]
 where $v_{\rm E}(r)$ is the speed at $r$ of a particle of energy $E$.  After
computing the Rosenbluth potentials with $f_{\rm iso}$, the appropriate
anisotropic DF $\overline{f}$ is used in equations~\eqref{eq:tildeF} for the
flux.

Fig.~\ref{fig:FJflow} shows for two finite values of $R_\ra$, namely
$R_\ra=2b$ and $4.2b$, and for the isotropic model, the resulting classical
flux vectors scaled as in Figs.~\ref{fig:ChrisFlux} and~\ref{fig:totFlux} for
a cluster of $N=10^5$ equal-mass stars under the assumption that
$\ln\Lambda=8.9$ (Section~\ref{sec:Coulomb}).
\begin{figure*}
\centerline{\includegraphics[height=.3\hsize]{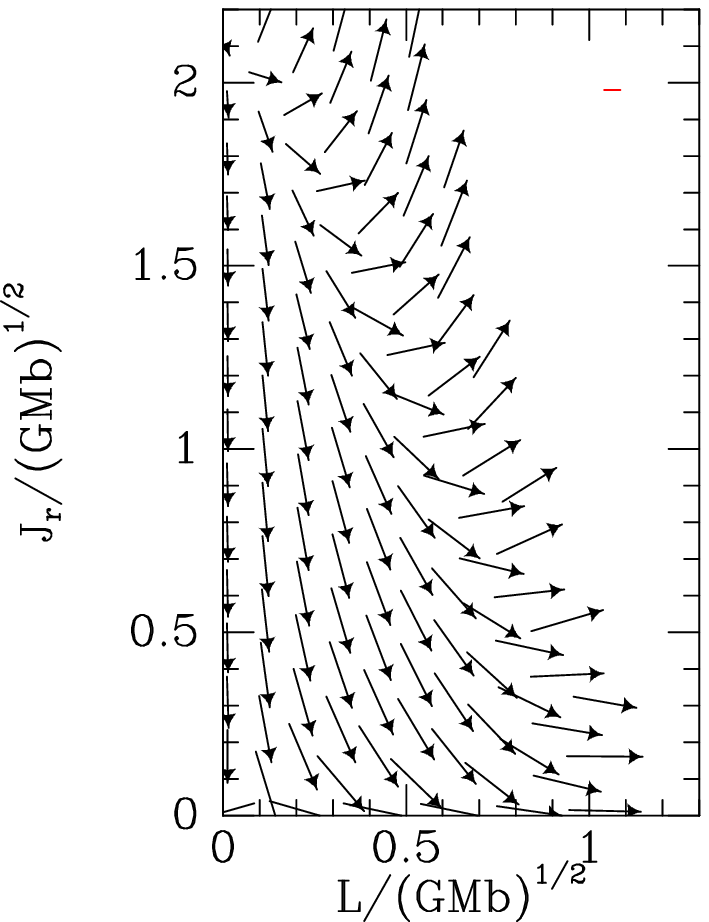}
\includegraphics[height=.3\hsize]{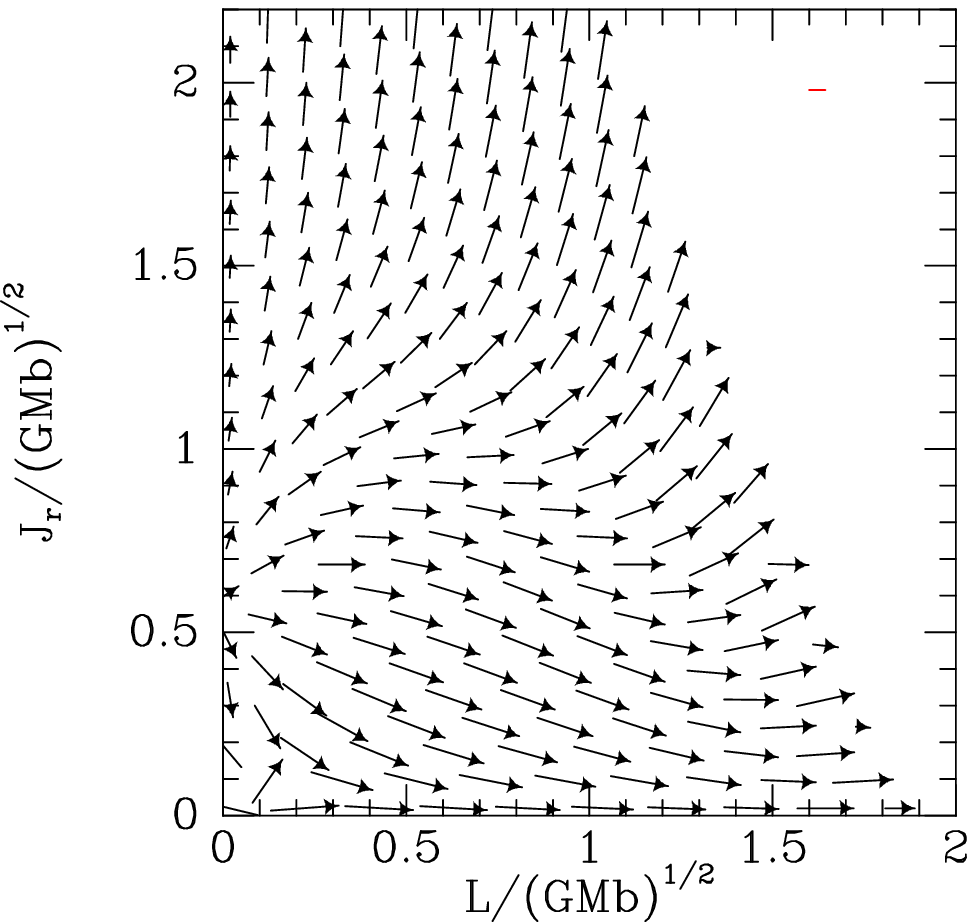}
\includegraphics[height=.3\hsize]{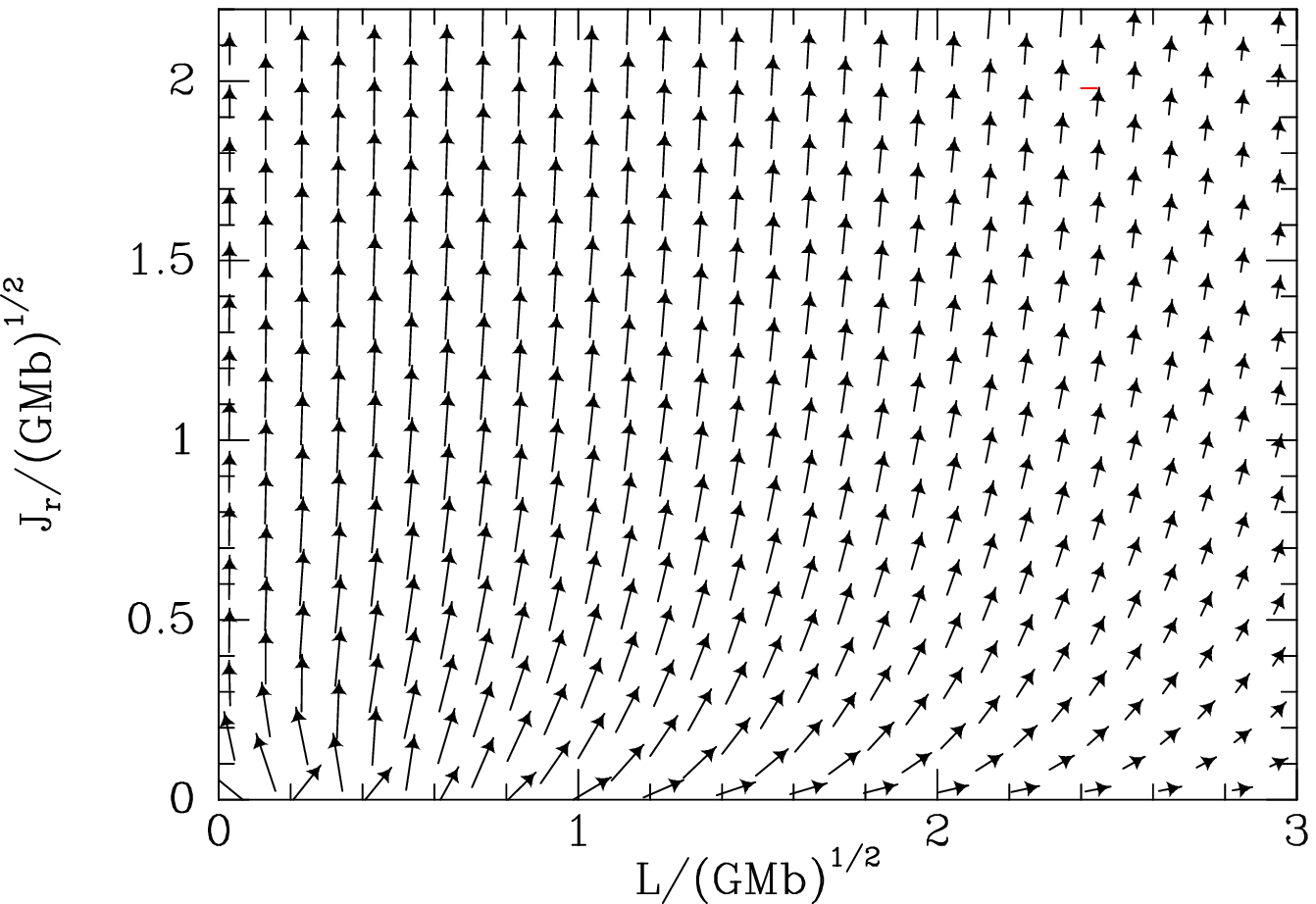}}
\caption{Flux through $\widetilde\vJ$ space, $\overline{\vF}(\widetilde \vJ)$, computed using the local
approximation for models with $R_\ra=2b$ (left), $4.2b$ (centre) and the
isotropic model
(right). The length of an arrow is proportional to
$\log_{10}[|\overline{\vF}|/(10^{-12}Mb^{-2})]$, so that an arrow of zero length corresponds to $|\overline{\vF}|=10^{-12}Mb^{-2}$
as in Fig.~\ref{fig:ChrisFlux}. The red bar at top right corresponds to a
hundred-fold increase in $ |\overline{\vF}|$.}
\label{fig:FJflow}
\end{figure*}
The flow is structured by stagnation points.  For $R_\ra=2b$ there is a single stagnation point that lies on the $J_r$ axis at $J_r\sim2.0\sqrt{GMb}$.  This point moves down the $J_r$ axis as $R_\ra$ increases, so that by
$R_\ra =4.2b$ it is at $J_r\sim0.6\sqrt{GMb}$.  The $R_\ra =4.2b$ plot also shows signs of another stagnation point on the $L$ axis, roughly at $L\sim 0.1(GMb)^{1/2}$ (see Appendix \ref{sec:numericaldetails}).  For larger $R_\ra$ we have a single stagnation point on the $L$ axis, where it sticks
as $R_\ra$ tends to infinity. Above the stagnation point and near the $J_r$
axis the flow is vertically upwards, while below the stagnation point the
flow is to lower $J_r$ and higher $L$.

The classical flow pattern for the model with $R_\ra=2b$ is not unlike that
obtained from the BL equation (left panel of Fig.~\ref{fig:totFlux}). Given
that in computing the BL fluxes we have only included the $\ell=1,2$ terms
and confined ourselves to wave vectors $\widetilde\vn$ with $\widetilde n_1\in[-2,2]$, the
extent of the agreement between the two panels on the left of
Fig.~\ref{fig:totFlux} and the corresponding panels of Fig.~\ref{fig:FJflow}
is remarkable. As we proceed to larger $R_\ra$, the differences between the
classical and the BL flows becomes pronounced. In the isotropic model the BL
flow is everywhere towards smaller $J_r$ whereas the classical flow is
towards larger $J_r$. The tendency of the classical flow to increase
$J_r$ is consistent with the well known tendency
of the halos of clusters to become radially biased. 

Fig.~\ref{fig:rate} shows the rate of change of the DF $\p \overline{f} / \p t$ (in units of $\sqrt{M/Gb^5}$) for
the models with $R_\ra/b=2$ and $4.2$ (left and centre) and the isotropic model (right),
all computed from the divergence of the classical flux vectors shown in Fig.~\ref{fig:FJflow}.
\begin{figure*}
\centerline{
\includegraphics[height=.27\hsize]{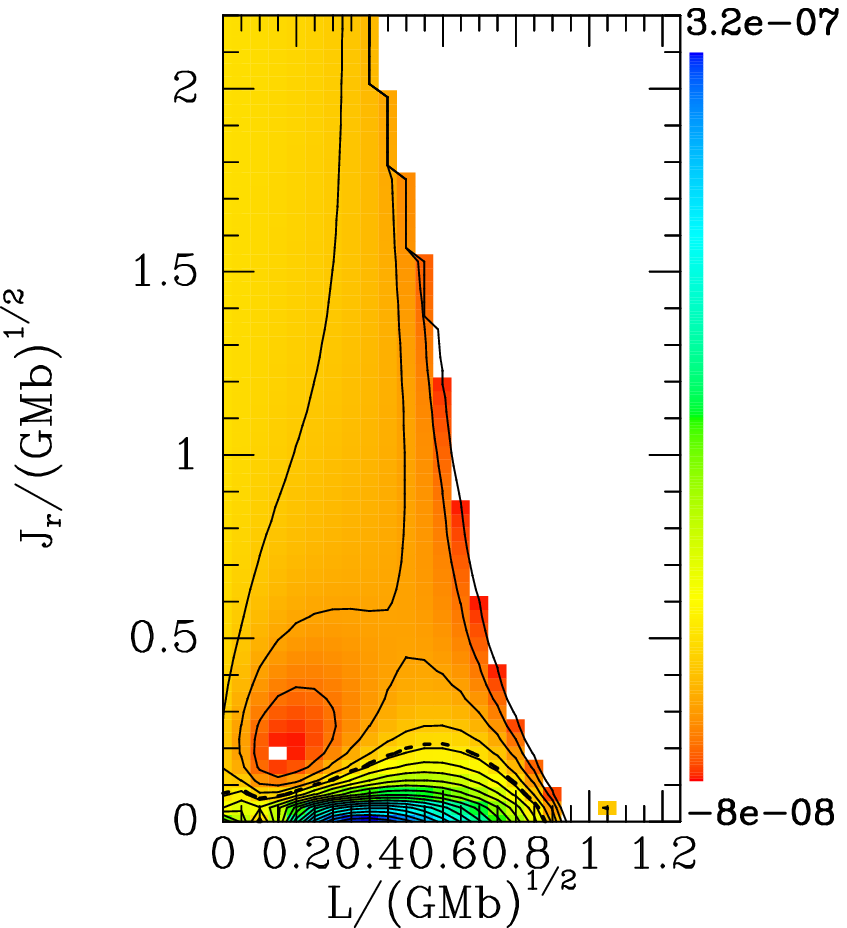}
\includegraphics[height=.27\hsize]{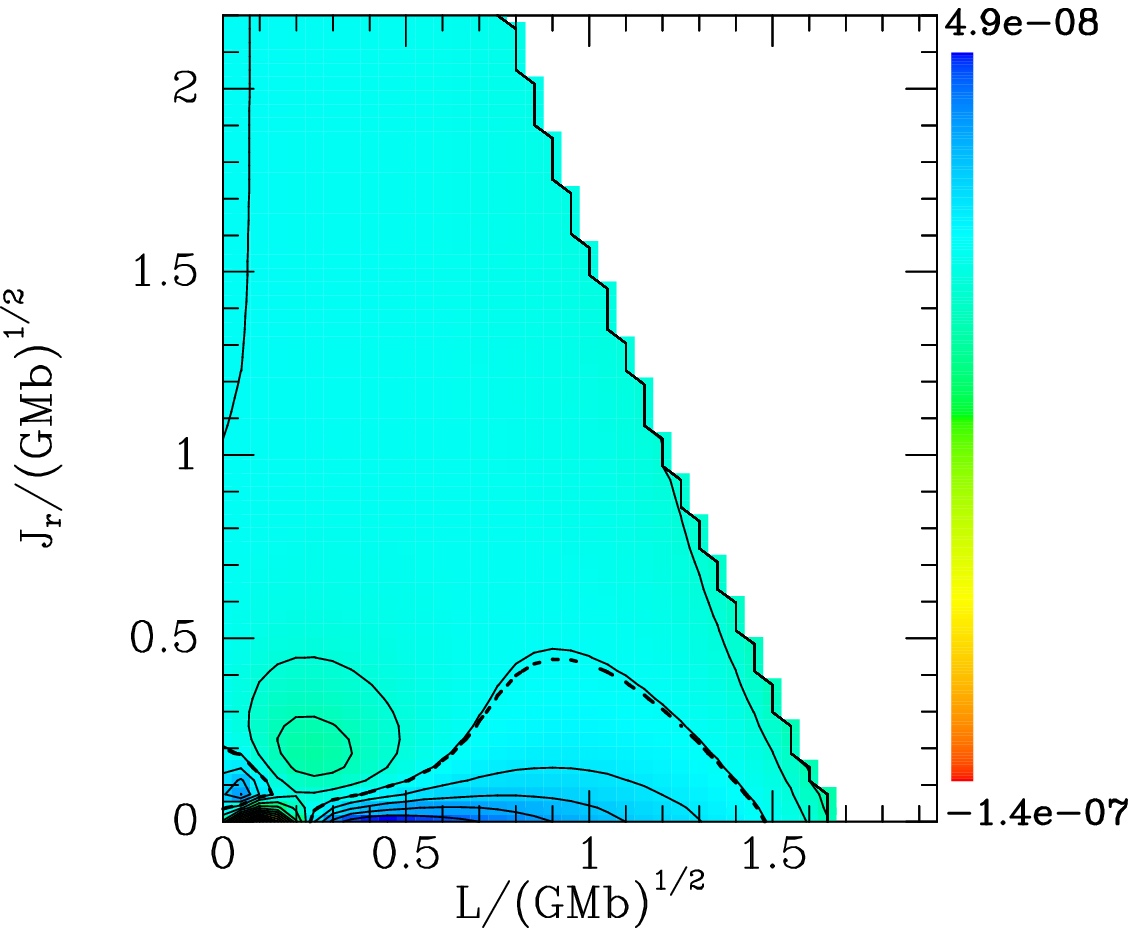}
\includegraphics[height=.27\hsize]{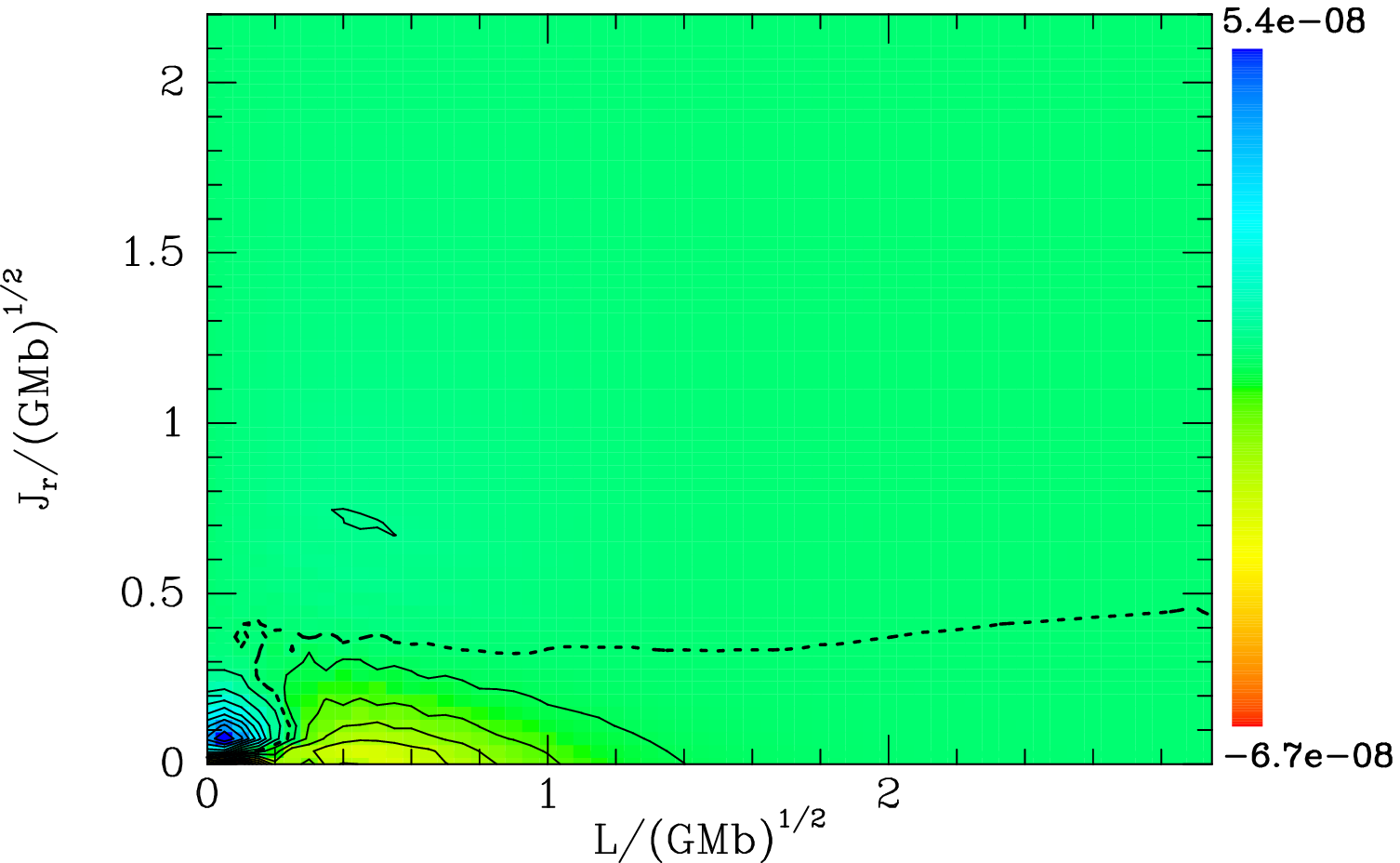} }
 \caption{The rate of change of the DF, $\p\overline{f}/\p
t=-\p/\p\widetilde\vJ \cdot \overline\vF$, computed in the local
approximation for the model with $R_\ra=2b$ (left), $R_\ra=4.2b$ (centre), and for the
isotropic model (right). In each case we plot 20 solid contours, spaced linearly from the minimum to maximum value of $(\partial \overline{f}/\p t)/\sqrt{M/Gb^5}$. The dashed contours divide regions of increasing and decreasing phase-space
density.}\label{fig:rate}
\end{figure*}
In each panel the dashed contour divides the regions in which the
DF is increasing from those at which it is decreasing.
Whereas in
the isotropic model $\overline\vF$ tends to shift stars from nearly circular
orbits to more eccentric ones, in the highly anisotropic model $R_\ra=2b$ the
portion of action space associated with eccentric orbits is depopulated to
feed growth in the region associated with nearly circular orbits.

Fig.~\ref{fig:ratio} compares the magnitudes of the BL fluxes in each panel
of Fig.~\ref{fig:totFlux} to the classical fluxes plotted in
Fig.~\ref{fig:FJflow}.
\begin{figure*}
\centerline{ \includegraphics[height=.27\hsize]{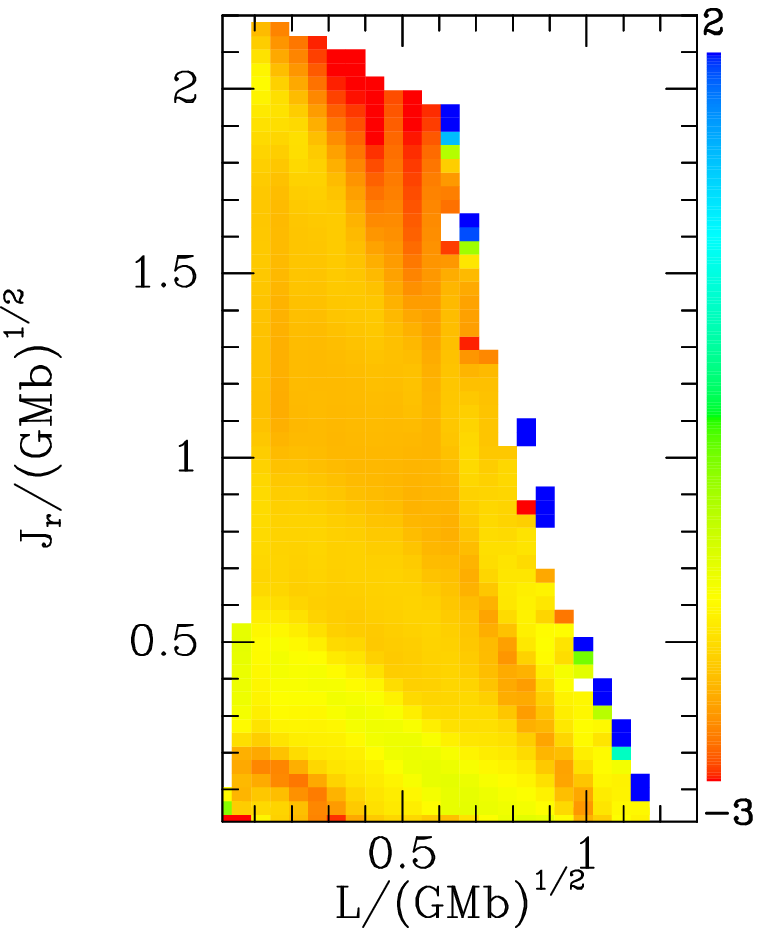}
\includegraphics[height=.27\hsize]{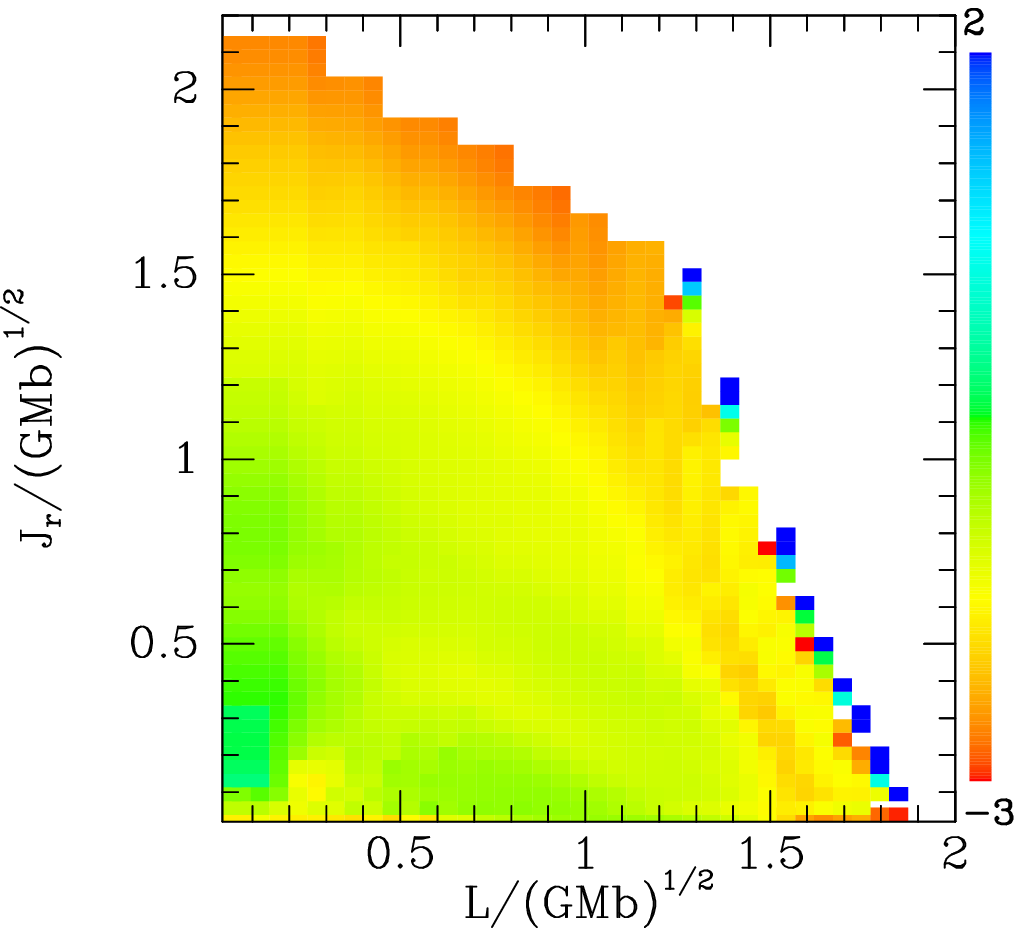}
\includegraphics[height=.27\hsize]{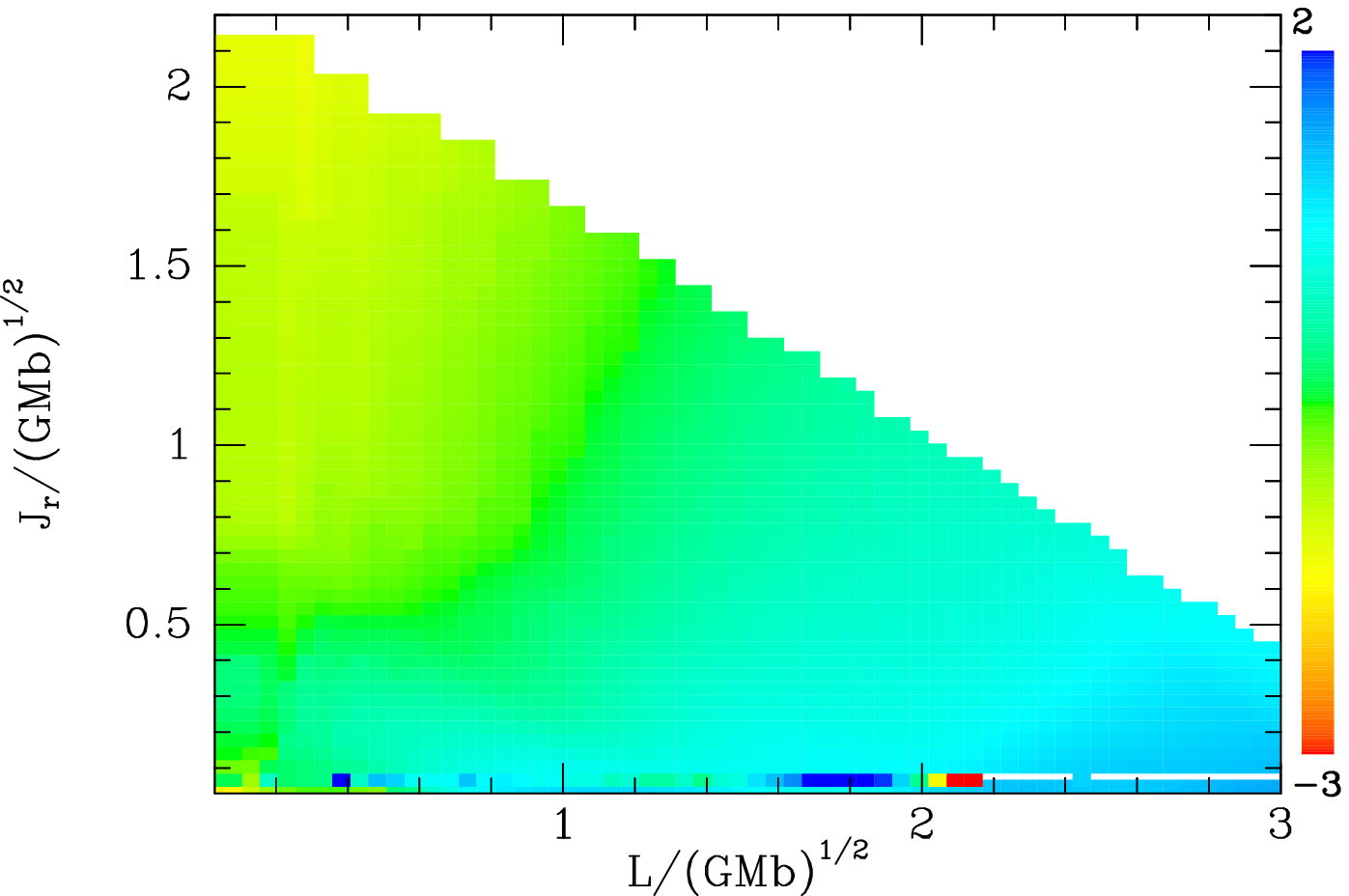} 
} 
\centerline{
\includegraphics[height=.27\hsize]{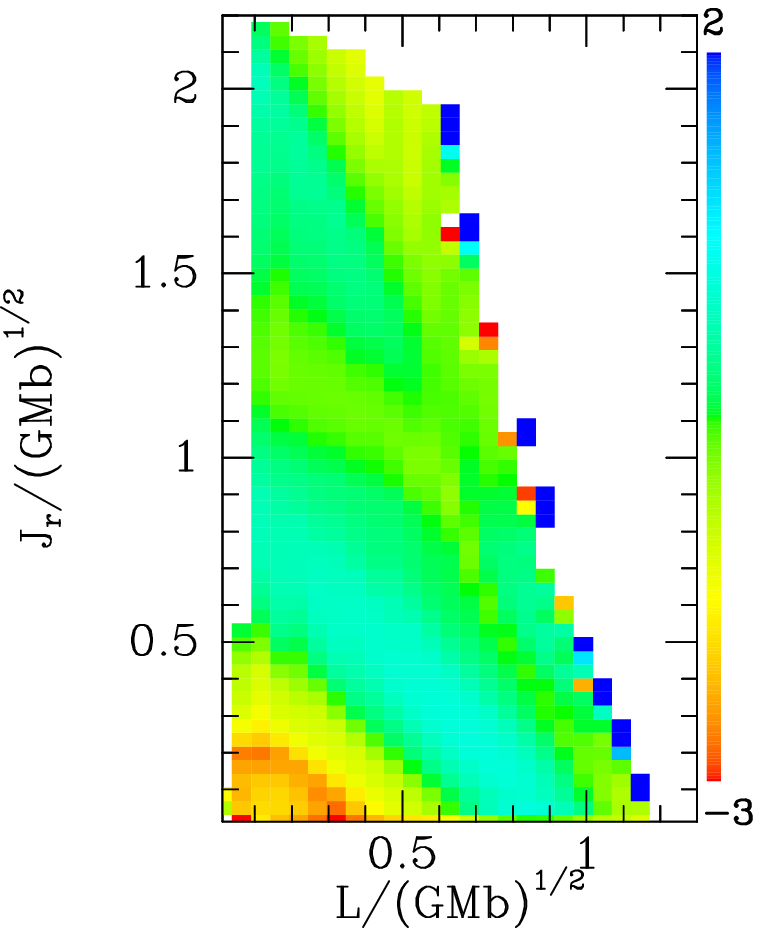}
\includegraphics[height=.27\hsize]{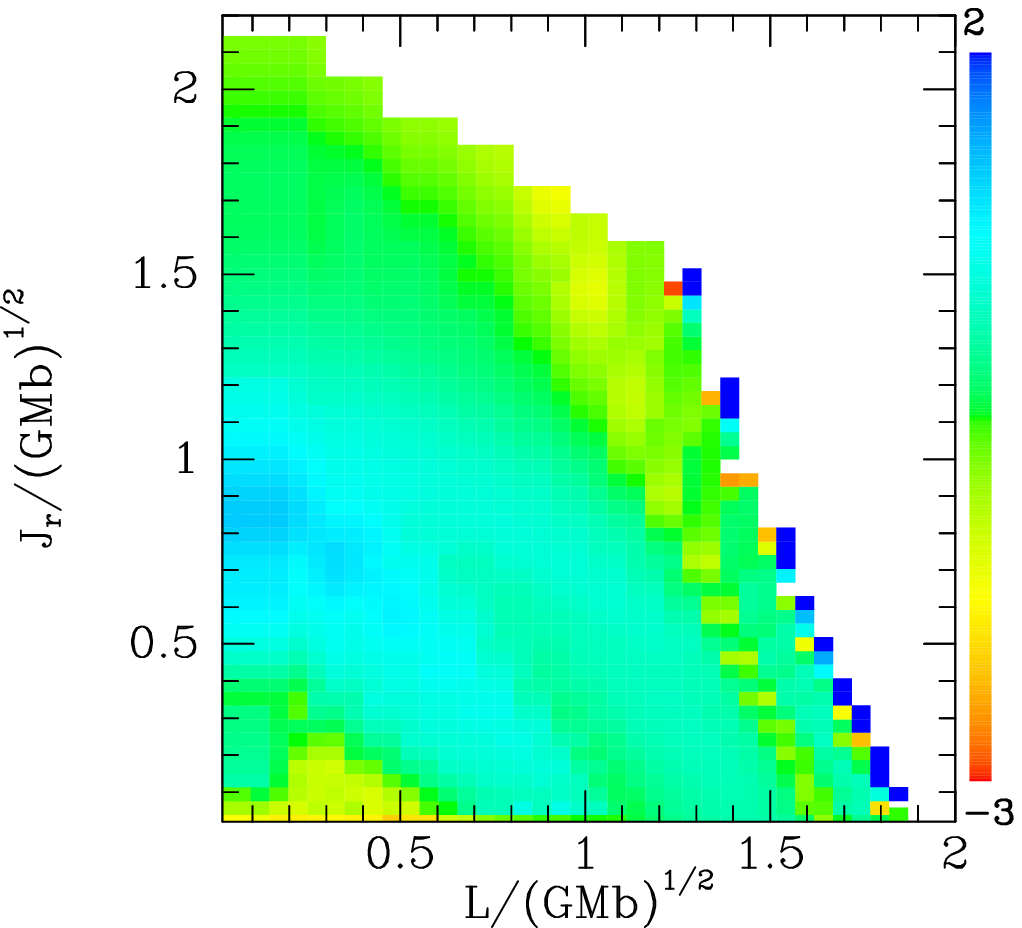}
\includegraphics[height=.27\hsize]{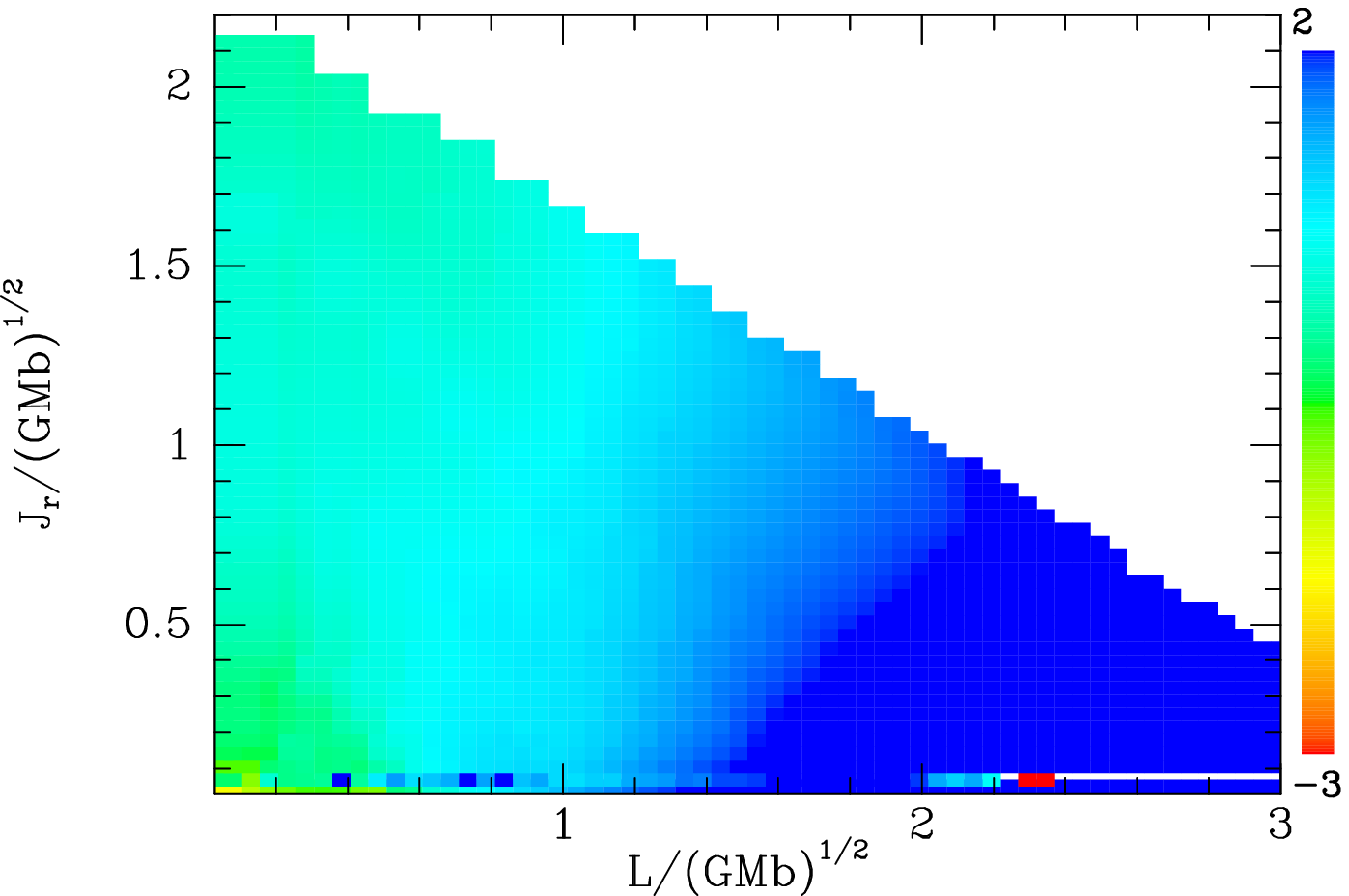} 
}
 \caption{Ratios of the magnitudes of BL fluxes to classical fluxes. Upper
row: bare fluxes; lower row: dressed fluxes. The left and central panels are
for the models with 
$R_\ra=2b$ and $4.2b$, respectively, while the right panel is for the
isotropic model. The colour scales shows $\log_{10}(|\vF_{\rm
BL}|/|\vF_{\rm Class}|)$.}
\label{fig:ratio}
\end{figure*}
The upper row is for the bare fluxes while the lower
row is for dressed fluxes. The colour scales indicate logarithms to base 10
of the ratio of the BL flux to the classical flux. Whereas the bare BL fluxes
are almost everywhere significantly smaller than the classical fluxes, outside the core the
dressed fluxes are as large as or larger than the classical fluxes. The BL fluxes, both bare
and dressed, tend to increase relative to the classical fluxes as $R_\ra$
increases.

\section{Discussion}
\label{sec:discuss}

\cite{Chavanis2013b}~\citep[see also][]{Chavanis2013} has shown that the BL
equation yields the classical theory of relaxation when one neglects
inhomogeneity and collective effects.  Indeed, in the limit of a homogeneous
system, the three components of velocity take on the role of actions, the
Cartesian coordinates $x_i$, which increase uniformly along unperturbed
trajectories, play the role of angles, and the sum over discrete vectors
$\vn$ becomes an integral over wavevectors.  Neglect of collective effects
enables one to execute this integral and recover the classical theory. Thus
there is little doubt that if we could sum over all resonating pairs
$(\widetilde\vn,\widetilde\vn')$ and all $\ell$, we would obtain a flux that embraced the
classical flux but went beyond it in that (i) our flux would include the
impact of large-scale  amplified fluctuations that are not properly handled in the
classical theory, and (ii) the flux would not depend on an ill-determined
parameter such as $\log\Lambda$.

Since collective effects should be unimportant on small scales, it follows
from Chavanis' work that for some choice of $\log\Lambda$, classical theory
should yield a good approximation to the portion of the diffusive flux that
arises from encounters at small impact parameters -- we shall refer to this
as the `small-scale' BL flux. With our current computational machinery it is
not feasible to compute the small-scale BL flux by summing to high $\ell$
and/or high $|\widetilde n_1|$ or $|\widetilde n_2|$. Hence our BL fluxes arise exclusively from the part of
the BL sum that returns the flux driven by interactions between stars for
which the local approximation of classical theory is certainly invalid.

For our comparison to the classical flux we have used $N=10^5$.  The BL flux is proportional to the mass $\mu$ of a star and therefore scales with the number of particles like $1/N$ (Section \ref{sec:BL}).  The classical flux is proportional to $\mu \log \Lambda$ (Appendix \ref{App:class}) and so scales like $\ln N / N$.  Therefore the ratio of BL to classical flux scales with particle number like $1/\ln N$.

Given the ill-defined value of the Coulomb logarithm, the magnitude of the
classical flux is uncertain. We have adopted $\log\Lambda=8.9$, to get an
idea of the magnitude of the small-scale BL flux. 
Our key finding
is that, when the cluster's self-gravity is taken into account, the
large-scale BL flux is as large as or even larger than the small-scale flux.
Including self-gravity dramatically increases the diffusion generated by
thermal excitation of the weakly damped, low frequency $\ell=1$ mode of
clusters. This finding makes perfect sense physically, because this mode
involves the cluster's core and envelope oscillating in antiphase and
periodically exchanging momentum. If the stars move in a fixed gravitational
field, this mode of communication is impractical because the core
then  oscillates at a much higher frequency than the halo, and each section
will exchange momentum with the source of the fixed gravitational field
rather than the other half of the cluster.

When there are many orbits that move between the core and the halo, it will
be harder for the core and the halo to oscillate in antiphase, so the
$\ell=1$ mode will be harder to excite. Hence for such modes it is natural that the
magnitude of the large-scale BL flux relative to that of  the small-scale
flux  is decreased by making the cluster radially anisotropic.

It is well known that as radial anisotropy increases the energy required to
make a cluster prolate decreases, and at a critical level of anisotropy it
vanishes so the cluster is liable to spontaneous bar formation. Our results
confirm the view of~\cite{MayB1986} that the critical level of anisotropy is
reached near $R_\ra=1.7b$. We find, as expected, that the $\ell=2$
contribution to the dressed BL flux associated with quadrupole symmetry
increases as $R_\ra$ decreases through thermal excitation of the quadrupole
mode.  We have not, however, computed a model that is sufficiently close to
the critical anisotropy for the enhancement by self-gravity to be comparable
to that found for the $\ell=1$ contribution to the flux in every model.

We have found that contributions to $\overline{\vF}$ from $\ell=0,3,4$ are
much smaller than contributions from $\ell=1,2$.  From these conclusions it
does not follow that it was legitimate to neglect these and other
$\overline{\vF}_\ell$,
nor does the large magnitude of our dressed BL fluxes justify limiting the
sums over wavevectors to small $\widetilde\vn$.  Indeed, classical theory,
which consists of an approximate summation over the neglected terms, strongly
suggests that the terms we have dropped are collectively important.  Each
wavevector $\widetilde\vn$ in the sum contributes a flux in its own unique
direction. The direction of our BL flux is set by the vectors $\widetilde\vn$
we have chosen to include, while the classical flux is the result of summing
over an infinite number of vectors, and in consequence it can point in a
radically different direction, as comparison of the panels on the right of Figs.~\ref{fig:totFlux} and~\ref{fig:FJflow} shows to be the case
in the isotropic model.

This discussion suggests that the partial BL
summation on the one hand and the classical theory on the other hand
provide distinct approximations to the flux. 
A summation over the leading discrete $\widetilde\vn$,  combined with
summation over a myriad of other $\widetilde\vn$ that make small individual
contributions to the flux may eventually prove sufficient~\citep{Weinberg86}. Most of these small contributions are expected to
come from small-scale fluctuations in the potential that are probably
adequately modelled by the orbit-averaged classical theory because neither inhomogeneity nor
collective effects are important on small scales. Consequently, a proper
procedure may involve computing the dominant terms in the dressed BL expressions for the
diffusion coefficients and  approximating the remaining infinite sum by
the classical diffusion coefficients using a smaller value of $\log\Lambda$.
The remaining problems are (i)
expediting the computation of the BL terms, which is unreasonably costly with
our current code and  (ii) estimating the correct value of $\log\Lambda$.

\subsection{N-body simulations}

If classical theory is so misleading, why have N-body simulations of cluster
relaxation not exposed this fact?  A likely answer has two parts. First,
classical theory has the free parameter $\log\Lambda$, and second analysis of
N-body simulations is hard and the action-space flux $\overline{\vF}$ is
rarely computed. Rather the rate at which the model's radial density profile
evolves is computed, and $\log\Lambda$ is adjusted to make the classically
predicted rate agree with the measured rate. \cite{Theuns1996} did compute
the rate at which particles in N-body clusters (in this case isotropic King models) diffused in energy.  He
concluded that on average they diffused about twice as fast as classical theory
predicted, and that outside the core classical theory underestimated the size of the the dynamical friction coefficient by up to a factor of $10$.  These conclusions are roughly in agreement with the lower-right panel of our Fig.~\ref{fig:ratio}.   Additionally, the excess power due to large scale modes has been seen in N-body simulations (for example those of \cite{Weinberg1998}; see also \cite{Weinberg2001}). The patterns of the weakly damped modes themselves might simply be too hard to see in real space because they have such low frequencies and are soon washed out by noise \citep{Weinberg2001}. \citet{Sellwood2015} performed N-body simulations of homogeneous spheres and found a scaling of the energy relaxation with $N$ which strongly suggested the dominance of large-scale collective modes driven by Poisson noise.  On the other hand, \citet{Sellwood2015} also simulated Hernquist and Plummer spheres and found fairly good agreement with the classical picture.  The changes to the relaxation rate that can result from tuning $\log \Lambda$ to any sensible value are typically only $\sim 10-20 \%$ and so this alone cannot account for the consistent agreement which has been found between N-body and Fokker-Planck models.  Finally we note that \cite{Kim2008} measured a significantly smaller discrepancy in energy diffusion rate from the classical expectation than \cite{Theuns1996}.  

The validity of our conclusions should be tested by direct N-body simulations
of clusters. A suitable programme would comprise an ensemble of short
simulations of isochrone clusters.  Individual simulations could be integrated
for times significantly shorter than the half-mass relaxation time, so their
mean-field potentials would remain close to that of the isochrone, and the
actions of particles could be readily computed. By comparing actions computed
at successive timesteps, fluxes across a grid of lines in $LJ_r$ space could
be determined. Moreover, the difference between the actual potential and the
isochrone potential, projected onto a set of basis functions $\Phi_{n\ell m}$
could be monitored so one could estimate how much of the flux is generated by
large-scale fluctuations that are significantly enhanced by self-gravity and
cannot be well described by classical theory. Good statistics would be
accumulated by stacking results from simulations that differed only in their
initial conditions. We plan to present the results of such a study in the
near future.

\section{Conclusions}
\label{sec:conclude}

The classical theory of relaxation in star clusters has for decades formed
one of the pillars of stellar dynamics. Yet it is well known to rest on shaky
foundations in that its predictions all involve integrals that diverge
logarithmically at large impact parameters. Conventionally this divergence is
mastered by imposing an upper limit on impact parameters. Predictions then
become proportional to the logarithm of this cutoff, so the theory has a free
parameter. 

A further weakness of the classical theory is that it neglects the cluster's
self-gravity. In Section~\ref{sec:Nfluct} we gave a back-of-the-envelope
argument that by accounting for Poisson fluctuations alone one can recover the classical expression for the relaxation time.  Since self-gravity makes a star cluster more compressible than an ideal gas it follows that slow, large-scale fluctuations in density are liable to drive a cluster to relax faster than classical theory predicts.

In two important papers~\cite{Weinberg1993,Weinberg1998}
presented arguments that the self-gravity significantly speeds the relaxation
of stellar systems through  excitation of large-scale modes. However, he did
not provide an equation for the secular evolution of the DF when large-scale
modes and self-gravity are properly handled. Such an equation, derived from a
rigorous, self-consistent theory of relaxation that contains no arbitrary
constants, appeared first in~\cite{Heyvaerts2010}. The structure of this `BL
equation' suggests a physical picture of how systems relax that is
radically different from the classical picture of successive stellar
encounters. In Section~\ref{sec:BL} we summarised this theory, and in
Section~\ref{sec:sphere} we modified it so it could be applied to spherical
systems, which have a degeneracy that precludes application of the unmodified
equation. 

According to the new theory, relaxation does not occur through localised
encounters between stars but through pairs of stars communicating with each
other by shaking the entire system at a specific frequency $\omega$. The
stars do not need to be near each other in real space, although proximity
will make communication easier. The effectiveness of communication between
stars depends also on whether they can communicate at a frequency that lies
near a weakly damped mode of the entire system. Self-gravity is important
because it determines the modal structure.
\cite{Fouvry2015b} applied the
new theory to stellar discs and showed that it could for the first time
explain the emergence of strong spiral structure in  N-body simulations of
linearly stable discs. In particular,  they showed that when self-gravity is
properly included, realistic stellar discs can relax $\sim1000$ times faster
than classical theory predicts. Moreover, through a mechanism proposed by
\cite{Sellwood2014}, the resonant nature of the disc's relaxation steers
the disc to a configuration in which it is unstable as a collisionless
system, so it eventually develops a strong bar on a dynamical 
timescale.

In Section~\ref{sec:isochrone} we put forward the first implementation of  the BL theory to three spherical clusters
that all have the spatial structure of H\'enon's isochrone but distribution
functions with different degrees of radial velocity anisotropy.  The problem
one encounters is the need to sum over infinitely many resonant interactions,
each defined by a pair of two-dimensional vectors
$\widetilde\vn,\widetilde\vn'$ with integer components.  We tried to
identify, and then sum, the interactions that make the largest contributions.
The sum can be partitioned into contributions from each angular momentum
quantum number $\ell$, and we considered $\ell=0,\ldots,4$.  We found that
the sums for $\ell=1,2$ dominate strongly.  However, the number of vector
pairs that could contribute significantly grows rapidly with $\ell$ and the
size of the vectors, and it  is possible  for some configurations
 that large numbers of interactions
that individually contribute little together contribute as much as the most
important interactions~\citep{Weinberg86}. This being so,   our partial sum may
under-estimate the magnitude of the BL flux through phase space. In all three
clusters, the partial BL fluxes are much larger when self-gravity is included
than otherwise. This is to be expected because the terms we have included
arise from large-scale distortions of the cluster, which are precisely the
fluctuations     enhanced by self-gravity.

In Appendix \ref{App:class} we recovered the fluxes predicted by classical theory. In
Section~\ref{sec:compare} we compared these fluxes with our partial BL
fluxes. In the isotropic cluster, the fluxes have quite different directions,
and the classical flux is often larger than the bare partial BL flux,
consistent with our having missed the contributions of myriads of
interactions that individually contribute little. The dressed partial BL
fluxes are, by contrast, often larger than the classical flux, suggesting
that the classical flux significantly under-estimates the rate of relaxation.
The dominance of the dressed BL flux over the classical flux is most
pronounced in the isotropic cluster and at large $L$.

We conclude that classical theory is liable to under-estimate the rate of
relaxation through neglect of self-gravity and non-local resonances. Whereas this
can lead to the relaxation rate of a stellar disc being under-estimated by a
factor $\sim1000$, for a spherical system the magnitude of the relaxation rate predicted by (large scale) BL theory is typically $\sim$ a few times that of classical theory (but is often in a completely different direction and can be significantly larger depending on the anisotropy of the cluster).
 However, brute-force application
of the BL equation may not yet provide a  fully viable alternative to classical theory
because of the evaluation of the infinite sum over interaction vectors. 
A way must be found to combine a sum over the most important
resonance vectors, which are associated with large-scale fluctuations and the
important effects of self-gravity, with an integral like that executed in
classical theory to include the effects of small-scale fluctuations, to which
self-gravity contributes little. However, we should first confirm our  finding that classical theory misses
out some important physics through  analysis of tailored  N-body
simulations.

\section*{Acknowledgements}
JJB is supported by 
the European Research Council under the European Union's Seventh Framework
Programme (FP7/2007-2013)/ERC grant agreement no.~321067.
JBF acknowledges support from Program number HST-HF2-51374 which was provided
by NASA through a grant from the Space Telescope Science Institute, which is
operated by the Association of Universities for Research in Astronomy
Incorporated, under NASA contract NAS5--26555. This research was in part
carried out within the framework of the Spin(e) collaboration
(ANR-13-BS05-0005 \url{http://cosmicorigin.org})

\bibliographystyle{mn2e}
\bibliography{references}

\appendix
\def\exf#1{\langle#1\rangle_{\rm f}}

\section{Tangentially biased distribution functions}
\label{App:Tangential}

The self-consistent isochrone DF introduced in equation~\eqref{barrydf} is radially
anisotropic. Following~\cite{AnEvans2006}, a tangentially biased DF for the isochrone
potential is given by
\begin{equation}
f_{0} (E , L) = p (E) + q (E) L^{2} ,
\label{f0Tan}
\end{equation}
where
\begin{align}
p(E) = & \, \frac{3M(1\!-\!E')^{-4}}{2^{15/2}\pi^3(GMb)^{3/2}} \left( \!(16E'^2 \!-\! 12E' \!+\! 3) \frac{15 \sin^{-1} \!\sqrt{E'}}{\sqrt{1-E'}} \right.
\nonumber
\\
& \, \left. - (45-150E'+144E'^2-208E'^3+64E'^4)\sqrt{E'} \right),
\nonumber
\end{align}
and
\begin{align}
q(E) = & \, \frac{3M(1\!-\!E')^{-5}}{2^{15/2}\pi^3(GMb)^{3/2}} \left( \! (40E'^2 \!-\! 24E' \!+\! 5) \frac{15 \sin^{-1} \!\sqrt{E'}}{\sqrt{1-E'}} \right.
\nonumber
\\
& \!\!\!\!\!\!\!\!\!\!\!\!\! \left. - (75 \!-\! 310E' \!+\! 400E'^2 \!-\! 928E'^3 \!+\! 576E'^4 \!-\! 128E'^5) \sqrt{E'} \right),
\nonumber
\end{align}
and ${ E' \equiv -b E / G M }$. As emphasised in the BL equation~\eqref{eq:Fbar2}, 
we may consider the averaged DF, $\overline{f} (J_{r} , L) = 2 L f_{0}$. In Fig.~\ref{fig:TangentialDiv},
we illustrate the respective contributions of the $\ell=1$ and $\ell=2$ to the diffusion flux,
with and without collective effects. 
\begin{figure*}
\centerline{
\includegraphics[height=.3\hsize]{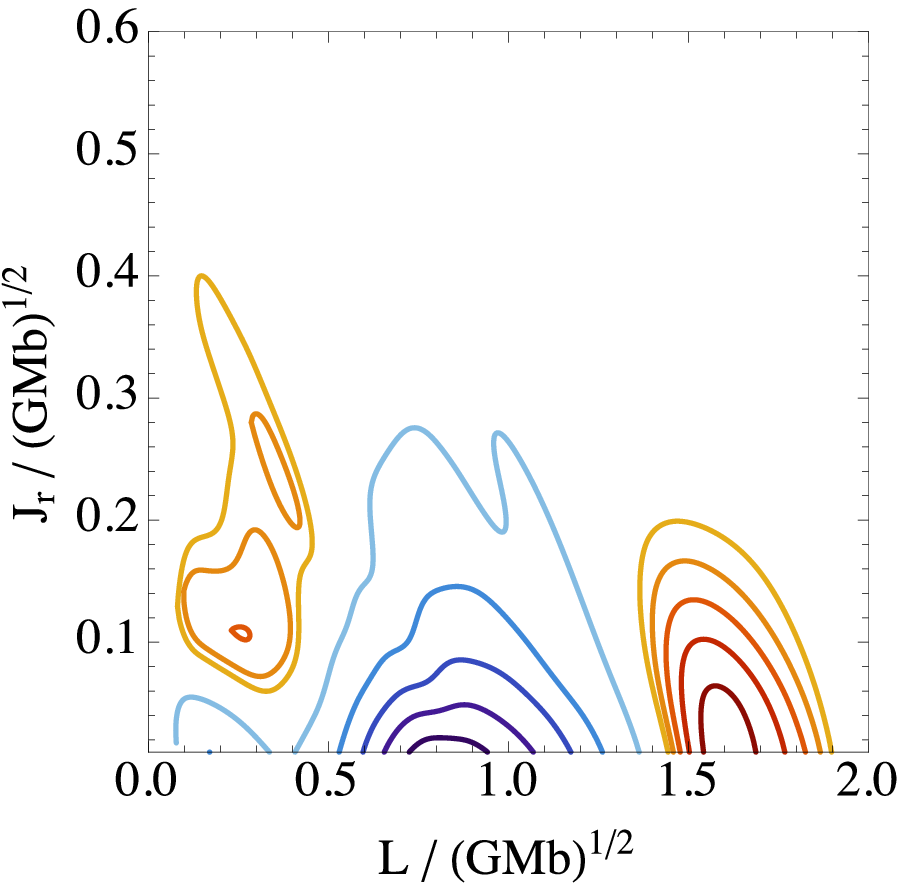}
\includegraphics[height=.3\hsize]{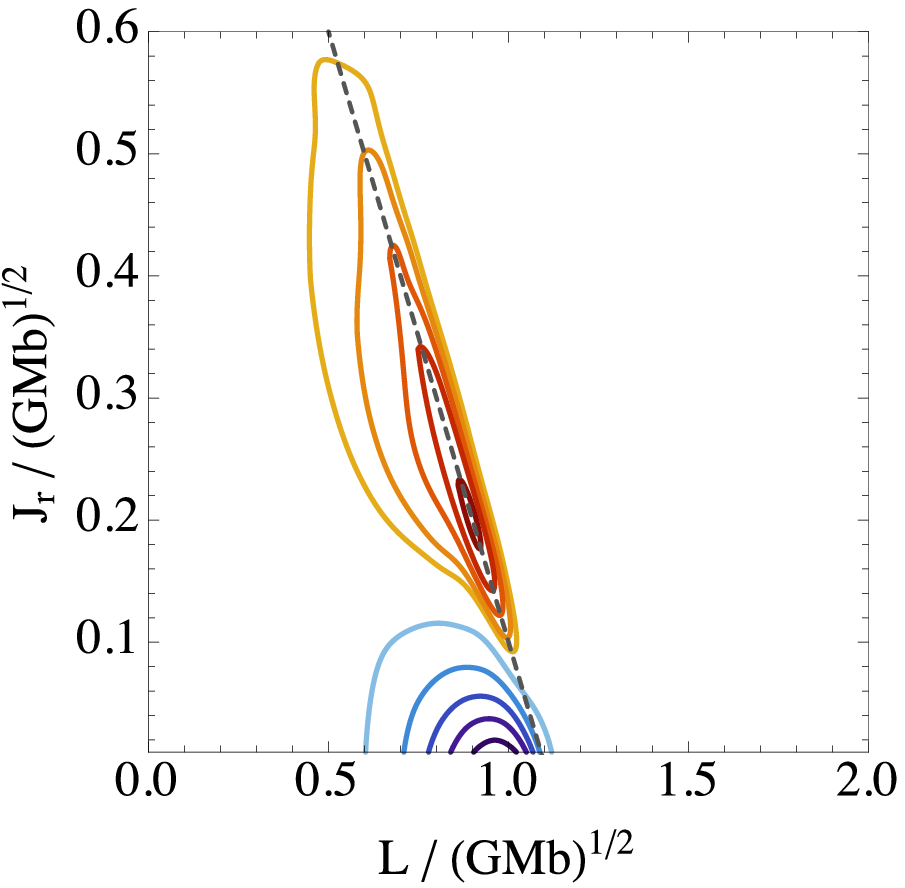}
}
\centerline{
\includegraphics[height=.3\hsize]{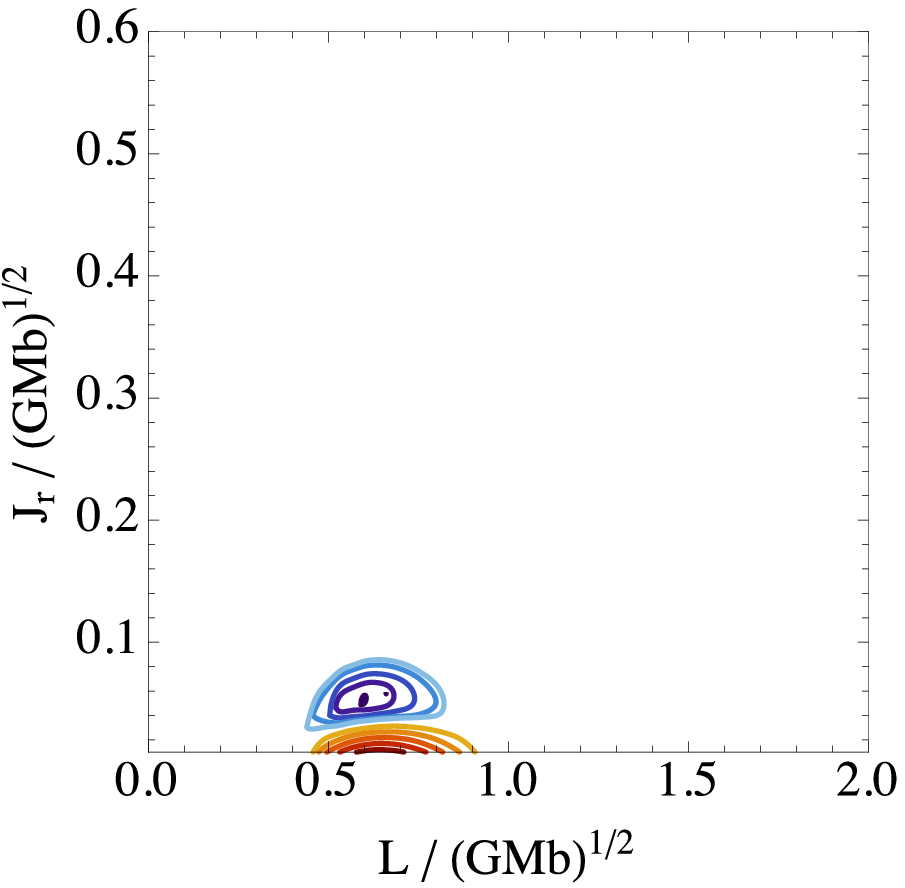}
\includegraphics[height=.3\hsize]{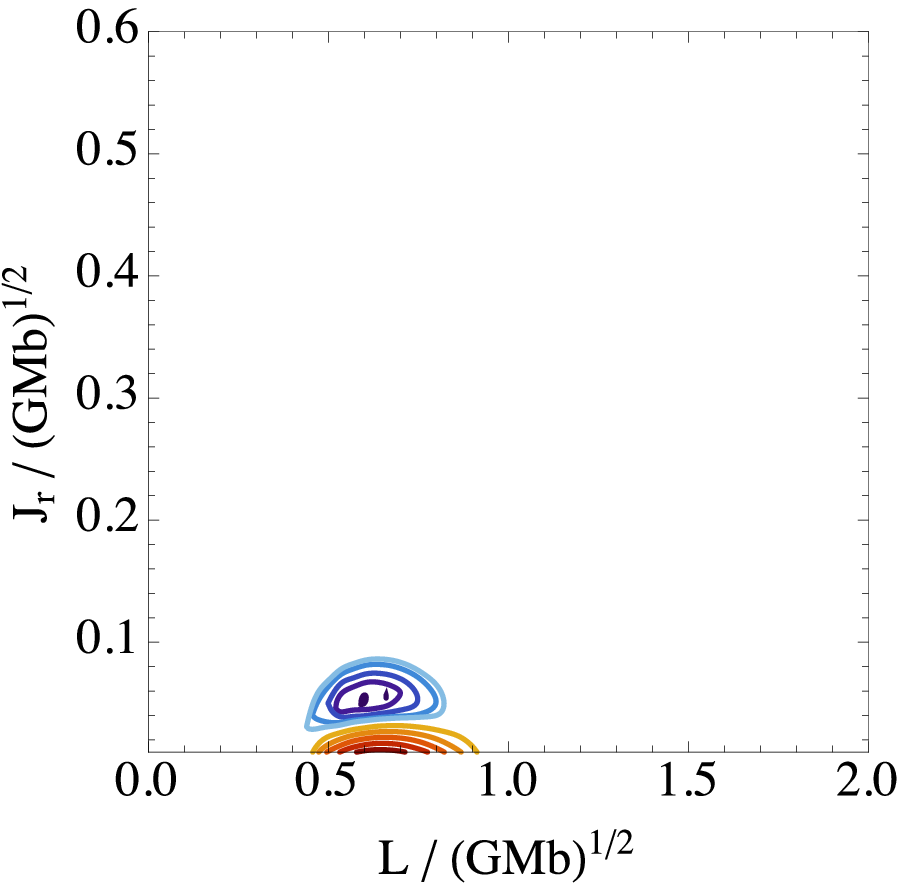}
}
 \caption{The rate of change of the DF ${ \partial \overline{f}/\partial t = - \partial \overline\vF_{\ell}/\partial \widetilde{\mathbf{J}}  }$
 for the tangentially anistropic DF from equation~\eqref{f0Tan}, for $\ell=1$ (top row) and $\ell=2$ (bottom row),
 without collective effects (left column) and with collective effects (right column). The convention used is the same
 as in Fig.~\ref{fig:divFBL}. In the top-right panel, the dashed gray line is aligned with direction $(-1,1)$.
}\label{fig:TangentialDiv}
\end{figure*}

In the first row of Fig.~\ref{fig:TangentialDiv}, we note the importance of collective
effects in amplifying the flux and altering its direction for the $\ell=1$ component, as induced by
the existence of a weakly damped $\ell=1$ mode in these systems~\citep{Weinberg1994}.
In particular, in the top-right panel, the diffusion takes place predominantly along the
direction $(-1,1)$. In the tangentially anisotropic model, stars are being picked up
and moved along the $(-1,1)$ line. Indeed, for perfectly circular orbits, one has
$(-1,1)\cdot\widetilde\vOmega = 0$. As there are numerous almost-circular orbits
in this model, many stars will have a characteristic frequency $\widetilde\vn\cdot\widetilde\vOmega \simeq 0$.
Hence, many pairs of stars will have matching frequencies, allowing them to interact resonantly.
The direction of the flux is then along the the most strongly-amplified resonance vector $\widetilde\vn$,
which in this case is $(-1,1)$. On the contrary, in the bottom row of Fig.~\ref{fig:TangentialDiv},
we note that collective effects do
not boost the amplitude of the $\overline\vF_{2}$ flux,
because the tangential model has so few radial orbits,
i.e. so few orbits prone to undergo a strong self-gravitating amplification via the $\ell=2$ response.

\section{Classical diffusion coefficients}
\label{App:class}

We use classical Spitzer--Chandrasekhar theory to compute the
diffusive  action-space flux. Our strategy is as follows. At each phase of an
orbit of given $(J_r,L)$ we compute the probability  flux and then average
this flux over orbital phase. We obtain the flux at a given location by
transforming the classical velocity-space Fokker-Planck equation to
curvilinear  coordinates for velocity-space that include $J_r$ and $L$.  The requisite tensor calculus is found in e.g.~\citet{Carroll2004}.

We use as our velocity-space coordinates $\vP=(J_r,L,\psi)$,
where $\psi$ is the angle between the tangential component of the velocity
and a fixed direction $\ve_x$ in the plane perpendicular to the current
position vector $\vr$. In terms of the new velocity coordinates the
Cartesian components of velocity are
\begin{align}
(v_r,v_x,v_y)&=\left(\sqrt{2(E-\Phi)-L^2/r^2},{L\over r}\cos\psi,
{L\over r}\sin\psi\right),
\nonumber
\end{align}
where $E=E(J_r,L)$, so
\[
\vL=(0,-L\sin\psi,L\cos\psi).
\]
Given that $\d E=\Omega_r\d J_r+\Omega_\vartheta\d L$ it follows that
\[
\d v_r={\Omega_r\d J_r+(\Omega_\vartheta-L/r^2)\d L\over v_r}.
\]
Hence,
\[\label{eq:ddP}
{\p\vv\over\p\vP}=
\begin{pmatrix}
{\displaystyle{\Omega_r\over v_r}}&{\displaystyle{\Omega_\vartheta-L/r^2\over
v_r}}&0\cr
0&{\displaystyle{\cos\psi\over r}}&-v_t\sin\psi\cr
0&{\displaystyle{\sin\psi\over r}} &v_t\cos\psi
\end{pmatrix}.
\]
To compute the inverse matrix we note that 
in any potential we have 
\[\label{eq:gradL}
{\p L\over\p\vv}={\vL\times\vr\over L}
\quad {\rm and} \quad
{\p\psi\over\p\vv}={r\over L^2}\vL ,
\]
and
in the case of the isochrone there is a simple expression for $\p
J_r/\p\vv$. By applying the operator $\p/\p\vv$ to
\[
J_r={GM\over\sqrt{-2H}}-\fracj12\bigl(L+\sqrt{L^2+4GMb}\bigr) ,
\]
and using $\p H/\p\vv=\vv$ and equation~\eqref{eq:gradL} one easily
shows that
\[\label{eq:gradJr}
{\p J_r\over\p\vv}={\vv-\Omega_\vartheta \vL\times\vr/L\over\Omega_r} .
\]
In matrix notation
\begin{align}\label{eq:dPdv}
{\p P^i\over\p v^j}&=
\begin{pmatrix}
{\displaystyle{v_r\over\Omega_r}}&{\displaystyle{1\over\Omega_r}\left(v_x-{\Omega_\vartheta L_yr\over L}\right)}
&{\displaystyle{1\over\Omega_r}\left(v_y+{\Omega_\vartheta L_xr\over L}\right)}
\cr
0&{L_yr/L}&-L_xr/L\cr
0&L_x r/L^2&L_yr/L^2\cr
\end{pmatrix}\cr
&=
\begin{pmatrix}
{\displaystyle{v_r\over\Omega_r}}&{\displaystyle{1\over\Omega_r}\left(v_x-{\Omega_\vartheta L_yr\over L}\right)}
&{\displaystyle{1\over\Omega_r}\left(v_y+{\Omega_\vartheta L_xr\over L}\right)}
\cr
0&r\cos\psi&r\sin\pi\cr
0&-(r/L)\sin\psi&(r/L)\cos\psi
\end{pmatrix}.
\nonumber
\end{align}
The invariant distance in velocity space is
\begin{align}
\d s^2&=\d v_r^2+
\d v_x^2+\d v_y^2\cr
&={\Omega_r^2\over v_r^2}\,\d J_r^2+
\biggl[{(\Omega_\vartheta-L/r^2)^2\over v_r^2}+{1\over r^2}\biggr]\,\d L^2\cr
&\qquad
+{2\Omega_r\over v_r^2}(\Omega_\vartheta-L/r^2)\d J_r\d L+{L^2\over r^2}\,\d\psi^2
\end{align}
so the non-vanishing components of the  metric are
\begin{align}
g_{JJ}&={\Omega_r^2\over v_r^2},\qquad
g_{LL}=\biggl[{(\Omega_\vartheta-L/r^2)^2\over v_r^2}+{1\over r^2}\biggr],\cr
g_{LJ}&={\Omega_r\over v_r^2}(\Omega_\vartheta-L/r^2),\qquad
g_{\psi\psi}={L^2\over r^2}.
\end{align}
 We will need
\begin{align}\label{eq:surdg}
g&\equiv{\hbox{det}\, g_{ij}}\cr
&={L^2\over r^2}\biggl\{
{\Omega_r^2\over v_r^2}\biggl[{(\Omega_\vartheta-L/r^2)^2\over
v_r^2}+{1\over r^2}\biggr]
-\biggl[{\Omega_r\over v_r^2}(\Omega_\vartheta-L/r^2)\biggr]^2\biggr\}\cr
&={L^2\over r^4}{\Omega_r^2\over v_r^2}.
\end{align}

In Cartesian velocity coordinates the Fokker-Planck equation is
\[\label{eq:descart}
{\p f\over\p t}=-\p_i\left[\exf{\delta v^i}f-\fracj12\p_j(\exf{\delta
v^i\delta v^j}f)\right],
\]
 where the Einstein summation convention is used, $\p_i\equiv{\p/\p v^i}$,
and we use superscripts for the indices of velocity components in preparation
for the introduction of a non-trivial metric. To obtain the corresponding
equation in curvilinear coordinates $P^i$, we have to transform the vector
$\exf{\delta v^i}$ and the tensor $\exf{\delta v^i\delta v^j}$ according to
\begin{align}\label{eq:bigD}
 D^i&\equiv{\p P^i\over\p v^j}\exf{\delta v^j}\cr
 D^{ij}&\equiv{\p P^i\over\p v^m}{\p P^j\over\p
v^n}\exf{\delta v^m\delta v^n}
\end{align}
and to convert the partial derivatives in equation~\eqref{eq:descart} into
covariant derivatives. We exploit the standard results
\begin{align}
\nabla_k A^k&={1\over\surd g}\p_k(\surd gA^k)\cr
\nabla_k B^{kn}&={1\over\surd g}\p_k(\surd gB^{kn})+\Gamma^n_{kj}B^{kj},
\end{align}
to deduce that in a general coordinate system $f$ obeys
\citep{RosenbluthMJ1957}
\begin{align}\label{eq:descartone}
{\p f\over\p t}
&=-{1\over\surd g}\p_i\left\{\surd g\left[D^if
-{1\over2\surd g}\p_j(\surd
gD^{ij}f)-\fracj12\Gamma^i_{jn}D^{jn}f\right]\right\}\cr
&=-{1\over\surd g}\p_i\left\{\surd
g\left[D^i-\fracj12\Gamma^i_{jn}D^{jn}\right]f -\fracj12\p_j(\surd
gD^{ij}f)\right\}.
\end{align}
When this equation is multiplied by the infinitesimal phase-space volume
$\d^3\vx\,\d^3\vv=\surd g\,\d^3\vx\,\d^3\vP$, the left side becomes the rate
of change of the stellar mass in that element. Integrating this over all
variables except $J_r$ and $L$ we obtain the rate of change of stellar mass
with actions within $(\d J_r\,\d L)$ of $(J_r,L)$:
\begin{align}\label{eq:descarttwo}
{\p N(J_r,L)\over\p t}&=\int\d^3\vx\,\d\psi\surd g{\p f\over\p t}\cr
&=-\int\d^3\vx\,\d\psi\,\p_i\biggl\{\surd
g\Bigl[D^i-\fracj12\Gamma^i_{jn}D^{jn}\Bigr]f\cr
&\qquad-\fracj12\p_j(\surd
gD^{ij}f)\biggr\}.
\end{align}
Since $\p_3=\p_\psi$ and we are integrating $\psi$ from $0$ to $2\pi$, the
sum over $i$ can be restricted to $i=1,2$. So, taking the
remaining differentials out of the integral, we can rewrite equation~\eqref{eq:descarttwo} in the form
\[\label{eq:curvy}
{\p N\over\p t}=-(2\pi)^3\sum_{i=1}^2\p_i(\overline F_1^i + \overline F_2^i),
\]
where we have broken the flux in $(J_r,L)$ space into two parts
\begin{align}
\overline F_1^i&\equiv (2\pi)^{-3}f
\int\d^3\vx\,\d\psi\,\surd
g\Bigl[D^i-\fracj12\Gamma^i_{jn}D^{jn}\Bigr]\cr
\overline F_2^i&\equiv-\fracj12(2\pi)^{-3}\p_j\biggl(f\int\d^3\vx\,\d\psi\,\surd
gD^{ij}\biggr).
\end{align}
and the sum over $j$ can be restricted to $j=1,2$.

We obtain an alternative expression for $N(\widetilde\vJ)$ by integrating $f$
over $\vtheta$ and $L_z$ and recalling the definition~\eqref{eq:def_fbar}:
\[\label{eq:descartthree}
N(\widetilde\vJ)=(2\pi)^32Lf(\widetilde\vJ)=(2\pi)^3\overline{f}(\widetilde\vJ).
\]
Eliminating $N$ between equations~\eqref{eq:descarttwo} and~\eqref{eq:descartthree} we have
\[
{\p\overline{f}\over\p t}=-\sum_{i=1}^2\p_i(\overline F_1^i + \overline F_2^i).
\]

We use spherical polar coordinates to execute the spatial integrals in
the definition of $\overline\vF_\alpha$. Since
the system is spherically symmetric, the angular integrals produce a factor
$4\pi$. The radial integrals run between the peri- and apo-centre implied by
$(J_r,L)$ and are conveniently turned into integrals over radial phase using
$\d r=v_r\d\theta_r/\Omega_r$. Using $\surd g=L\Omega_r/(r^2v_r)$ from
equation~\eqref{eq:surdg}, we then find
\begin{align}\label{eq:FtildeChandra}
\overline F_1^i&=(2\pi)^{-2} \overline{f}
\int\d\theta_r\,\d\psi\,\Bigl[D^i-\fracj12\Gamma^i_{jn}D^{jn}\Bigr]\cr
\overline F_2^i&=-\fracj12(2\pi)^{-2}\p_j\biggl(\overline{f}\int\d\theta_r\,\d\psi\,D^{ij}\biggr).
\end{align}

From~\cite{Binney2008} \S7.4.4 we have
\begin{align}
\exf{\delta\vv}&=4\pi G(m+m')\log\Lambda{\p h\over\p\vv}\cr
\exf{\delta v_i\delta v_j}&=4\pi Gm'\log\Lambda{\p^2 g\over\p
v_i\p v_j},
\end{align}
 where $\log\Lambda$ is the Coulomb logarithm (Section~\ref{sec:Coulomb}) and
the Rosenbluth potentials are
\begin{align}
h(\vv)&\equiv Gm\int\d^3\vv'\,{n(\vv')\over|\vv-\vv'|}\cr
g(\vv)&\equiv Gm\int\d^3\vv'\,n(\vv')|\vv-\vv'|.
\end{align}
Here $\int\d^3\vv\,n(\vv)$ is the number density of stars in real space.
If we restrict ourselves to an ergodic DF, so
\[
n(\vv')=n(|\vv'|)=f(H)/m
\]
 the
Rosenbluth potentials only depend on the magnitude of $\vv$. Integrating over
all directions of $\vv'$ we obtain
\begin{align}
h(v)&=4\pi G\biggl\{{1\over v}\int_0^{v}\d
v'\,v^{\prime2}f(v')+\int_v^\infty\d v'\,v'f(v')\biggr\}\cr
g(v)&={4\pi G\over3}\biggl\{\int_0^{v}\d v'\,f(v'){v^{\prime2}\over
v}(v^{\prime2}+3v^2)\cr
&\qquad+\int_{v}^\infty\d v'\,f(v')v'(v^2+3v^{\prime2})\biggr\}.
\end{align}
Hence the averages over field stars are
\begin{align}\label{eq:expofv}
\exf{\delta\vv}&=4\pi G(m+m')\log\Lambda{\d h\over\d v}{\vv\over v}\cr
&\equiv h_1(v)\vv\cr
\exf{\delta v_i\delta v_j}&=4\pi Gm'\log\Lambda\biggl\{{\d^2 g\over\d
v^2}{v_iv_j\over v^2}+{1\over v}{\d g\over\d
v}\biggl(\delta_{ij}-{v_iv_j\over v^2}\biggr)\biggr\}\cr
&\equiv[g_2(v)-g_1(v)]{v^iv^j\over v^2}+g_1(v)\delta_{ij},
\end{align}
 where 
\begin{align}
h_1(v)&\equiv4\pi G(m+m'){\log\Lambda\over v}{\d h\over\d v}\cr
g_1(v)&\equiv4\pi Gm'\log\Lambda{1\over v}{\d g\over\d v}\cr
g_2(v)&\equiv4\pi Gm'\log\Lambda{\d^2 g\over\d v^2}.
\end{align}
and the required derivatives are
\begin{align}
\!\!\!\!\!\!\!\!\!\!\! {\d h\over\d v}&=-{4\pi G\over v^2}\int_0^v\d v'\,v^{\prime2}f(v')\cr
\!\!\!\!\!\!\!\!\!\!\! {\d g\over\d v}&={4\pi G\over3}\biggl\{\int_0^v\d
v'\,v^{\prime2}f(v')\left(3-{v^{\prime2}\over v^2}\right)\cr
&\qquad+2v\int_v^\infty\d v'\,v'f(v')\biggr\}\cr
\!\!\!\!\!\!\!\!\!\!\! {\d^2g\over\d v^2}&={8\pi G\over3}\biggl\{{1\over v^3}\int_0^v\!\!\d
v'\,v^{\prime4}f(v')\!+\!\int_v^\infty\!\!\d v'\,v'f(v')\biggr\} .
\end{align}

When we insert equations~\eqref{eq:expofv} into equations~\eqref{eq:bigD} we
encounter several instances of 
\[\label{eq:vdotgradP}
\vv\cdot{\p P^i\over\p\vv}=
\left({v^2-\Omega_\vartheta L\over\Omega_r},L,0\right)
\]
from which we construct
\begin{align}
\vM&\equiv\vv\cdot{\p P^i\over\p\vv}\vv\cdot{\p P^j\over\p\vv}\cr
&=\begin{pmatrix}
{\displaystyle{(v^2-\Omega_\vartheta L)^2\over\Omega_r^2}}&{\displaystyle{v^2-\Omega_\vartheta
L\over\Omega_r}L}&0\cr
{\displaystyle{v^2-\Omega_\vartheta
L\over\Omega_r}L}&L^2&0\cr
0&0&0
\end{pmatrix}.
\end{align}
We also encounter
\begin{align}
&\vN\equiv{\p P^i\over\p\vv}\cdot{\p P^j\over\p\vv}\cr
&=\begin{pmatrix}
{\displaystyle {v^2-2\Omega_\vartheta
L+\Omega_\vartheta^2r^2\over\Omega_r^2}}&{\displaystyle{L-\Omega_\vartheta r^2\over
\Omega_r}}&0\cr
{\displaystyle{L-\Omega_\vartheta r^2\over
\Omega_r}}&r^2&0\cr
0&0&{\displaystyle {r^2\over L^2}}\cr
\end{pmatrix}.
\end{align}
 With these definitions  the $(J_r,L,\psi)$ diffusion coefficients are
\begin{align}\label{eq:bigDtwo}
 D^i&=h_1(v)\left({v^2-\Omega_\vartheta L\over\Omega_r},L,0\right)\cr
 D^{ij}&=
\Bigl[{g_2(v)-g_1(v)\over v^2}\vM
+g_1(v)\vN\Bigr].
\end{align}

Before the right-hand side of equation~\eqref{eq:descartone} can be evaluated, it
remains only to compute the Christoffel symbols that appear in it. This is
most easily done by exploiting the fact that they vanish in the Cartesian
system, so in the $(J_r,L,\psi)$ system they are given by
\[
\Gamma^i_{jn}={\p P^i\over\p v^k}{\p^2 v^k\over\p P^j\p P^n}
\] 
The matrix of first derivatives is given by equations~\eqref{eq:gradL} to~\eqref{eq:gradJr}.
Differentiating the matrix~\eqref{eq:ddP}, we find the
second derivatives are
\[
{\p^2\vv\over\p J_r\p\vP}=
\begin{pmatrix}
{\displaystyle{1\over v_r} {\p\Omega_r\over\p J_r}-{\Omega_r^2\over v_r^3}} &0&0\cr
{\displaystyle{1\over v_r}{\p\Omega_\vartheta\over\p
J_r}-(\Omega_\vartheta-L/r^2){\Omega_r\over v_r^3}}&0&0\cr
0&0&0
\end{pmatrix}^T
\]
\begin{align}
{\p^2\vv\over\p L\p\vP}=
\begin{pmatrix}
{\displaystyle{1\over v_r}{\p\Omega_r\over\p L}
-(\Omega_\vartheta-L/r^2){\Omega_r\over v_r^3}}&0&0\cr
{\displaystyle{1\over v_r}\biggl({\p\Omega_\vartheta\over\p
L}-{1\over r^2}\biggr)-{(\Omega_\vartheta-L/r^2)^2\over v_r^3}}&0&0\cr
0&{\displaystyle-{\sin\psi\over r}}&{\displaystyle{\cos\psi\over r}}
\end{pmatrix}^T \nonumber
\end{align}
\[
{\p^2\vv\over\p\psi\p\vP}=
\begin{pmatrix}
0&0&0\cr
0&{\displaystyle-{\sin\psi\over r}}&-v_t\cos\psi\cr
0&{\displaystyle{\cos\psi\over r}}&-v_t\sin\psi
\end{pmatrix} .
\]
The matrix formed by a Christoffel with a given lower left index is obtained by
premultiplying each of these matrices by $\p\vP/\p\vv$.

\subsection{The Coulomb logarithm}
\label{sec:Coulomb}

The isochrone sphere provides natural units $M$ and $b$ of mass and length, and
from them we derive the  natural unit of time
\[\label{eq:defsTI}
T_{\rm I}=\sqrt{{b^3\over GM}}.
\]

The Coulomb logarithm is traditionally defined to be
\[
\log\Lambda=\ln\left({b_{\rm max}\sigma^2\over2G\mu}\right),
\]
 where $b_{\rm max}$ is the largest impact parameter we should consider when
one star encounters another and $\sigma$ is the typical speed of stars. For
the isochrone the
virial theorem yields
\[
\sigma^2={GM\over b}\left(\fracj14\pi-\fracj23\right).
\]
Hence 
\[
\log\Lambda=\ln\left(N\left(\fracj14\pi-\fracj23\right){b_{\rm max}\over 2b}\right).
\]
 In the core one might reasonably argue that $b_{\rm max}$ is of order $b$,
while at some radius $r$ outside the core $b_{\rm max}$ could be a significant fraction of
$r$. We will assume $b_{\rm max}\sim b$ so
$\log\Lambda=\ln(0.059N)\simeq8.7$ for $N=10^5$.

\subsection{Numerical details} \label{sec:numericaldetails}

Numerical evaluation of the diffusive flux is quite delicate because
components of the Christoffel symbols diverge in the limits $L\to0$ and
$v_r\to0$. The components that diverge as $v_r\to0$ cancel when $\Gamma$
is multiplied into $D$. The divergence at $L\to0$ permits $\overline\vF_1$,
the definition of which (equation~\eqref{eq:FtildeChandra}) has a leading factor
$L$, to remain finite as $L\to0$. Fig.~\ref{fig:killit} shows why it is
important that $\overline\vF_1\ne0$ at $L=0$ by plotting $\overline\vF_1$ and
$\overline\vF_2$ separately.
\begin{figure}
\centerline{\includegraphics[width=.8\hsize]{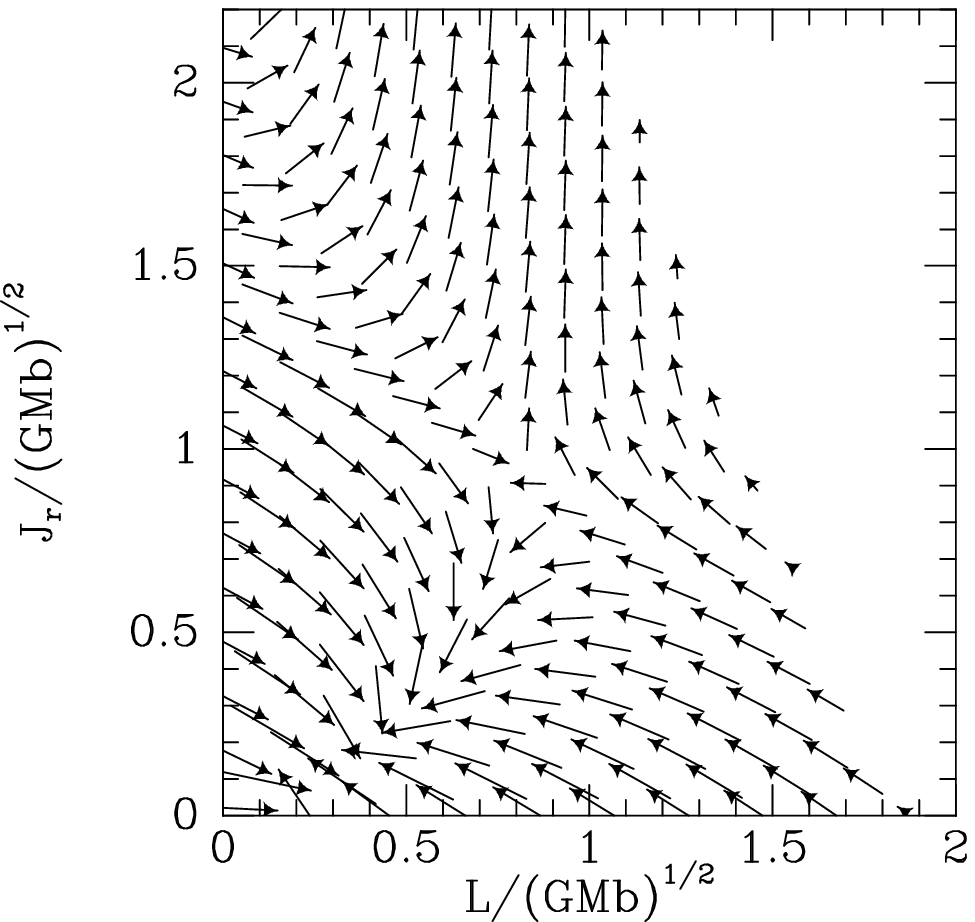}}
\centerline{\includegraphics[width=.8\hsize]{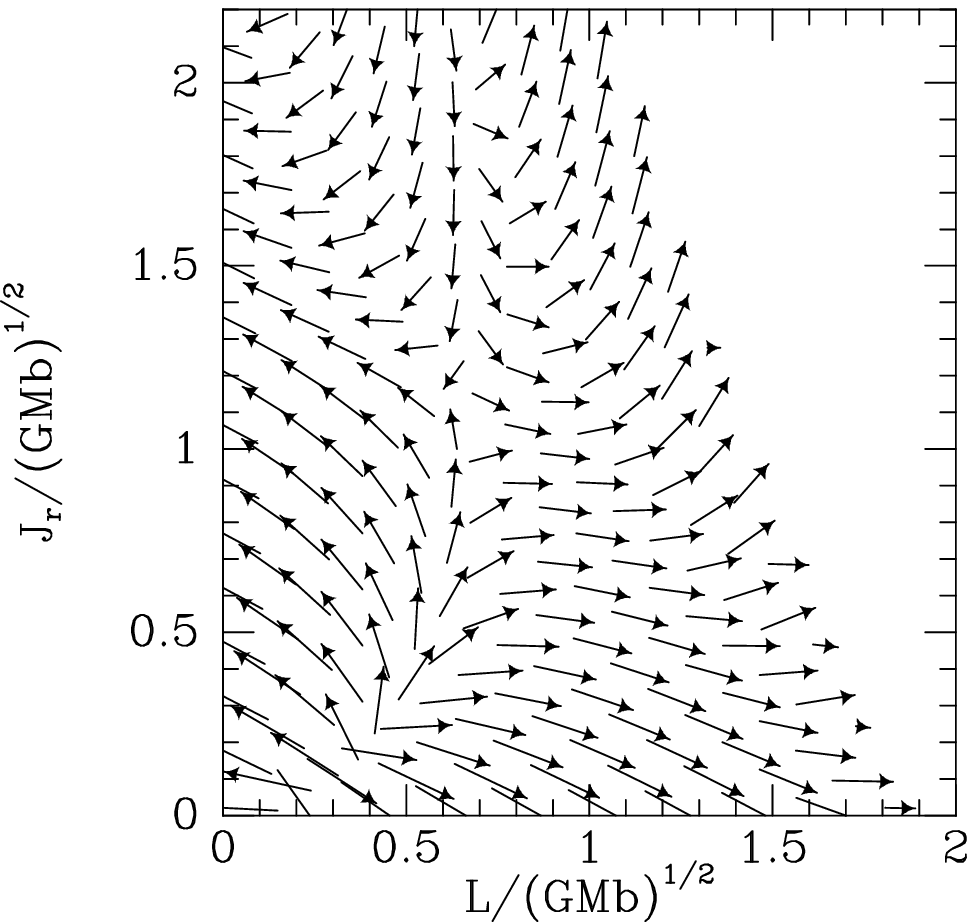}}
 \caption{Lower panel: the contribution from the divergence
$\overline F_2^i=-\fracj12\p_j(Lf\int\d\theta_r\d\psi\,D^{ij})$ to the flow in $LJ_r$ space. Upper
panel: the complementing contribution to the flow, $\overline \vF_1$.}
\label{fig:killit}
\end{figure}
Both $\overline \vF_1$ and $\overline \vF_2$ indicate unphysical fluxes of stars across
the boundaries $L=0$ and $J_r=0$. Physically the total flux, $\overline \vF_1 + \overline \vF_2$,
across these boundaries has to vanish, and it does so because the components
of $\overline \vF_1$ and $\overline \vF_2$ perpendicular to the boundaries cancel to a good
approximation. Any error in the numerical scheme is liable to spoil this
cancellation, so the general structure of the flow shown in
Fig.~\ref{fig:FJflow} gives confidence in the soundness of the numerics.

However, along the bottom of Fig.~\ref{fig:FJflow} the arrows suggest that
stars are flowing into $LJ_r$ space across the $L$ axis.
Fig.~\ref{fig:FJsmall} shows this region for the case $R_\ra=4.2b$ at higher
resolution, for $\overline \vF_1 + \overline \vF_2$ (top) and for
$\overline \vF_1$ and $\overline \vF_2$ below.
\begin{figure}
\centerline{\includegraphics[width=.9\hsize]{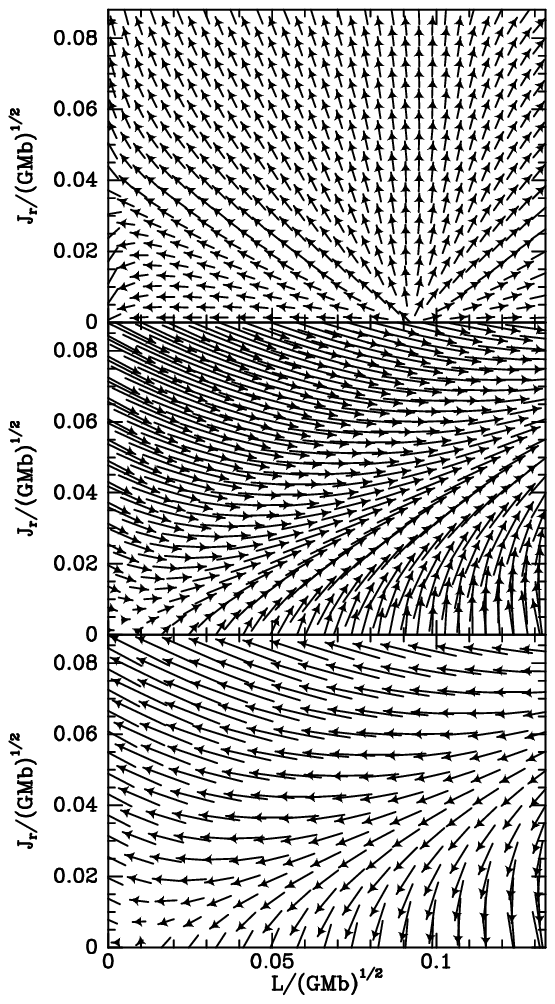}}
\caption{Top: overall flow $\overline\vF_1+\overline\vF_2$ in the region of the singular point of the model
with $R=4.2b$. Centre:
the flow $\overline\vF_1$ driven by dynamical friction. Bottom: the flow
$\overline\vF_2$ that depends on  grad $f$.}\label{fig:FJsmall}
\end{figure}
We see that the net flow is
away from a singular point at $(L,J_r)\simeq(0.095,0)\sqrt{GMb}$ while neither
$\overline \vF_1$ nor $\overline\vF_2$ has such a singular point and that
away from the singular point the net flow is parallel to the $L$ axis because
the vertical components of $\overline\vF_1$ and $\overline\vF_2$ precisely
cancel. The singular point in the net flow is almost certainly a stagnation
point, so by virtue of remarkable cancellations no stars pass over either
axis.

\end{document}